\documentclass[a4paper,fleqn,usenatbib]{mnras}

\usepackage{newtxtext,newtxmath}

\usepackage[T1]{fontenc}
\usepackage{ae,aecompl}

\usepackage{graphicx}	
\usepackage{amsmath}	
\usepackage{amssymb}	
\usepackage{multicol}        
\usepackage{longtable}
\usepackage{pdflscape}	
\usepackage{hyperref}

\title[LOSS Photometry of 93 SNe~Ia]{Lick Observatory Supernova Search Follow-Up Program: Photometry Data Release of 93 Type Ia Supernovae}

\author[B. E. Stahl et al.]
{Benjamin E. Stahl,$^{1,2}$\thanks{E-mail: benjamin\_stahl@berkeley.edu}\thanks{Marc J. Staley Graduate Fellow}
WeiKang Zheng,$^{1}$
Thomas de Jaeger,$^{1}$\thanks{Bengier Postdoctoral Fellow}
Alexei V. Filippenko,$^{1,3}$
\newauthor
Andrew Bigley,$^{1}$
Kyle Blanchard,$^{1}$
Peter K. Blanchard,$^{4}$
Thomas G. Brink,$^{1}$
\newauthor
Samantha K. Cargill,$^{1}$
Chadwick Casper,$^{1}$
Sanyum Channa,$^{1}$
Byung Yun Choi,$^{1}$
\newauthor
Nick Choksi,$^{1}$
Jason Chu,$^{5}$
Kelsey I. Clubb,$^{1}$
Daniel P. Cohen,$^{6}$
Michael Ellison,$^{1}$
\newauthor
Edward Falcon,$^{1}$
Pegah Fazeli,$^{1}$
Kiera Fuller,$^{1,6}$
Mohan Ganeshalingam,$^{7}$
\newauthor
Elinor L. Gates,$^{8}$
Carolina Gould,$^{1}$
Goni Halevi,$^{1,9}$
Kevin T. Hayakawa,$^{8}$
\newauthor
Julia Hestenes,$^{1}$
Benjamin T. Jeffers,$^{1}$
Niels Joubert,$^{10}$
Michael T. Kandrashoff,$^{1}$
\newauthor
Minkyu Kim,$^{1}$
Haejung Kim,$^{1}$
Michelle E. Kislak,$^{1,11}$
Io Kleiser,$^{12}$
Jason J. Kong,$^{1}$
\newauthor
Maxime de Kouchkovsky,$^{1}$
Daniel Krishnan,$^{1}$
Sahana Kumar,$^{1,13}$
Joel Leja,$^{4}$
\newauthor
Erin J. Leonard,$^{1,14}$
Gary Z. Li,$^{15}$
Weidong Li,$^{1}$\thanks{Deceased 2011 December 12}
Philip Lu,$^{2,6}$
Michelle N. Mason,$^{16}$
\newauthor
Jeffrey Molloy,$^{1}$
Kenia Pina,$^{1}$
Jacob Rex,$^{1}$
Timothy W. Ross,$^{1}$
Samantha Stegman,$^{1}$
\newauthor
Kevin Tang,$^{1}$
Patrick Thrasher,$^{1}$
Xianggao Wang,$^{17}$
Andrew Wilkins,$^{1}$
\newauthor
Heechan Yuk,$^{18}$
Sameen Yunus,$^{1}$
and Keto Zhang$^{1}$
\\
\\
$^{1}$Department of Astronomy, University of California, Berkeley, CA 94720-3411, USA\\
$^{2}$Department of Physics, University of California, Berkeley, CA 94720-7300, USA\\
$^{3}$Miller Senior Fellow, Miller Institute for Basic Research in Science, University of California, Berkeley, CA 94720, USA\\
$^{4}$Harvard-Smithsonian Center for Astrophysics, 60 Garden Street, Cambridge, MA 02138, USA\\
$^{5}$Gemini Observatory, 670 N. Aohoku Place, Hilo, HI, USA\\
$^{6}$Department of Physics and Astronomy, University of California, Los Angeles, CA 90095, USA\\
$^{7}$Lawrence Berkeley National Laboratory, 1 Cyclotron Rd, Berkeley, CA 94720, USA\\
$^{8}$Lick Observatory, P.O. Box 85, Mount Hamilton, CA 95140, USA\\
$^{9}$ Department of Astrophysical Sciences, Princeton University, 4 Ivy Lane, Princeton, NJ 08544, USA\\
$^{10}$Department of Computer Science, Stanford University, 353 Serra Mall, Stanford, CA 94305, USA\\
$^{11}$Netflix, Inc., 100 Winchester Cir, Los Gatos, CA 95032, USA\\
$^{12}$NASA Jet Propulsion Laboratory, 4800 Oak Grove Dr, Pasadena, CA 91109, USA\\
$^{13}$Department of Physics, Florida State University, Tallahassee, FL 32306, USA\\
$^{14}$Department of Earth, Planetary, and Space Sciences, University of California, Los Angeles, 90025, USA\\
$^{15}$Department of Mechanical and Aerospace Engineering, University of California, Los Angeles, 90025, USA\\
$^{16}$Department of Physics and Astronomy, University of Wyoming, 1000 E. University, Dept 3905, Laramie, WY 82071, USA\\
$^{17}$Department of Physics, Guangxi University, Nanning 530004, China\\
$^{18}$Department of Physics and Astronomy, University of Oklahoma, 440 W. Brooks St., Norman, OK 73019, USA\\
\vspace{5cm}
}

\date{Accepted XXX. Received YYY; in original form ZZZ}

\pubyear{2019}

\begin{document}

\label{firstpage}
\pagerange{\pageref{firstpage}--\pageref{lastpage}}
\maketitle

\begin{abstract}
We present \emph{BVRI} and unfiltered light curves of 93 Type Ia supernovae (SNe~Ia) from the Lick Observatory Supernova Search (LOSS) follow-up program conducted between 2005 and 2018. Our sample consists of 78 spectroscopically normal SNe~Ia, with the remainder divided between distinct subclasses (three SN~1991bg-like, three SN~1991T-like, four SNe~Iax, two peculiar, and three super-Chandrasekhar events), and has a median redshift of 0.0192. The SNe in our sample have a median coverage of 16 photometric epochs at a cadence of 5.4 days, and the median first observed epoch is $\sim 4.6$ days before maximum \emph{B}-band light. We describe how the SNe in our sample are discovered, observed, and processed, and we compare the results from our newly developed automated photometry pipeline to those from the previous processing pipeline used by LOSS. After investigating potential biases, we derive a final systematic uncertainty of 0.03~mag in \emph{BVRI} for our dataset. We perform an analysis of our light curves with particular focus on using template fitting to measure the parameters that are useful in standardising SNe~Ia as distance indicators. All of the data are available to the community, and we encourage future studies to incorporate our light curves in their analyses.
\end{abstract}




\section{Introduction}
\label{sec:introduction}

Type Ia supernovae (SNe~Ia) are objects of tremendous intrigue and consequence in astronomy. As individual events, SNe~Ia --- especially those at the extremes of what has been previously observed \citep[e.g.,][]{91T,91bg,Iax} --- present interesting case studies of high-energy, transient phenomena. Collectively, SNe~Ia are prized as ``cosmic lighthouses'' with luminosities of several billion Suns, only a factor of 2--3 lower than an $L^*$ host galaxy of $\sim 10^{10}\,L_\odot$. The temporal evolution of the luminosity of a SN~Ia, which is powered largely by the radioactive decay chain $^{56}{\rm Ni} \to$ $^{56}{\rm Co} \to$ $^{56}{\rm Fe}$, is codified by light curves (typically in several broadband filters). With some variation between filters, a SN~Ia light curve peaks at a value determined primarily by the mass of $^{56}{\rm Ni}$ produced and then declines at a rate influenced by its spectroscopic/colour evolution \citep{2007ApJ...656..661K}. With the advent of empirical relationships between observables (specifically, the rate of decline) and peak luminosity~\citep[e.g.,][]{Phillips93,Riess96,Jha07,Zheng18}, SNe~Ia have become immensely valuable as cosmological distance indicators. Indeed, observations of nearby and distant SNe~Ia led to the discovery of the accelerating expansion of the Universe and dark energy \citep{Riess98,Perlmutter99}, and they continue to provide precise measurements of the Hubble constant \citep{Riess16,Riess19}.

The aforementioned light-curve ``width-luminosity'' relations form the basis for the use of SNe~Ia as cosmological distance indicators. To further refine these relationships as well as understand their limitations, extensive datasets of high-precision light curves are required. At low redshift, multiple groups have answered the call, including the Cal\'an/Tololo Supernova Survey with \emph{BVRI} light curves of 29 SNe~Ia \citep{Hamuy96}, the Harvard-Smithsonian Center for Astrophysics (CfA) Supernova Group with $> 300$ multiband light curves spread over four data releases \citep[][henceforth CfA1-4, respectively]{CfA1,CfA2,CfA3,CfA4}, the Carnegie Supernova Project (CSP) with $>100$ multiband light curves \citep[][henceforth CSP1, CSP1a, CSP2, and CSP3, respectively]{CSP1,CSP1a,CSP2,CSP3}, and our own Lick Observatory Supernova Search (LOSS) follow-up program with \emph{BVRI} light curves of 165 SNe~Ia \citep[][henceforth G10]{Ganeshalingam2010}. More recently, the Foundation Supernova Survey has published its first data release of 225 low redshift SN~Ia light curves derived from Pan-STARRS photometry \citep{Foundation}. Despite these extensive campaigns, there exist many more well-observed light curves for high redshift ($z \gtrsim 0.1$) SNe~Ia than for those at low redshift \citep{Betoule14}. As low-redshift SNe~Ia are used to calibrate their high-redshift counterparts, a larger low-redshift sample will be useful for further improving width-luminosity relations, gauging systematic errors arising from the conversion of instrumental magnitudes to a uniform photometric system, and for investigating evolutionary effects over large timescales.

The LOSS follow-up program has been in continuous operation for over 20 years. The result is an extensive database of SN~Ia photometry from images obtained with the 0.76~m Katzman Automatic Imaging Telescope (KAIT) and the 1~m Nickel telescope, both located at Lick Observatory. G10 released SN~Ia light curves from the first 10 years of the LOSS follow-up campaign, and in this paper we publish the corresponding dataset for the following 10 years (2009--2018). We also include several earlier SNe~Ia that were omitted from the first publication. In aggregate, our dataset includes \emph{BVRI} light curves of 93 SNe~Ia with a typical cadence of $\sim5.4$ days drawn from a total of 21,441 images.

Our dataset overlaps with those of CfA3, CfA4, and CSP3. In particular, we share 7 SNe with CfA3 and 16 with CfA4; however, we expect the upcoming CfA5 release to have considerable overlap with ours, as it will be derived from observations over a similar temporal range. With regard to CSP3, we have 16 SNe in common. Accounting for overlaps, 28 SNe in our sample have been covered by at least one of these surveys, thus leaving 65 unique SNe in our sample.

The remainder of this paper is organised in the following manner. Section~\ref{sec:observations} details our data acquisition, including how our SNe are discovered and which facilities are employed to observe them. In Section~\ref{sec:data-reduction} we discuss our data-reduction procedure, with particular emphasis placed on our automated photometry pipeline. Section~\ref{sec:results} presents our results, including comparisons with those in the literature that were derived from the same KAIT and Nickel images, when such an overlap exists. We derive and discuss the properties of our light curves in Section~\ref{sec:discussion}, and our conclusions are given in Section~\ref{sec:conclusion}.

\section{Observations}
\label{sec:observations}

\subsection{Discovery}
\label{ssec:discovery}

Many of the SNe~Ia presented here were discovered and monitored by LOSS using the robotic KAIT \citep[][see G10 for remarks on SN~Ia discovery with LOSS]{Li00,Filippenko01}. We note that the LOSS search strategy was modified in early 2011 to monitor fewer galaxies at a more rapid cadence, thus shifting focus to identifying very young SNe in nearby galaxies \citep[e.g.,][]{2012cg}. Consequently, the proportion of our sample discovered by LOSS is less than in that presented by G10. Those SNe in our sample that were not discovered with KAIT were sourced from announcements by other groups in the SN community, primarily in the form of notices from the Central Bureau of Electronic Telegrams (CBETs) and the International Astronomical Union Circulars (IAUCs). Whenever possible and needed, we spectroscopically classify and monitor newly discovered SNe~Ia with the Kast double spectrograph \citep{Kast} on the 3~m Shane telescope at Lick Observatory. Discovery and classification references are provided for each SN in our sample in Table~\ref{tab:sample-information}.

While the focus in this paper is on SNe~Ia, we have also built up a collection of images containing SNe~II and SNe~Ib/c \citep[see][for a discussion of SN spectroscopic classification]{Filippenko97}. These additional datasets have been processed by our automated photometry pipeline and will be made publicly available pending analyses (T. de Jaeger et al. 2019, in prep., \& W. Zheng et al. 2019, in prep.; for the SN~II and SN~Ib/c datasets, respectively).

\subsection{Telescopes}
\label{ssec:telescopes}

The images from which our dataset is derived from were collected using the 0.76~m KAIT ($\sim86\%$ of the total) and the 1~m Nickel telescope ($\sim14\%$ of the total), both of which are located at Lick Observatory on Mount Hamilton near San Jose, CA. The seeing at this location averages $\sim2^{\prime \prime}$, with some variation based on the season.

KAIT is a Ritchey-Chr\'etien telescope with a primary mirror focal ratio of $f/8.2$. Between 2001 September~11 and 2007 May~12 the CCD used by KAIT was an Apogee chip with $512 \times 512$ pixels, and henceforth it has been a Finger Lakes Instrument (FLI) camera with the same number of pixels. We refer to these as KAIT3 and KAIT4, respectively\footnote{G10 use KAIT1 and KAIT2 for earlier CCD/filter combinations. Our use of KAIT3 and KAIT4 is consistent with theirs.}. Both CCDs have a scale of $0\overset{\prime \prime}{.}8$~pixel$^{-1}$, yielding a field of view of $6\overset{\prime}{.}7 \times 6\overset{\prime}{.}7$. As a fully robotic telescope, KAIT follows an automated nightly procedure to acquire data. Observations of a target are initiated by submitting a request file containing its coordinates as well as those of a guide star. A master scheduling program then determines when to perform the observations with minimal disruption to KAIT's SN search observations. Under standard conditions we use an exposure time of 1--6 min in \emph{B} and 1--5 min in each of \emph{VRI}. 

The 1~m Nickel is also a Ritchey-Chr\'etien telescope, but with a primary mirror focal ratio of $f/5.3$. Since 2001 April~3 its CCD has been a thinned, Loral, $2048 \times 2048$~pixel chip located at the $f/17$ Cassegrain focus of the telescope. With a scale of $0\overset{\prime \prime}{.}184$ pixel$^{-1}$, the field of view is $6\overset{\prime}{.}3 \times 6\overset{\prime}{.}3$. In March of 2009 the filter set was replaced --- we refer to the period before as Nickel1\footnote{Our Nickel1 is referred to as Nickel by G10.} and after as Nickel2. Pixels are binned by a factor of two to reduce readout time. Since 2006, most of our Nickel observations have been performed remotely from the University of California, Berkeley campus. Our observing campaign with Nickel is focused on monitoring more distant SNe and supplementing (particularly at late times) data taken with KAIT. Under standard conditions, we use exposure times similar to those for KAIT.

\begin{figure}
	\includegraphics[width=\columnwidth]{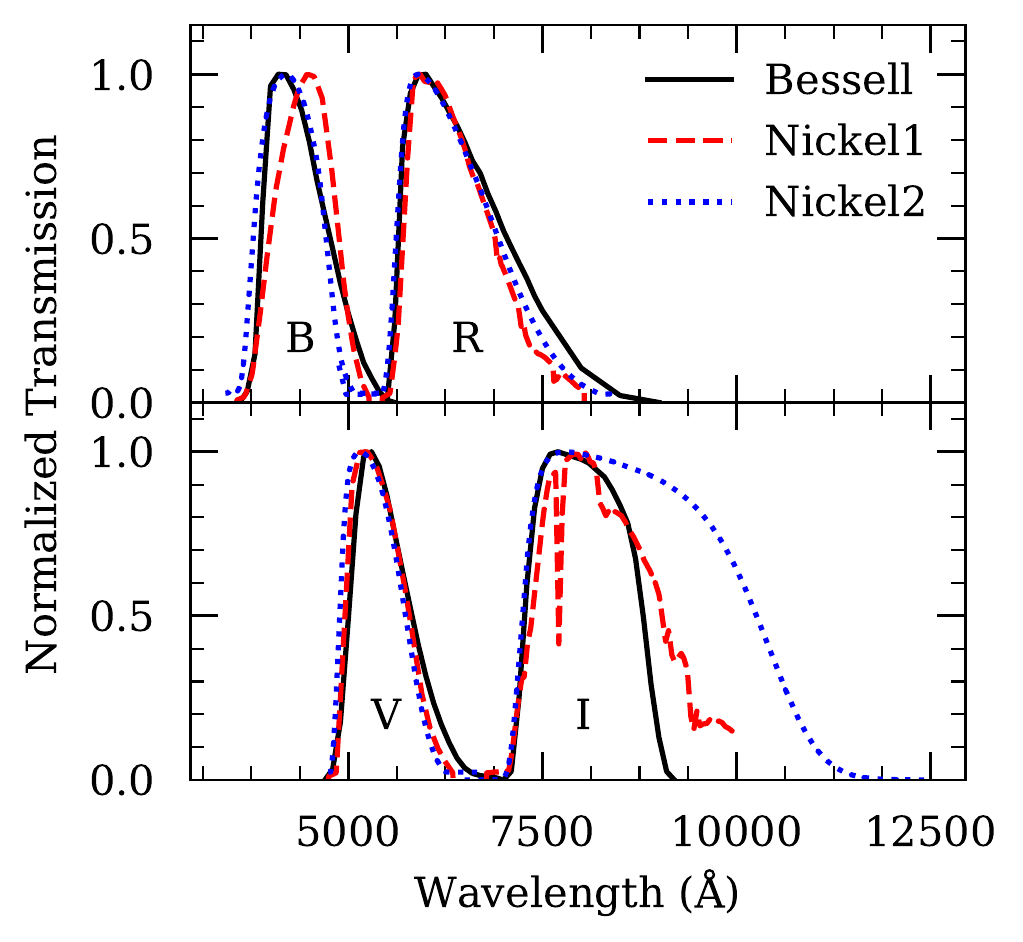}
	\caption{Transmission curves for the two Nickel 1~m configurations covered by our dataset compared with standard \citet{Bessell} \emph{BVRI} curves.\label{fig:transmission}}
\end{figure}

In Figure~\ref{fig:transmission} we compare the standard throughput curves of \citet{Bessell} to those of the two Nickel 1~m configurations covered by our dataset (G10 show the analogous curves for KAIT3 and KAIT4). We find good agreement between both Nickel1 and Nickel2 filter responses in the \emph{VR} bands with the corresponding Bessell curves. In \emph{B}, the agreement is good for Nickel2 but there is a noticeable discrepancy between the Nickel1 filter response compared to that of Bessell. The filter response in \emph{I} for both Nickel configurations shows the most substantial departures from the Bessell standard, with Nickel2 exhibiting the most egregious disagreement. Nevertheless, the transmission curve has been verified through repeated measurements.

\section{Data Reduction}
\label{sec:data-reduction}

With over 21,000 images spanning 93 SNe~Ia with a median of 16 observed epochs, our dataset is too large to manually process. We have therefore developed an automated photometry pipeline\footnote{\url{https://github.com/benstahl92/LOSSPhotPypeline}} to calculate light curves from minimally preprocessed\footnote{Preprocessing consists of removing bias and dark current, flatfielding, and determining an astrometric solution.} KAIT and Nickel images (those from other telescopes could be incorporated with minimal modifications). Although it makes use of distinct software packages and utilises components written in several different programming languages, the pipeline is wrapped in a clean Python interface. It automatically performs detailed logging, saves checkpoints of its progress, and can be run interactively if desired --- thus, in cases where the data require special care, the user is able to perform each processing step manually with increased control. We detail the primary steps performed by the pipeline in the following sections.

\subsection{Start Up and Image Checking}
\label{ssec:start-up-and-image-checking}

At a minimum, the pipeline requires four pieces of information to run: the coordinates of the target (right ascension and declination), the name of an image to use for selecting candidate calibration stars (henceforth, the ``reference image''), and a text file containing the name of each image to process. In the absence of additional information, the pipeline will make sensible assumptions in setting various parameters during the start up process.

Processing commences by performing several checks on the specified images to see if any should be excluded. The first removes any images collected through an undesired filter, and the second excludes those collected outside a certain range of dates. In processing our dataset, we allow only unfiltered (referred to as ``{\it Clear}'') images and those collected through standard \emph{BVRI} filters between 60 days prior to, and 2 yr after, discovery as specified on the Transient Name Server (TNS)\footnote{\url{https://wis-tns.weizmann.ac.il/}}, to continue to subsequent processing steps.

\subsection{Selection of Calibration Star Candidates}
\label{ssec:calibration-star-candidate-selection}

In the next processing step, candidate calibration stars are identified in the reference image using a three-stage process. First, all sources above a certain threshold in the image are identified and those that are farther than 8$^{\prime \prime}$ from that target are retained.

Next, a catalog of potential calibration stars in the vicinity of the SN is downloaded (in order of preference) from the archives of Pan-STARRS~\citep[PS1;][]{PS1}, the Sloan Digital Sky Survey~\citep[SDSS;][]{SDSS}, or the AAVSO Photometric All-Sky Survey~\citep[APASS;][]{APASS}. The 40 brightest stars common to the reference image and the catalog are then retained. If the pipeline is being run interactively, the user can visually inspect the positions of these stars against the reference image and remove any that should not be used (such as those that are not well-separated from the target's host galaxy).

Finally, the magnitudes (and associated uncertainties) of the selected catalog stars are converted to the Landolt system \citep{Landolt83,Landolt92} using the appropriate prescription\footnote{The transformation given by \citet{tonry2012} is used for PS1 catalogs, whereas SDSS and APASS catalogs are treated with the prescription of Robert Lupton in 2005 (\url{https://www.sdss.org/dr12/algorithms/sdssUBVRITransform/})}, and subsequently to the natural systems of the various telescope/CCD/filter sets that are spanned by our dataset as discussed in Section~\ref{ssec:telescopes}. Conversion from the Landolt system to the aforementioned natural systems is accomplished using equations of the form
\begin{subequations}
\begin{align}
b &= B + C_B (B - V) + {\rm constant},\\
v &= V + C_V (B - V) + {\rm constant},\\
r &= R + C_R (V - R) + {\rm constant}, \text{ and}\\
i &= I + C_I (V - I) + {\rm constant},
\end{align}
\end{subequations}
where lower-case letters represent magnitudes in the appropriate natural system, upper-case letters represent magnitudes in the Landolt system, and $C_X$ is the linear colour term for filter $X$ as given in Table~\ref{tab:color-terms}. The KAIT3, KAIT4, and Nickel1 colour terms were originally given by G10, while those for Nickel2 are presented here for the first time. We derive the Nickel2 colour terms (and atmospheric correction terms, $k_i$; see Section~\ref{sssec:evolution-of-atmospheric-terms}) as the mean values of the appropriate terms measured over many nights using steps from the calibration pipeline described by G10.

\begin{table}
\caption{Summary of Colour Terms \label{tab:color-terms}}
\begin{tabular}{lrrrr}
\hline
\hline
System & $C_B$ & $C_V$ & $C_R$ & $C_I$\\
\hline
KAIT3 & $-0.057$ & $0.032$ & $0.064$ & $-0.001$\\
KAIT4 & $-0.134$ & $0.051$ & $0.107$ & $0.014$\\
Nickel1 & $-0.092$ & $0.053$ & $0.089$ & $-0.044$\\
Nickel2 & $0.042$ & $0.082$ & $0.092$ & $-0.044$\\
\hline
\end{tabular}
\end{table}

\subsection{Galaxy Subtraction}
\label{ssec:galaxy-subtraction}

A large proportion of SNe occur near or within bright regions of their host galaxies. It is therefore necessary to isolate the light of such a SN from that of its host prior to performing photometry. This is accomplished by subtracting the flux from the host at the position of the SN from the measured flux of the SN. To measure such host fluxes for the SNe in our sample needing galaxy subtraction (as determined by visual inspection and consideration of the offsets given in Table~\ref{tab:sample-information}), we obtained template images using the 1~m Nickel telescope (for \emph{BVRI} images) and KAIT (for unfiltered images) after the SNe had faded beyond detection, or from prior to the explosions if available in our database. Template images selected for use in galaxy subtraction are preprocessed identically to science images as described above.

\begin{figure}
	\includegraphics[width=\columnwidth]{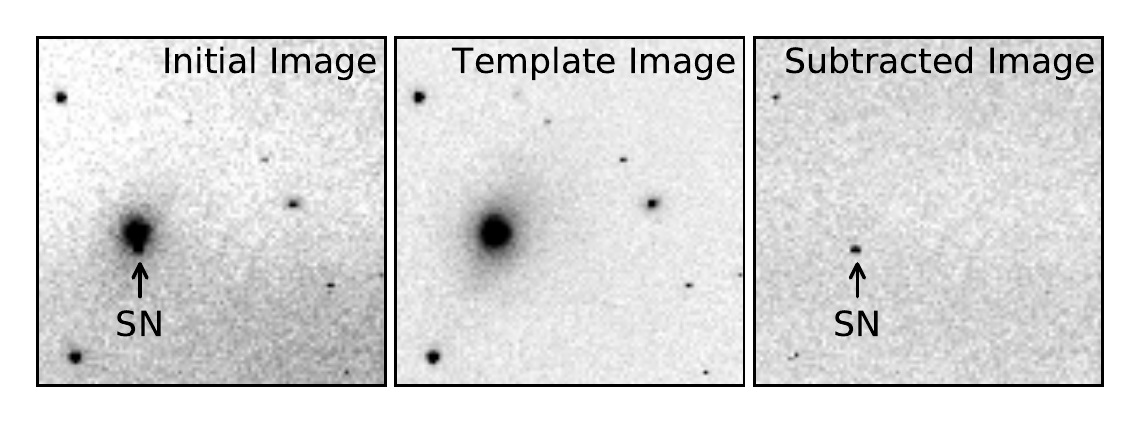}
	\caption{Example of our galaxy-subtraction procedure. The left image shows SN~2013gq on 2013 March~25 UT, with the SN flux clearly contaminated by the host galaxy. The centre image is the host-galaxy template used for subtraction, and the right image is the result of our galaxy-subtraction procedure.\label{fig:galaxy-subtraction}}
\end{figure}

The first step in our subtraction procedure is to align each science image to its corresponding template image. We do this by warping each template such that the physical coordinates of its pixels match those of the science image. Next, we perform the subtraction using the ISIS package~\citep{ISIS,ISIS2}, which automatically chooses stars in both images and uses them to compute the convolution kernel as a function of position. We use ten stamps in the $x$ and $y$ directions to determine the spatial variation in the kernel. ISIS matches the seeing between the warped template image and the science image by convolving the one with better seeing and then subtracts the images. An example image with subtraction applied is shown in Figure~\ref{fig:galaxy-subtraction}.

Some SNe in our dataset occurred sufficiently far from the nuclei of their host galaxies to not suffer significant contamination from galaxy light. In these cases, we did not perform galaxy subtraction. Table~\ref{tab:sample-information} includes a column that indicates whether host-galaxy subtraction was performed for each SN in our sample.

\subsection{Photometry}
\label{ssec:photometry}

After galaxy subtraction has been performed (or skipped if not needed), the pipeline performs photometry on the target SN and each selected calibration star. For images that have been galaxy subtracted, photometry is only performed on the SN (as the calibration stars will have been subtracted out), and photometry of the calibration stars is measured from the unsubtracted images. This requires the user to take care when doing calibration (see Section~\ref{ssec:calibration}) to ensure that the calibration stars used are not themselves contaminated by light from the SN's host galaxy.

By default, both point-spread function (PSF) and aperture photometry (through multiple apertures), along with standard photometry uncertainty calculations for each, are performed using procedures from the IDL Astronomy User's Library\footnote{\url{https://idlastro.gsfc.nasa.gov/homepage.html}}. Henceforth, we consider only PSF photometry.

The pipeline automatically keeps track of failures and removes the associated images from further processing. The user can easily track such failures and subsequently investigate each problematic image in more detail.

\subsection{Calibration to Natural Systems}
\label{ssec:calibration}

In the next step, the pipeline calibrates measured photometry to magnitudes in the appropriate natural system as follows. For each unsubtracted image, the mean magnitude of the selected calibration stars in the natural system appropriate to the image (from the catalog downloaded and converted according to the specifications in Section~\ref{ssec:calibration-star-candidate-selection}) is computed. Next, the mean \emph{measured} magnitude of the same set of reference stars is computed for each aperture. The difference between the former and the latter yields a set of offsets (one for each aperture) to add to the measured magnitudes such that, in the current image, the average magnitude of the selected calibration stars matches that from the catalog. These offsets are also applied to the \emph{measured} SN photometry from the image (and if it exists, the SN photometry from the associated galaxy-subtracted image). Standard techniques of error propagation are applied through these operations to determine the uncertainty in all derived natural system magnitudes, accounting for uncertainties in the calibration catalog and photometry.

This procedure is clearly sensitive to which calibration stars are used, and so several steps are employed in an attempt to make an optimal decision. First, calibration is performed on each image using all available calibration stars. Any calibration stars that are successfully measured in $< 40$\% of images are removed and calibration is run again using the remaining calibration stars. Next, any images in which $< 40$\% of the calibration stars are successfully measured are removed from further consideration. After these two preliminary quality cuts are performed, an iterative process is used to refine and improve the calibration. Each iteration consists of a decision that changes which calibration stars are used or which images are included and a recalibration based on that decision.

When run interactively, the pipeline provides the user with extensive information to consider when making this decision. In each iteration, the reference image is displayed with the current calibration stars and the SN identified. It also provides tables for each passband which include, for each calibration star: the median measured and calibration magnitudes as well as the median of their differences, the standard deviation of the measured magnitudes, and the proportion of all images in the current passband for which the calibration star's magnitude was successfully measured. The user can remove certain calibration stars, or all that (in any passband) exceed a specific tolerance on the median magnitude difference. Other options and diagnostics are available, and thus an experienced user will develop certain decision-making patterns when performing interactive calibration, but further discussion is beyond the scope of this description.

The automated pipeline makes the decision as follows. Any image containing a reference star that differs by the greater of 3 standard deviations or 0.5~mag from the mean measured magnitude of that reference star in the relevant filter/system is removed and logged internally for later inspection. If no such discrepant images are identified, then the calibration star whose median difference between measured and reference magnitudes is most severe is removed, so long as the difference exceeds 0.05~mag. If neither of these two criteria is triggered, then the calibration process has converged and iteration exits successfully. However, if a point is reached where only two reference stars remain, the tolerance of 0.05~mag is incremented up by 0.05~mag and iteration continues. If the tolerance is incremented beyond 0.2~mag without iteration ending successfully, the calibration process exits with a warning.

The process described above tends to lead to robust results, but it is still possible for individual measurements to be afflicted by biases. Because of this, we visually inspect our results after automated calibration and in some cases interactively recalibrate and/or remove certain images if they are suspected of contamination or are of poor quality.

\subsection{Landolt System Light Curves}
\label{ssec:light-curves}

The final stage of processing involves collecting each calibrated (natural system) magnitude measurement of the SN under consideration to form light curves (one for each combination of aperture and telescope system). Prior to transforming to the Landolt system, several steps are applied to these ``raw'' light curves. First, magnitudes in the \emph{same} passbands that are temporally close ($< 0.4$ days apart) are averaged together. Next, magnitudes in \emph{distinct} passbands that are similarly close in temporal proximity are grouped together so that they all have an epoch assigned as the average of their individual epochs. These steps result in a light curve for each telescope system used in observations, with magnitudes in the associated natural system.

Next, these light curves are transformed to the Landolt system by inverting the equations of Section~\ref{ssec:calibration-star-candidate-selection} and using the appropriate colour terms from Table~\ref{tab:color-terms}. Finally, the transformed light curves are combined into a final, standardised light curve which represents all observations of the SN.

\subsection{Uncertainties}
\label{ssec:errors}

To quantify the uncertainties in results derived from our processing routine, we inject artificial stars of the same magnitude and PSF as the SN in each image and then reprocess the images. We use a total of 30 artificial stars to surround the SN with five concentric, angularly offset hexagons of increasing size. The smallest has a ``radius'' of $\sim25^{\prime \prime}$ (exactly 20 KAIT pixels) and each concentric hexagon increases this by the same additive factor. We assign the scatter in the magnitudes of the 30 recovered artificial stars to be the uncertainty in our measurement of the SN magnitude. This is then added in quadrature with the calibration and photometry uncertainties and propagated through all subsequent operations, leading to the final light curve.

This method has the advantage of being an (almost) end-to-end check of our processing, and it can still be used effectively when certain steps (namely, host-galaxy subtraction) are not necessary. We note that by treating uncertainties in this way, we are making the assumption that the derived magnitude and PSF of the SN are correct. If this assumption is not met, the artificial stars we inject into each image will not be an accurate representation of the profile of the SN, and thus we cannot be assured that the distribution in their recovered magnitudes is a reasonable approximation to that of the SN. Furthermore, errors will be substantially overestimated when an injected star overlaps with a true star in the image. When this happens (as verified by a visual inspection) we do not inject a star at this position and thus in some cases the uncertainty estimate is made with slightly fewer than 30 stars.

\begin{figure*}
  \includegraphics[width=0.8\textwidth]{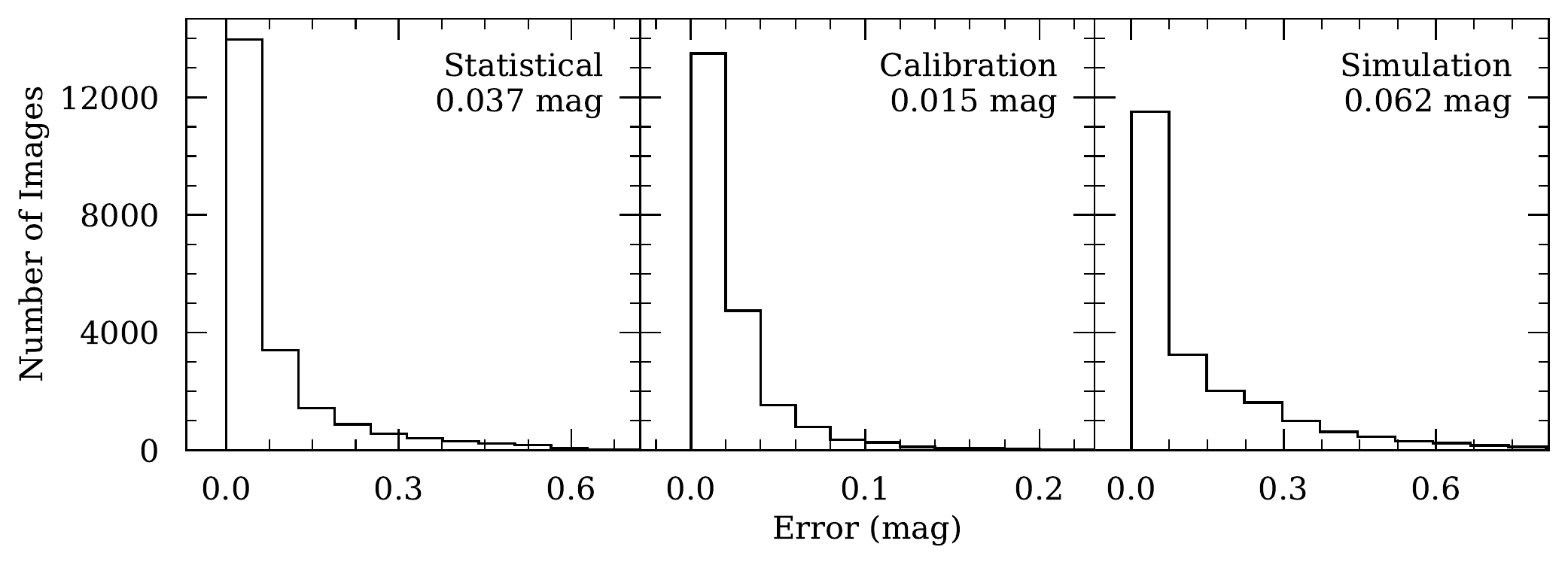}
  \caption{Distribution of uncertainties arising from statistical, calibration, and simulation sources. All magnitudes are instrumental magnitudes, and the median uncertainty from each source is printed.\label{fig:errors}}
\end{figure*}

Altogether, the final uncertainty on each magnitude in our light curves is derived by propagating three sources of uncertainty through our calculations. These sources are (1) ``statistical'' (e.g., scatter in sky values, Poisson variations in observed brightness, uncertainty in sky brightness), (2) ``calibration'' (e.g., calibration catalog, derived colour terms), and (3) ``simulation'' (as described in the preceding paragraphs). In terms of instrumental magnitudes, we find median uncertainties from these sources of 0.037~mag, 0.015~mag, and 0.062~mag, respectively. We show the distribution of each in Figure~\ref{fig:errors}.

\subsection{Systematic Errors}
\label{ssec:systematic-errors}

In order to combine or compare photometric datasets from different telescopes, one must understand and account for systematic errors. In this section, we consider sources of possible systematic errors and quantify their impact on our final photometry. As three of the four telescope/detector configurations spanned by our dataset are already extensively considered by G10, our goal here is primarily to extend their findings to cover the fourth configuration, Nickel2.

\subsubsection{Evolution of Colour Terms}
\label{sssec:evolution-of-color-terms}

\begin{figure}
	\includegraphics[width=0.8\columnwidth]{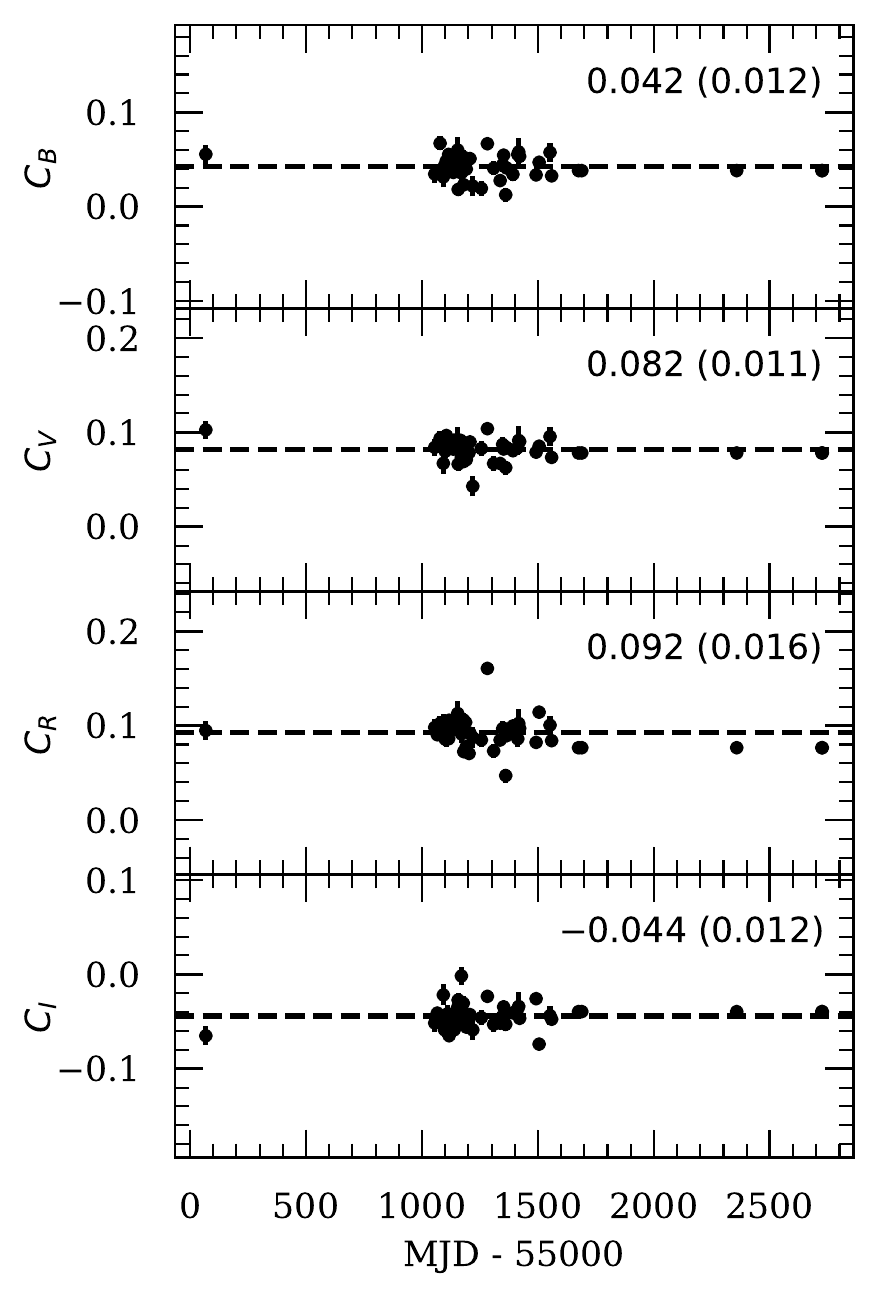}
	\caption{Nickel2 colour terms as a function of time. The mean and standard deviation in each passband are printed.\label{fig:C-Nickel}}
\end{figure}

The Nickel2 colour terms given in Table~\ref{tab:color-terms} are the average colour terms from observations of Landolt standards over many nights. Any evolution in the derived colour terms as a function of time introduces errors in the final photometry that are correlated with the colour of the SN and reference stars. To investivate this effect, we plot the Nickel2 colour terms as a function of time in Figure~\ref{fig:C-Nickel}, but find no significant evidence for temporal dependence. This conclusion is in line with the findings of G10 for KAIT3, KAIT4, and Nickel1.

\subsubsection{Evolution of Atmospheric Terms}
\label{sssec:evolution-of-atmospheric-terms}

\begin{figure}
	\includegraphics[width=0.8\columnwidth]{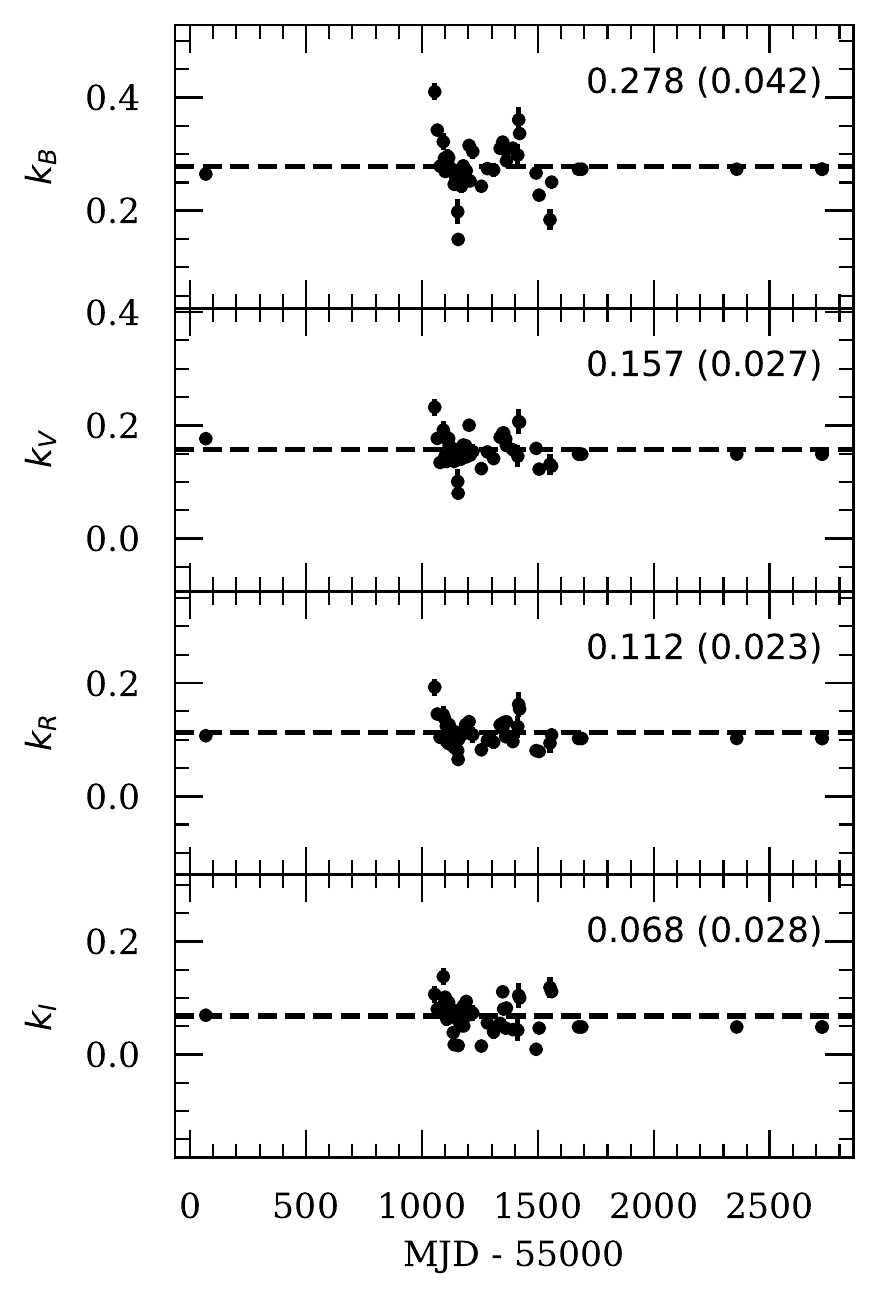}
	\caption{Same as Figure~\ref{fig:C-Nickel}, except for atmospheric correction terms used to transform Nickel2 natural-system magnitudes to the Landolt system.\label{fig:k-Nickel}}
\end{figure}

For the same set of nights for which we compute the colour terms which constitute Figure~\ref{fig:C-Nickel}, we also derive atmospheric correction terms. Because we source calibration stars from established catalogues (as outlined in Section~\ref{ssec:calibration-star-candidate-selection}), our derived atmospheric correction terms affect processing only indirectly (i.e., in the determination of colour terms). As such, we discuss them here only as a stability check. Figure~\ref{fig:k-Nickel} shows their evolution as a function of time. We do not find significant evidence for temporal dependence, which is consistent with the findings of G10 for KAIT3, KAIT4, and Nickel1. It is also worth noting that our derived terms ($k_B = 0.278$, $k_V = 0.157$, $k_R = 0.112$, and $k_I = 0.068$)) are similar to those derived for Nickel1 by G10 (0.277, 0.171, 0.120, and 0.078, respectively).

\subsubsection{Combining KAIT and Nickel Observations}
\label{sssec:kait-and-nickel-observations}

\begin{figure*}
	\includegraphics[width=0.65\textwidth]{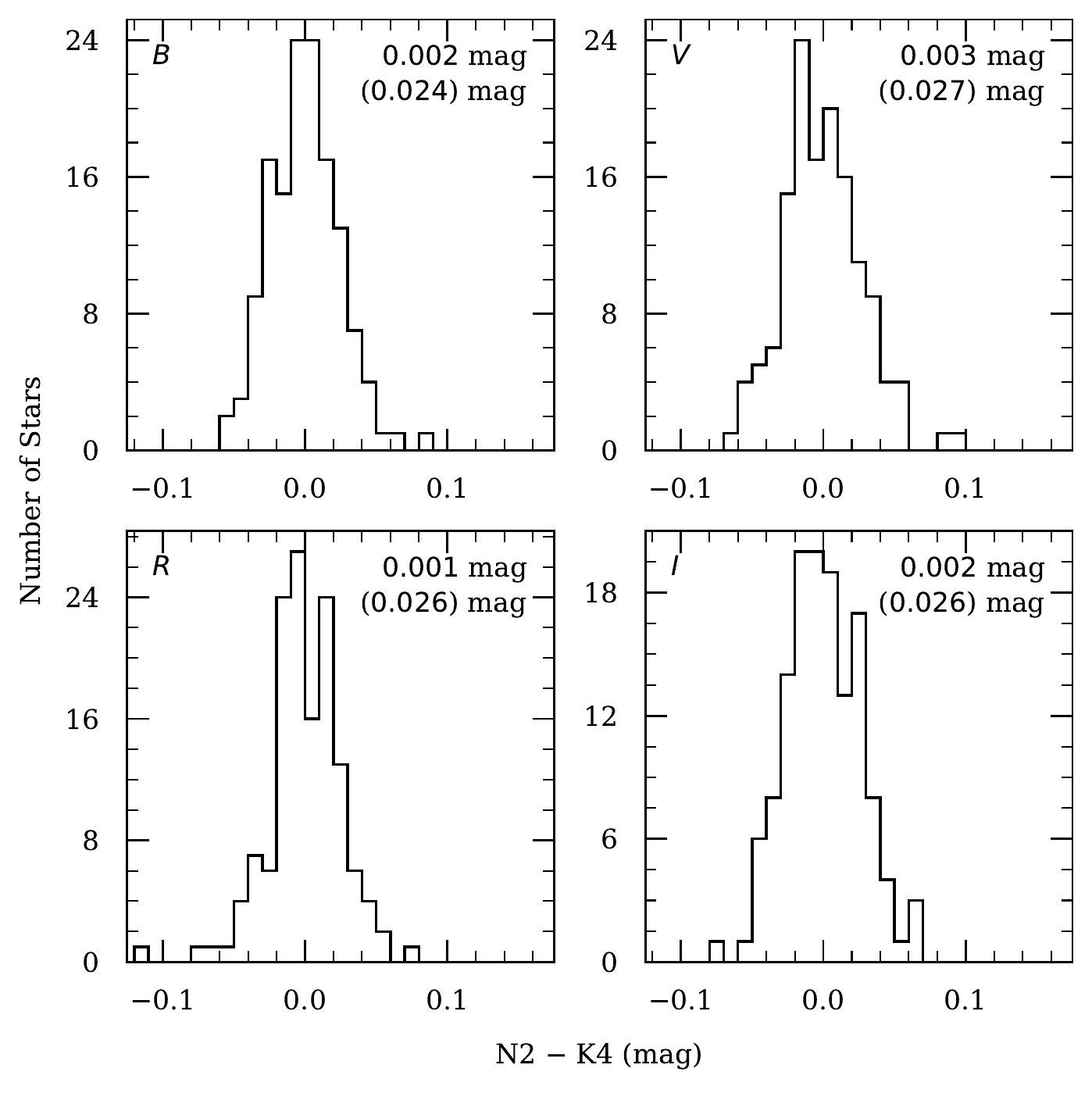}
	\caption{Distributions of the residuals of the mean derived magnitude of each calibration star used in determining final photometry for SNe in our dataset covered by the KAIT4 and Nickel2 systems. The distributions reveal negligible offset between these two systems in all bands with a scatter $<0.03$~mag. The median and standard deviation of the residuals are printed for each passband.\label{fig:N2K4-comparison}}
\end{figure*}

Another potential source of systematic error arises when combining observations from different configurations (e.g., KAIT4 and Nickel2). Any systematic differences between configurations introduces an error when observations from various systems are combined. To search for and investigate such differences, we compare the mean derived magnitude of each calibration star used in determining our final photometry for unique combinations of passband and system. In this investigation, we only consider instances where a calibration star was observed using two different systems. Figure~\ref{fig:N2K4-comparison} shows the distribution of differences in each passband for the common set of calibration stars between the KAIT4 and Nickel2 systems, which have the largest overlap. Similar distributions were constructed for all other system combinations, and in all cases we find a median offset of $\lesssim 0.003$ mag\footnote{The only exception is the median \emph{I}-band offset between Nickel1 and KAIT3, which is 0.008~mag.} with scatter $\sigma \lesssim 0.03$~mag in each filter.

\subsubsection{Galaxy Subtraction}
\label{sssec:error-galaxy-subtraction}

When subtracting host-galaxy light, the finite signal-to-noise ratio (S/N) of the images used as templates can limit measurements of the magnitude of a SN, thereby introducing a correlated error between epochs of photometry. To investigate the severity of this effect, G10 stacked images to obtain a deeper set of template images with increased S/N for SN 2000cn, a SN~Ia from their sample. By reprocessing their data with the new template images, G10 were able to probe the influence of host-galaxy templates derived from single images. Unsurprisingly, they found that the correlated error introduced by using a single image for a template is not negligible, but that it is appropriately accounted for by their error budget. As the modest differences between the Nickel1 and Nickel2 systems should not manifest any substantial differences with regard to galaxy subtraction in this manner, and because the error budget of G10 is similar to our own (as laid out in Section~\ref{ssec:errors}), we see no need for repetition of this test.

\subsubsection{Total Systematic Error}
\label{sssec:total-systematic-error}

Based on the preceding discussion, we assign a systematic uncertainty of 0.03~mag in \emph{BVRI} to our sample, consistent with G10. This uncertainty is not explicitly included in our photometry tables or light curve figures (e.g., Tables~\ref{tab:phot-sample}~\&~\ref{tab:phot-sample-natural} and Figure~\ref{fig:lc-sample}), but must be accounted for when combining our dataset with others.

\section{Results}
\label{sec:results}

In this section we present the results obtained by running our photometry pipeline on SNe~Ia from LOSS images collected from 2009 through 2018, with several earlier SNe~Ia also included. Basic information and references for each SN in our sample are provided in Table~\ref{tab:sample-information}. The NASA/IPAC Extragalactic Database (NED)\footnote{The NASA/IPAC Extragalactic Database (NED) is operated by the Jet Propulsion Laboratory, California Institute of Technology, under contract with the National Aeronautics and Space Administration (NASA).} and the TNS were used to source many of the given properties.

Figure~\ref{fig:lc-sample} shows our light curves, each shifted such that time is measured relative to the time of maximum $B$-band brightness as determined by {\tt MLCS2k2} \citep{Jha07} fits or Gaussian Process interpolations \citep{Lochner16} for peculiar SNe (see Sections~\ref{sssec:mlcs}~\&~\ref{ssec:light-curve-properties}, respectively). An example of our photometry is given in Table~\ref{tab:phot-sample}. In addition to leaving out the systematic 0.03~mag uncertainty derived in Section~\ref{sssec:total-systematic-error}, we choose to provide light curves without considering corrections such as Milky Way (MW) extinction, $K$-corrections \citep{oke68,hamuy93,kim96}, or $S$-corrections \citep{scorr}. This provides future studies the opportunity to decide which corrections to apply and full control over how they are applied. Because of the low redshift range of our dataset (see the right panel of Figure~\ref{fig:histograms}) and the similarity between systems, the $K$- and $S$-corrections will be quite small in any case. Though magnitudes in Figure~\ref{fig:lc-sample} and Table~\ref{tab:phot-sample} are given in the Landolt system, we also make our dataset available in natural-system magnitudes for those that would benefit from the reduced uncertainties (see Appendix~\ref{app:nat-sys-light-curves}). Our entire photometric dataset (Landolt and natural-system magnitudes) is available online from the Berkeley SuperNova DataBase\footnote{\url{http://heracles.astro.berkeley.edu/sndb/info\#DownloadDatasets(BSNIP,LOSS)}} \citep[SNDB;][]{BSNIP,SNDB2}.

\begin{table*}
\caption{Photometry of SN 2008ds.\label{tab:phot-sample}}
\begin{tabular}{lccccccr}
\hline
\hline
SN & MJD & $B$ (mag) & $V$ (mag) & $R$ (mag) & $I$ (mag) & {\it Clear} (mag) & System\\
\hline
 2008ds &  54645.47 &                 ... &                 ... &                 ... &                 ... &  $15.700 \pm 0.033$ &    kait4 \\
 2008ds &  54646.47 &                 ... &                 ... &                 ... &                 ... &  $15.574 \pm 0.024$ &    kait4 \\
 2008ds &  54647.46 &  $15.613 \pm 0.012$ &  $15.630 \pm 0.010$ &  $15.593 \pm 0.012$ &  $15.744 \pm 0.018$ &  $15.501 \pm 0.010$ &    kait4 \\
 2008ds &  54650.47 &  $15.503 \pm 0.014$ &  $15.487 \pm 0.010$ &  $15.475 \pm 0.013$ &  $15.766 \pm 0.016$ &                 ... &    kait4 \\
 2008ds &  54653.13 &  $15.483 \pm 0.009$ &  $15.474 \pm 0.005$ &  $15.413 \pm 0.006$ &  $15.756 \pm 0.008$ &                 ... &  nickel1 \\
 2008ds &  54653.44 &  $15.492 \pm 0.018$ &  $15.470 \pm 0.010$ &  $15.435 \pm 0.011$ &  $15.828 \pm 0.017$ &                 ... &    kait4 \\
 2008ds &  54655.13 &  $15.570 \pm 0.008$ &  $15.512 \pm 0.006$ &  $15.451 \pm 0.007$ &  $15.826 \pm 0.009$ &                 ... &  nickel1 \\
 2008ds &  54655.48 &  $15.567 \pm 0.016$ &  $15.507 \pm 0.012$ &  $15.467 \pm 0.015$ &  $15.925 \pm 0.023$ &                 ... &    kait4 \\
 2008ds &  54658.13 &  $15.704 \pm 0.008$ &  $15.606 \pm 0.006$ &  $15.542 \pm 0.006$ &  $15.962 \pm 0.008$ &                 ... &  nickel1 \\
 2008ds &  54662.16 &  $15.995 \pm 0.012$ &  $15.773 \pm 0.005$ &                 ... &                 ... &                 ... &  nickel1 \\
 \hline
 \multicolumn{8}{p{15cm}}{\textbf{Note:} First 10 epochs of \emph{BVRI} + unfiltered photometry of SN 2008ds. This table shows the form and content organisation of a much larger table that covers each epoch of photometry for each SN in our dataset. The full table is available in the online version of this article.}
 \end{tabular}
 \end{table*}

\subsection{The LOSS Sample}

\begin{figure*}
\includegraphics[width=\textwidth]{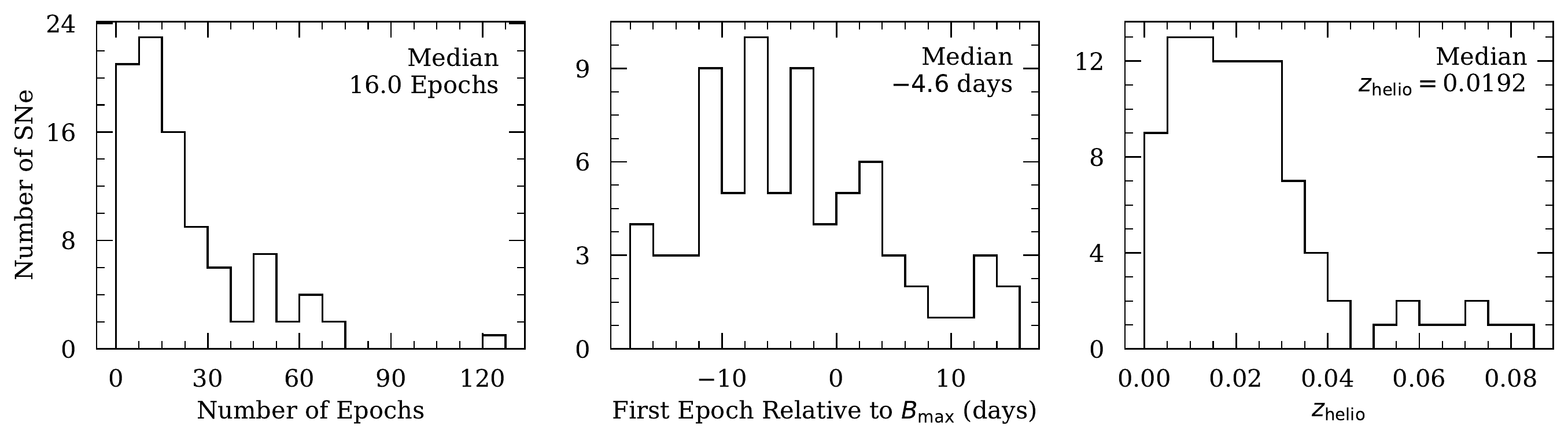}
\caption{Distributions of dataset parameters. The left panel is the number of epochs of photometry as measured from \emph{V}-band observations, centre is the first epoch of observation relative to time of maximum \emph{B}-band light, and right is redshift.\label{fig:histograms}}
\end{figure*}

\begin{figure}
	\includegraphics[width=\columnwidth]{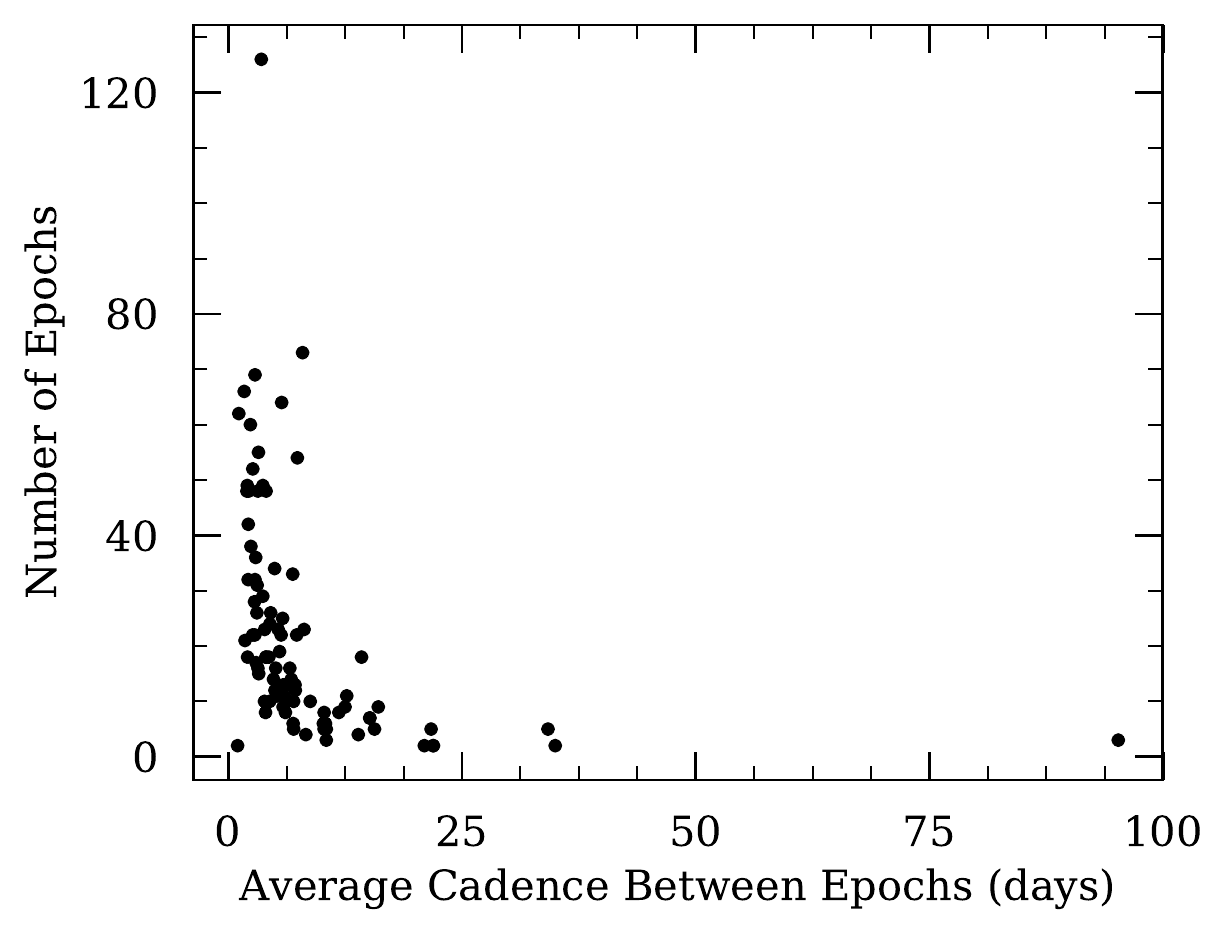}
	\caption{Scatterplot of the number of photometry epochs for each SN vs. the average cadence between epochs. The tight grouping with a lower average cadence and mid to high number of epochs indicates that our SNe are well sampled and cover a large portion of photometric evolution. The single SN with an average cadence in excess of 80 days is SN 2016ffh.\label{fig:cadence-vs-epochs}}
\end{figure}

In order to accurately measure and exploit the correlation between light-curve width and luminosity for SNe~Ia, thus allowing for precision measurements of cosmological parameters, densely sampled multicolour light curves which span pre- through post-maximum evolution are required. In Figure~\ref{fig:cadence-vs-epochs} we show the number of epochs of photometry for each SN in our sample versus the average cadence between epochs of photometry. The plot indicates that the majority of SNe in our sample have more than 10 epochs of observations with a cadence of fewer than 10 days, while a significant number of SNe were observed many more times at even higher frequency. These metrics confirm that on average, our light curves are well sampled and span a large range of photometric evolution.

The left panel of Figure~\ref{fig:histograms} presents a histogram of the total number of photometry epochs for all SNe in our sample, and we find a median of 16 epochs. SN~2011dz has just one epoch of photometry and five objects (SNe 2006ev, 2009D, 2009hp, 2012E, 2012bh) have two epochs each, while SN 2013dy has 126 (the most), followed by SN 2012cg and then SN 2017fgc. We begin photometric follow-up observations for the typical SN in our sample $\sim4.6$ days before maximum light in the \emph{B}-band, with 52 SNe having data before maximum brightness. The centre panel of Figure~\ref{fig:histograms} shows the distribution of first-observation epochs for our sample. The median redshift of our full sample is 0.0192, with a low of 0.0007 (SN 2014J) and a high of 0.0820 (SN 2017dws). We show the distribution of redshifts in the right panel of Figure~\ref{fig:histograms}. If we restrict to $z \geq 0.01$ (i.e., within the Hubble flow), our sample consists of 71 SNe with a median redshift of 0.0236.

\subsection{Comparison with Published LOSS Reductions}

For several of the SNe presented here, previous reductions of the photometry (usually performed with an earlier photometry pipeline, developed by G10) have been published. A comparison between these previous results and our own offers a useful efficacy check of our pipeline while avoiding the issues arising from comparisons between different telescopes or photometric systems. Wherever sufficient overlap between one of our light curves and that from a previous publication exists, we quantify the extent to which the datasets agree by computing the weighted mean residual. In some cases we further compare by considering the agreement between derived quantities such as the light-curve shape, $\Delta m_{15}(B)$, and the time of maximum brightness, $t_{B_{\rm max}}$. We emphasise that in general our results are derived from different sets of reference stars for calibration than those used to derive the results with which we compare, and that even when reference stars overlap, we may draw their magnitudes from different catalogs.

\subsubsection{SN 2005hk}

\citet{2005hk} published optical light curves from KAIT data for the Type Iax SN 2005hk. At the time of publication, no template images were available and so the authors acknowledged that their derived magnitudes for the SN, located $\sim18.5^{\prime \prime}$ from the nucleus of its host galaxy, were probably affected by the background light. In the prevailing time, we have obtained template images of the host and used them to separate its flux from that of the SN. Comparing results, both of which were obtained using PSF-fitting photometry, we find agreement to within 0.090~mag in \emph{BVRI}. It is worth noting that our measurements are generally fainter, especially when the SN is rising and declining. This suggests that host-galaxy subtraction is indeed necessary for this object. We also compare measurements of the light-curve shape parameter $\Delta m_{15} (B)$, and find strong agreement between our value (see Section~\ref{ssec:light-curve-properties} and Table~\ref{tab:gp-fits}) of $1.58 \pm 0.05$~mag and theirs of $1.56 \pm 0.09$~mag.

\subsubsection{SN 2009dc}

Our Nickel and KAIT images of the extremely slow-evolving SN 2009dc --- a super-Chandrasekhar candidate \citep[see][for a summary of the properties of this subclass of thermonuclear SNe]{SC} --- were initially processed and used to construct light curves by \citet{2009dc}. In both our reduction and theirs, PSF-fitting photometry was employed and galaxy subtraction was not performed owing to the large separation between the SN and its host galaxy. We find agreement to better that 0.020~mag in \emph{BVRI}. Furthermore, we derive $\Delta m_{15}(B) = 0.71 \pm 0.06$~mag, consistent with their result of $\Delta m_{15}(B) = 0.72 \pm 0.03$~mag.

\subsubsection{SN 2009ig}

Optical light curves of SN 2009ig were derived from KAIT data and published by \citet{2009ig}. Both our reduction procedure and theirs used PSF-fitting photometry after subtracting template images of the host galaxy. We find that our results agree to within 0.055~mag in \emph{BVRI}. It is worth adding that SN 2009ig is in a field with very few stars available for comparison when calibrating to natural-system magnitudes~---~\citet{2009ig} used only one star for comparison while we have used two. In light of these challenges, we are content with the similarity between our results, especially because we obtain a consistent value of $\Delta m_{15} (B)$\footnote{We find $\Delta m_{15} (B) = 0.85 \pm 0.12$~mag (the large uncertainty is mostly due to the uncertainty in the time of \emph{B} maximum), while \citet{2009ig} find $\Delta m_{15} (B) = 0.89 \pm 0.02$~mag.}. As an added check, we reprocessed our data for SN 2009ig using the same calibration star as \citet{2009ig} and find agreement to within $\sim0.025$~mag in \emph{BVRI}.

\subsubsection{SN 2011by}

KAIT \emph{BVRI} photometry of SN 2011by was published by \citet{2011by-silverman} and later studied in detail by \citet{Graham11fe}. In comparing our light curves (which have host-galaxy light subtracted) to theirs (which do not), we find agreement to within $\sim0.05$~mag. Furthermore, \citet{2011by-silverman} found $B_{\rm max} = 12.89 \pm 0.03$~mag and $\Delta m_{15}(B) = 1.14 \pm 0.03$~mag, which are consistent with our results of $B_{\rm max} = 12.91 \pm 0.02$~mag and $\Delta m_{15}(B) = 1.09 \pm 0.10$~mag.

\subsubsection{SN 2011fe}

SN 2011fe/PTF11kly in M101 is perhaps the most extensively observed SN~Ia to date \citep{Nugent11fe,Vinko11fe,RS11fe,Graham11fe,Zhang11fe}. Photometry derived from KAIT data has been published by \citet{Graham11fe} and \citet{Zhang11fe}, but we compare only with the latter. For the 20 epochs that overlap between our dataset and theirs, we find agreement of better than $\sim0.04$~mag in \emph{BVRI}.

\subsubsection{SN 2012cg}

SN 2012cg was discovered very young by LOSS, and KAIT photometry from the first $\sim2.5$ weeks following discovery was published by \citet{2012cg}. Because of the small temporal overlap between this early-time dataset and the much more expansive set presented herein, and because we have obtained template images and used them to remove the host-galaxy light, it is not instructive to quantitatively compare between our dataset and theirs. We note, however, that we find a similar time of \emph{B}-band maximum and that there is clear qualitative agreement between the two samples.

\subsubsection{SN 2013dy}

\citet{Zheng13dy} published early-time KAIT photometry of SN 2013dy and used it to constrain the first-light time, while \citet{2013dy} published extensive optical light curves. We compare the 85 overlapping epochs of our dataset with those of \citet{2013dy}, both of which were obtained using PSF-fitting photometry, and find agreement better than $\sim0.03$~mag in \emph{BVRI}.

\subsubsection{SN 2013gy}

KAIT \emph{B} and \emph{V} observations were averaged in flux space to create so-called \emph{BV}.5-band photometry by \citet{2013gy}, who then used $S$-corrections to transform to the \emph{g} band on the Pan-STARRS1 photometric system. Because of the difference between our choice of photometric system and theirs, we opt only to compare derived light-curve properties. Our result for the time of \emph{B}-band maximum is within one day of theirs (consistent, given the uncertainties), and we find $\Delta m_{15}(B) = 1.247 \pm 0.072$~mag, nearly  identical to their result of $\Delta m_{15}(B) = 1.234 \pm 0.060$~mag.

\subsubsection{SN 2014J}

SN 2014J in M82 has been extensively studied --- unfiltered KAIT images were presented by \citet{2014J-zheng} and used to constrain the explosion time, and \citet{2014J-foley} published photometry from many sources, including a number of KAIT \emph{BVRI} epochs. A comparison between our results and theirs reveals substantial ($\sim0.2$~mag) discrepancies. The origin of this disagreement stems from differences in our processing techniques --- \citet{2014J-foley} calibrated instrumental magnitudes against reference-star magnitudes \emph{in the Landolt system} (thereby disregarding linear colour terms), while we have done calibrations with reference-star magnitudes \emph{in the natural system} appropriate to the equipment before transforming to the Landolt system. When we reprocess our data using the former approach in conjunction with the reference stars used by \citet{2014J-foley}, we find agreement between our non-host-galaxy subtracted light curve and theirs to within 0.01~mag in \emph{BVRI}. Our final light curve for SN 2014J reflects the latter approach (which is the default of our pipeline), and was derived using a different set of calibration stars after subtracting host-galaxy light.

\subsubsection{SN 2016coj}

SN 2016coj was discovered at a very early phase by LOSS, and \citet{2016coj} presented the first 40 days of our optical photometric, low- and high-resolution spectroscopic, and spectropolarimetric follow-up observations. Because our full photometric dataset encompasses a much broader time frame and \citet{2016coj} focused only on unfiltered photometry, a direct comparison is not possible. However, we note that our derived $\Delta m_{15}(B) = 1.33 \pm 0.03$~mag, $B_{\rm max} = 13.08 \pm 0.01$~mag, and $t_{B_{\rm max}} = 57547.15 \pm 0.19$ MJD are consistent with their preliminary reporting, based on photometry \emph{without} host-galaxy subtraction, of $1.25 \pm 0.12$~mag, $13.1 \pm 0.1$~mag, and $57547.35$ MJD, respectively.

\subsubsection{Summary of Comparisons}

We have compared the results of our photometry to the results derived from previous processing pipelines used by our group for ten SNe~Ia. Of these, five (SNe 2009dc, 2009ig, 2011fe, 2013dy, and 2014J) can be directly compared in the sense that identical processing steps (e.g., whether galaxy subtraction was performed) were used. For this subsample, we find excellent ($\lesssim 0.05$~mag) agreement except for the cases of SN 2009ig ($< 0.055$~mag) and SN 2014J ($\sim0.2$~mag). However, we are able to attain much stronger agreement ($\lesssim 0.025$~mag and $\lesssim 0.010$~mag, respectively) if we employ the same calibration procedures used in the original processing. For the remaining five, we find consistent results in derived light-curve parameters, and more generally, good qualitative agreement in the shape of the light curves.

\section{Discussion}
\label{sec:discussion}

The absolute peak brightness that a SN~Ia attains has been shown to be strongly correlated with the ``width'' of its light curve \citep[e.g.,][]{Phillips93}. Thus, given a model for this correlation and a measurement of the light-curve width of a SN~Ia, one can compute its intrinsic peak luminosity. By comparing this to its \emph{observed} peak brightness, the distance to the SN~Ia can be estimated. In this section, we examine the properties of the light curves in our sample in more detail. Specifically, in Section~\ref{ssec:light-curve-properties} we directly measure light-curve properties from interpolations, whereas in Section~\ref{ssec:light-curve-fitters} we model our light curves with light-curve fitting tools.

\subsection{Interpolated Light-Curve Properties}
\label{ssec:light-curve-properties}

Perhaps the most ubiquitous parametrisation of the width (or decline rate) of a SN~Ia light curve is $\Delta m_{15} (X)$, the difference in its magnitude at maximum light and 15 days later in passband $X$. We measure this quantity in $B$ and $V$ by interpolating the (filtered) light curves using Gaussian Processes, a technique that has proved useful in astronomical time series analysis due to its incorporation of uncertainty information and robustness to noisy or sparse data \citep{Lochner16}.

For each SN in our sample where the photometry in $B$ and/or $V$ encompasses the maximum brightness in that band, we employ the following approach using tools from the {\tt SNooPy}\footnote{\url{https://csp.obs.carnegiescience.edu/data/snpy/documentation/snoopy-manual-pdf}} package \citep{SNooPy}. First, we interpolate the light curve in each passband using Gaussian Processes, allowing us to determine the time at which that light curve peaks. With the phase information that this affords, the data are $K$-corrected using the spectral energy distribution (SED) templates of \citet{Hsiao}. We further correct the data for MW extinction \citep{Schlafly2011} and then perform a second interpolation on the corrected data. From this interpolation we measure $t_{X_{\rm max}}, X_{\rm max},$ and $\Delta m_{15}(X)$ --- the time of maximum brightness, maximum apparent magnitude, and light-curve width parameter (respectively) --- in filters $B$ and $V$. In measuring $\Delta m_{15}(X)$, we correct for the effect of time dilation. The final results of this fitting process are presented in Table~\ref{tab:gp-fits}.

\subsection{Applying Light-Curve Fitters}
\label{ssec:light-curve-fitters}

While interpolation is viable for well-sampled light curves, those that are more sparsely sampled or which do not unambiguously constrain the maximum brightness cannot be reliably treated with this technique. Furthermore, interpolation completely disregards the effects of host-galaxy extinction, which must be accounted for when estimating distances. 

Because of these limitations, we also employ two light-curve fitters to measure the properties of our sample. To the extent that the templates used by these fitters span the diversity in our dataset, this approach does not suffer from the same limitations as interpolation.

\subsubsection{SNooPy $E(B-V)$ Model}
\label{sssec:ebv}

\begin{figure*}
\includegraphics[width=0.5\textwidth]{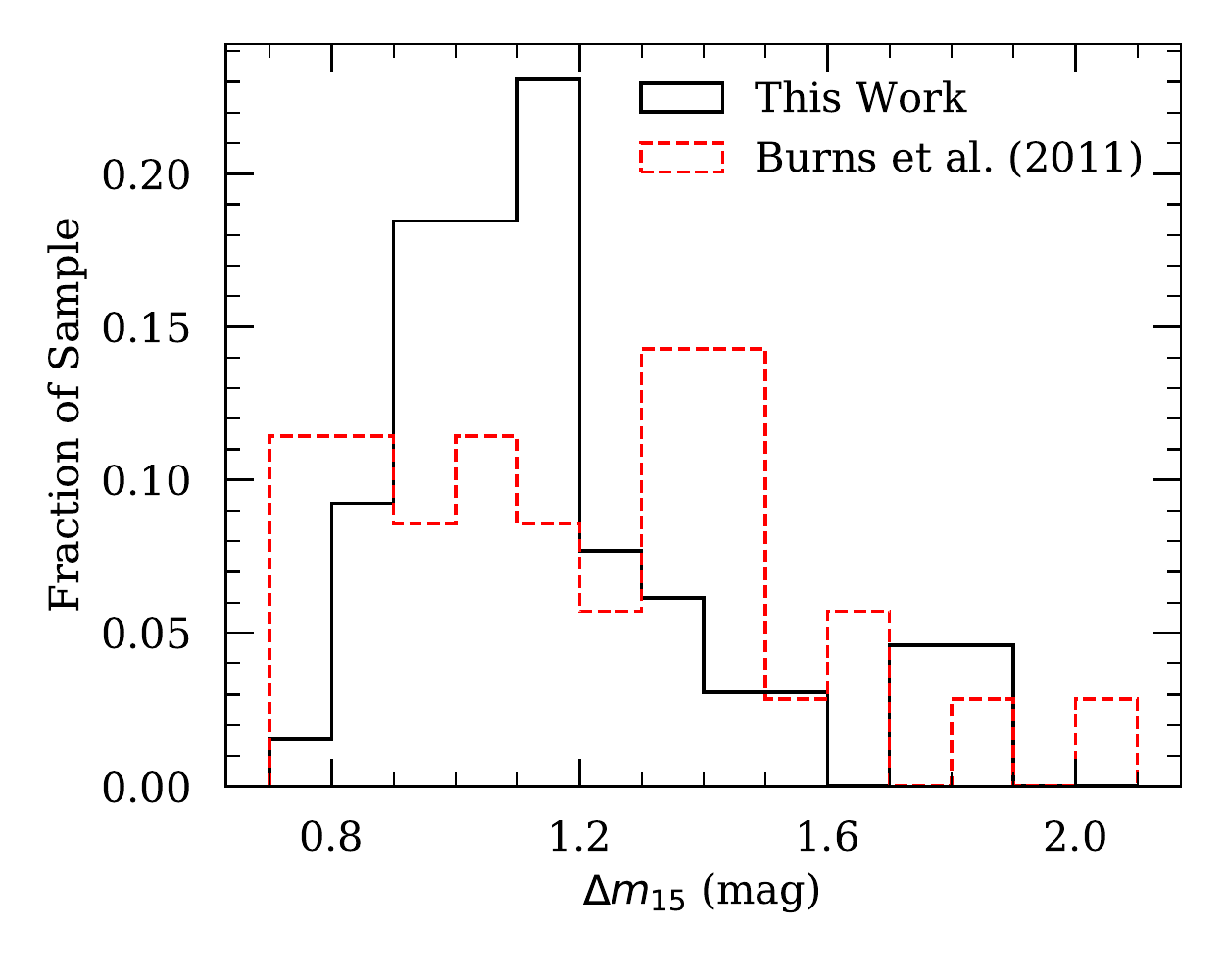}%
\includegraphics[width=0.5\textwidth]{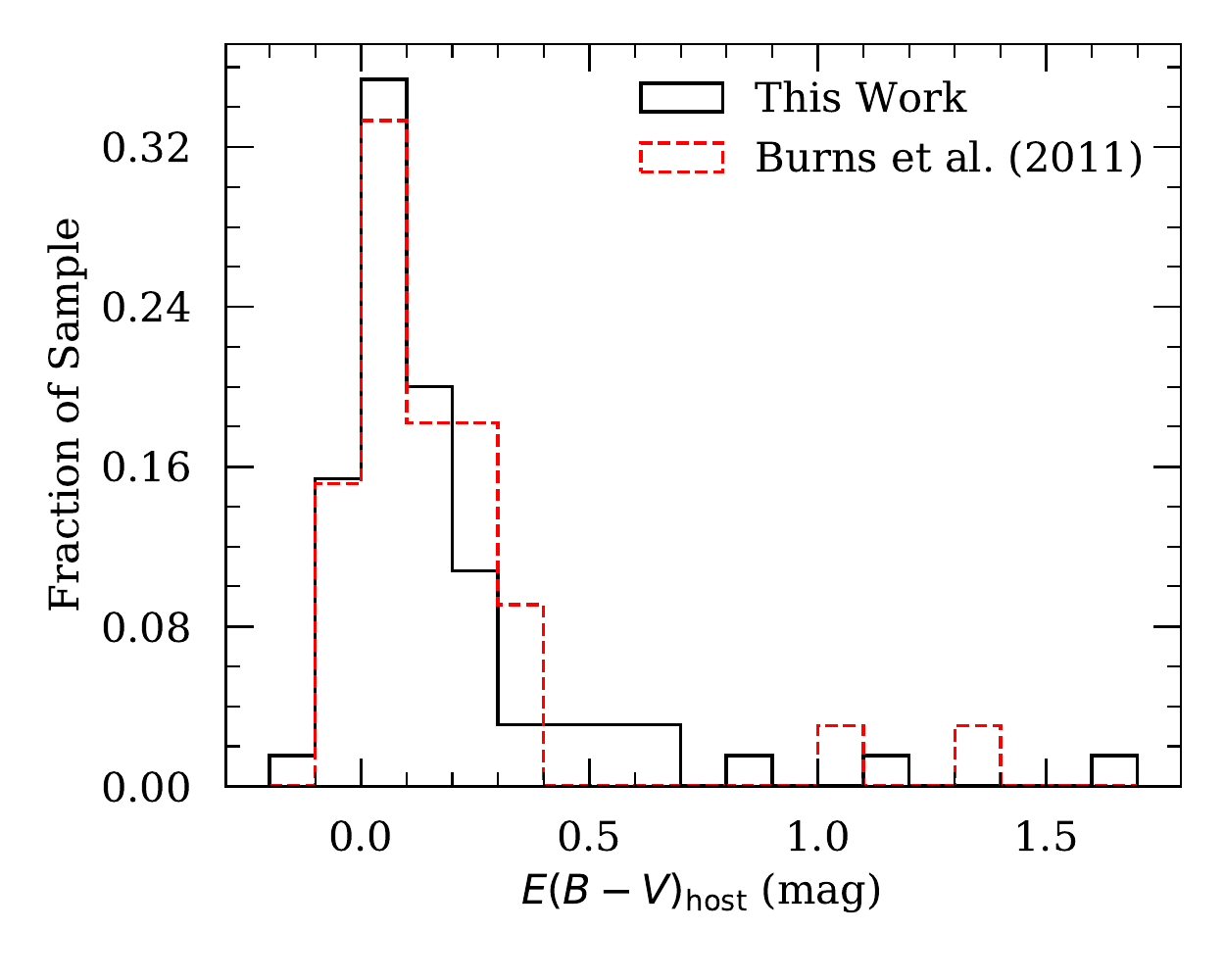}
\caption{Distributions of $\Delta m_{15}$ and $E(B - V)_{\rm host}$ from {\tt SNooPy} $E(B - V)$ model fits to the light curves in our dataset appear in black. We include the corresponding distributions derived from \citet{SNooPy} in red.\label{fig:ebv-hists}}
\end{figure*}

We use the so-called ``EBV\_model'' in {\tt SNooPy} to simultaneously fit the \emph{BVRI} light curves in our sample. In observed band $X$ and SN rest-frame band $Y$, the model takes on the mathematical form
\begin{align}
	\label{eqn:ebv}
	m_X(t - t_{\rm max}) = \; &T_Y\left(t_{\rm rel}, \Delta m_{15}\right) + M_Y(\Delta m_{15}) + \mu  + \nonumber \\
	&R_X E(B - V)_{\rm gal} + R_Y E(B - V)_{\rm host} + \nonumber \\
    &K_{X,Y}\left(z, t_{\rm rel}, E(B - V)_{\rm host}, E(B - V)_{\rm gal}\right),
\end{align}
where $m$ is the observed magnitude, $t_{\rm max}$ is the time of $B$-band maximum, $t_{\rm rel} = (t^\prime - t_{\rm max})/(1 + z)$ is the rest-frame phase, $M$ is the rest-frame absolute magnitude of the SN, $\mu$ is the distance modulus, $E(B - V)_{\rm gal}$ and $E(B - V)_{\rm host}$ are the reddening due to the Galactic foreground and host galaxy, respectively, $R$ is the total-to-selective absorption, and $K$ is the $K$-correction (which depends on the epoch and can depend on the host and Galactic extinction).

{\tt SNooPy} generates the template, $T(t,\Delta m_{15})$, from the prescription of \citet{Prieto2006}. As indicated, the light curve is parameterised by the decline-rate parameter, $\Delta m_{15}$, which is similar to $\Delta m_{15} (B)$. It is important to note, however, that these quantities are \emph{not} identical, and may deviate from one another randomly and systematically \citep[see Section 3.4.2 in][]{SNooPy}. The model assumes a peak $B$-band magnitude and $B - X$ colours based on the value of $\Delta m_{15}$, with six possible calibrations derived from CSP1a. We use calibration \#6, which is derived from the best-observed SNe in the sample, less those that are heavily extinguished.

The template-fitting process with {\tt SNooPy} consists of the following steps. First, an initial fit is made to determine the time of $B$-band maximum. This allows for initial $K$-corrections to be determined using the SED templates from \citet{Hsiao}. The $K$-corrected data are then fit again, allowing colours to be computed as a function of time. Next, improved $K$-corrections are computed, warping the SED such that it matches the observed colours. Last, a final fit is performed using the improved $K$-corrections. The results from fitting are $t_{\rm max}, \Delta m_{15}, E(B - V)_{\rm host},$ and $\mu$. We present these quantities for our dataset in Table~\ref{tab:model-fits}. We also visualise the distributions of $\Delta m_{15}$ and $E(B - V)_{\rm host}$ from our dataset in Figure~\ref{fig:ebv-hists}, with the corresponding distributions from \citet{SNooPy} overlaid for comparison. 

For $\Delta m_{15}$, we find a median value of 1.11~mag with a standard deviation of 0.26~mag, consistent with the respective values of 1.15~mag and 0.32~mag from the dataset of \citet{SNooPy}. And for $E(B - V)_{\rm host}$, we find a median of 0.10~mag with a dispersion of 0.29~mag for our sample, similar to their values of 0.12~mag and 0.29~mag, respectively. We stress that comparing these parameters between our dataset and that of \citet{SNooPy} is only to provide a diagnostic view of how our sample is distributed relative to another from the literature --- there is minimal overlap between the two samples, so we are \emph{not} looking for a one-to-one correspondence.

\begin{figure}
	\includegraphics[width=\columnwidth]{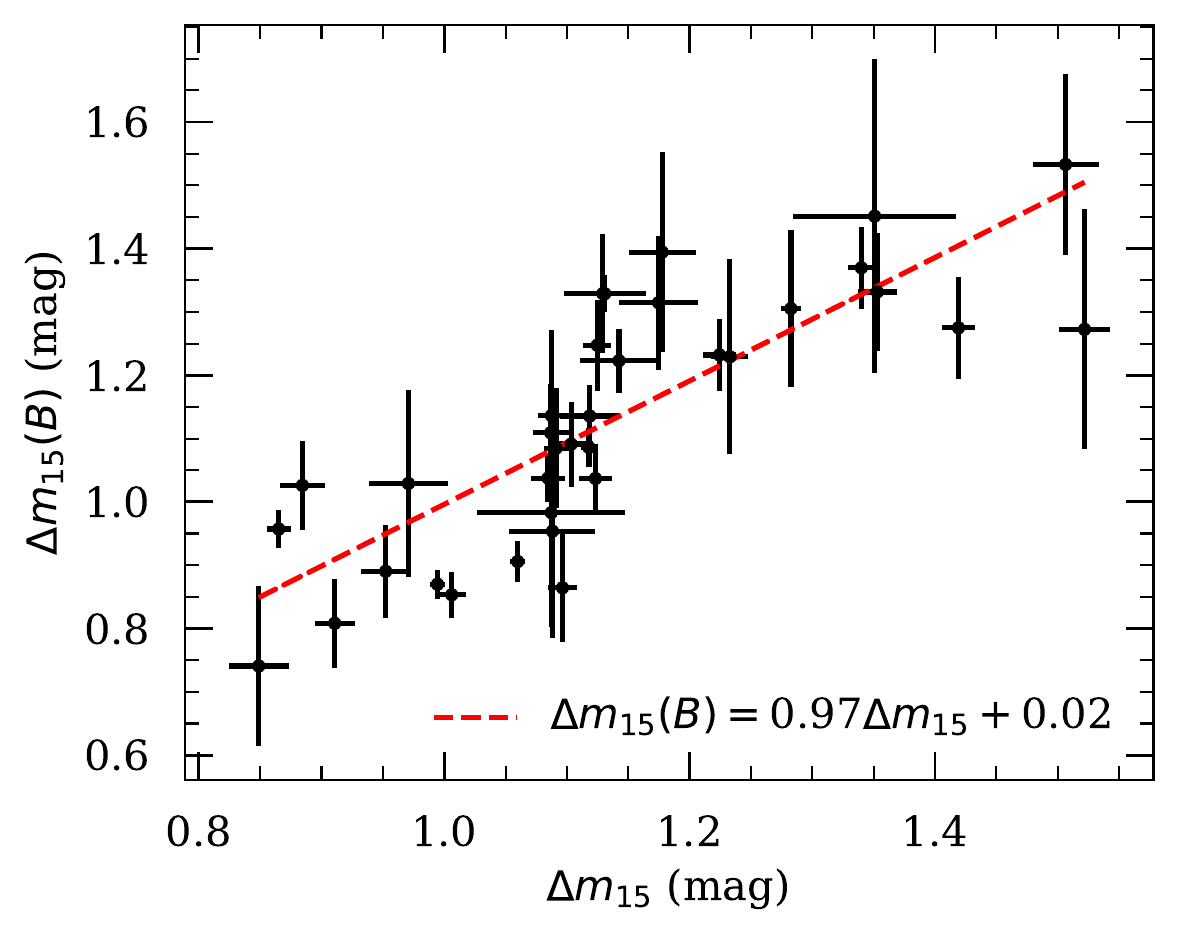}
	\caption{Comparison of the decline-rate parameter as measured from our Gaussian Process interpolations, $\Delta m_{15} (B)$, with that obtained directly from our {\tt SNooPy} $E(B - V)$ model fits, $\Delta m_{15}$.\label{fig:dm15-comp}}
\end{figure}

\begin{figure*}
\includegraphics[width=0.5\textwidth]{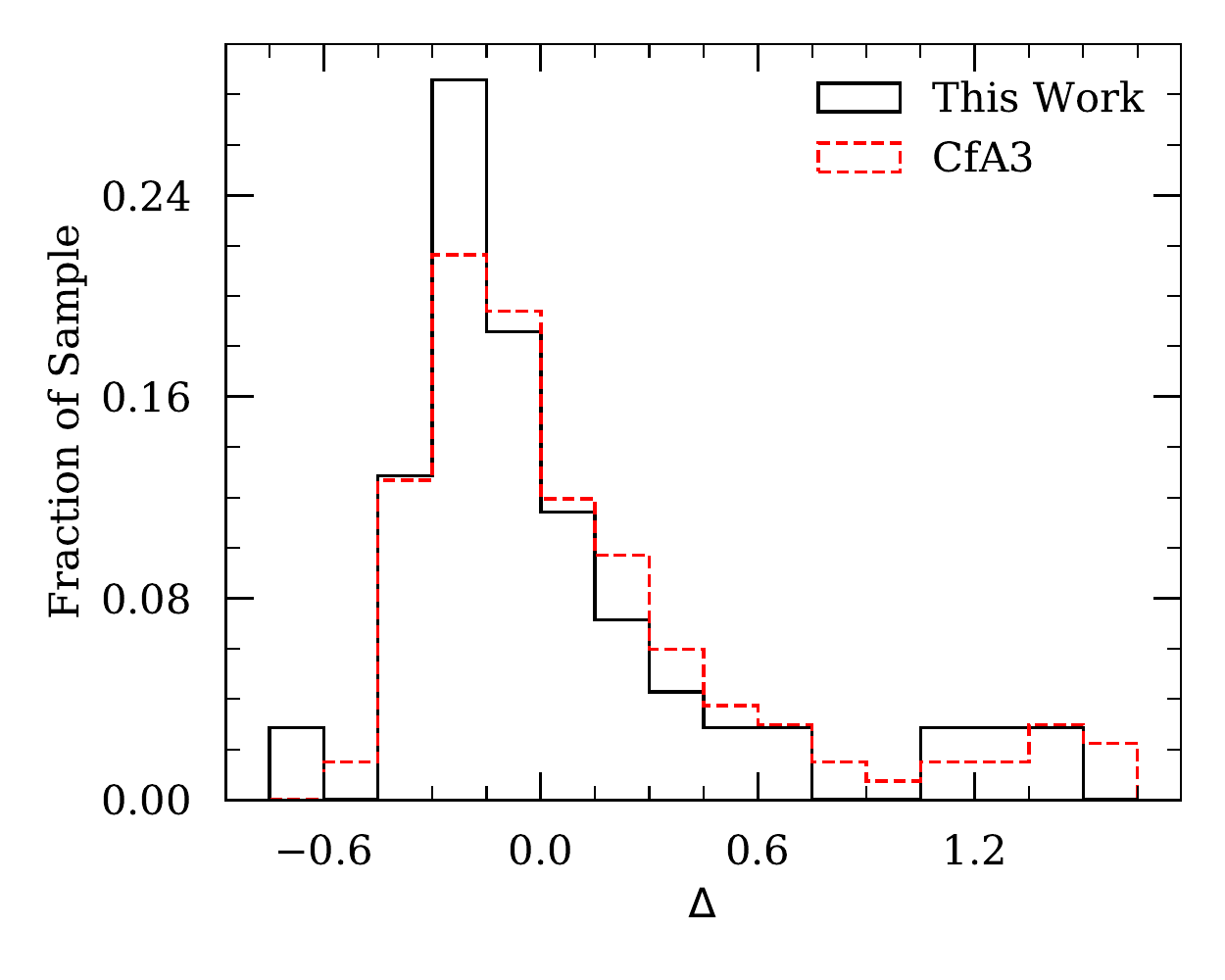}%
\includegraphics[width=0.5\textwidth]{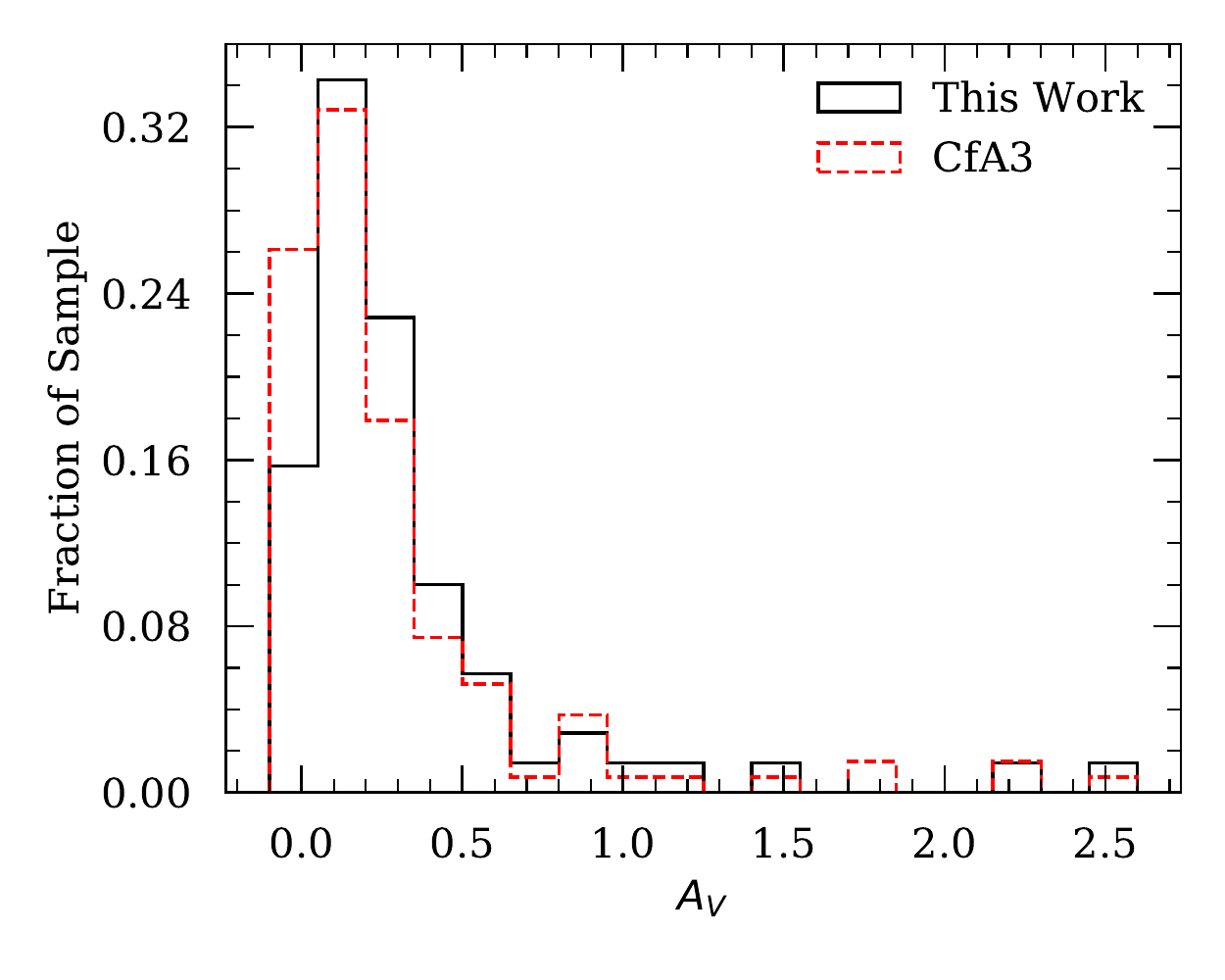}
\caption{Distributions of $\Delta$ and $A_V$ from {\tt MLCS2k2} model fits to the light curves in our dataset appear in black. We include the corresponding distributions derived from CfA3 in red.\label{fig:mlcs-hists}}
\end{figure*}

Furthermore, we can use the fitted model for each light curve to calculate other parameters of interest, such as those derived from direct interpolation. This gives a method by which we can check for consistency in our results. For example, we expect the time of maximum brightness in a given band to be the same, regardless of whether it was calculated from an interpolation or a fitted model. We employ Kolmogorov-Smirnoff tests on our calculated times of maximum (where we have results from both interpolation and template fitting) to quantify the likelihood that those from interpolation are drawn from the same distribution as those from template fitting. In both cases ($t_{B_{\rm max}}$ and $t_{V_{\rm max}}$), we find $p$-values of unity, indicating that our expectation is met.

Applying such tests for $B_{\rm max}$ and $V_{\rm max}$ is less straightforward because of the presence of systematic offsets between results derived from interpolation and those derived from fitting {\tt SNooPy}'s $E(B - V)$ model. While both methods provide peak magnitudes after performing $K$-corrections and correcting for MW reddening, only the $E(B - V)$ model fits account for host-galaxy reddening. With this caveat noted, it is still instructive to make comparisons, and in doing so we find $p$-values of 0.708 and 0.981 for $B_{\rm max}$ and $V_{\rm max}$, respectively. If we impose restrictions to make the comparison more legitimate --- namely to use only those SNe in our sample that are not heavily extinguished by their hosts ($|E(B - V)_{\rm host}| < 0.1$~mag), that are spectroscopically normal (as given in Table~\ref{tab:sample-information}), and for which {\tt SNooPy} measures $\Delta m_{15} < 1.7$~mag --- we find substantially improved agreement, with $p$-values of 0.956 and 1.000, respectively.

As noted above, $\Delta m_{15}$ does not exactly correspond to $\Delta m_{15} (B)$. In comparing them, \citet{SNooPy} found a linear relationship of $\Delta m_{15} (B) = 0.89 \Delta m_{15} + 0.13$. Performing an analogous comparison with our dataset subjected to the aforementioned light-curve shape restriction, we find $\Delta m_{15} (B) = (0.97 \pm 0.12) \Delta m_{15} + (0.02 \pm 0.14)$. Figure~\ref{fig:dm15-comp} shows our derived linear relationship within the context of our data.

\subsubsection{MLCS2k2}
\label{sssec:mlcs}

In addition to the methods described above, we have run {\tt MLCS2k2.v007} \citep{Jha07} on our sample of light curves. {\tt MLCS2k2} parameterises the absolute magnitude of a SN in terms of $\Delta$, which quantifies how luminous a SN is relative to a fiducial value. By using a quadratic dependence on $\Delta$, intrinsic variations in peak magnitude are modeled without introducing a parameter for intrinsic colour. In order to do this, {\tt MLCS2k2} corrects for MW reddening and \emph{attempts} to correct for reddening due to the host galaxy by employing a reddening law, $R_V$, to obtain the host-galaxy extinction parameter, $A_V$, after employing a prior on $E(B - V)$.

{\tt MLCS2k2} yields four fitted parameters for each \emph{BVRI} light curve: the distance modulus ($\mu$), the shape/luminosity parameter ($\Delta$), the time of $B$-band maximum ($t_0$), and the host-galaxy extinction parameter ($A_V$). In running {\tt MLCS2k2} on our dataset, we fix $R_V$ to 1.7 and use the default host-reddening prior, which consists of a one-sided exponential with scale length $\tau_{E(B - V)} = 0.138$~mag. We use the SED templates of \citet{Hsiao}, and following \citet{Hicken09} we use {\tt MLCS2k2} model light curves trained using $R_V = 1.9$. We present the results of running {\tt MLCS2k2.v007} on our sample in Table~\ref{tab:model-fits} and the distributions of $\Delta$ and $A_V$ in Figure~\ref{fig:mlcs-hists}. We find a median and standard deviation for $\Delta$ of $-0.11$ and 0.46, and for $A_V$ of 0.20 and 0.45. Comparing these to the corresponding parameters from CfA3 we find reasonable agreement, with $-0.04$ and 0.48, and 0.13 and 0.44, respectively. Our dataset only shares minimal overlap with that of CfA3, so these comparisons serve to reveal how our dataset is distributed relative to another low-$z$ sample.

\subsection{Comparison of Light-Curve Fitter Results}
\label{ssec:lc-fit-comparison}

\begin{figure*}
\includegraphics[width=0.5\textwidth]{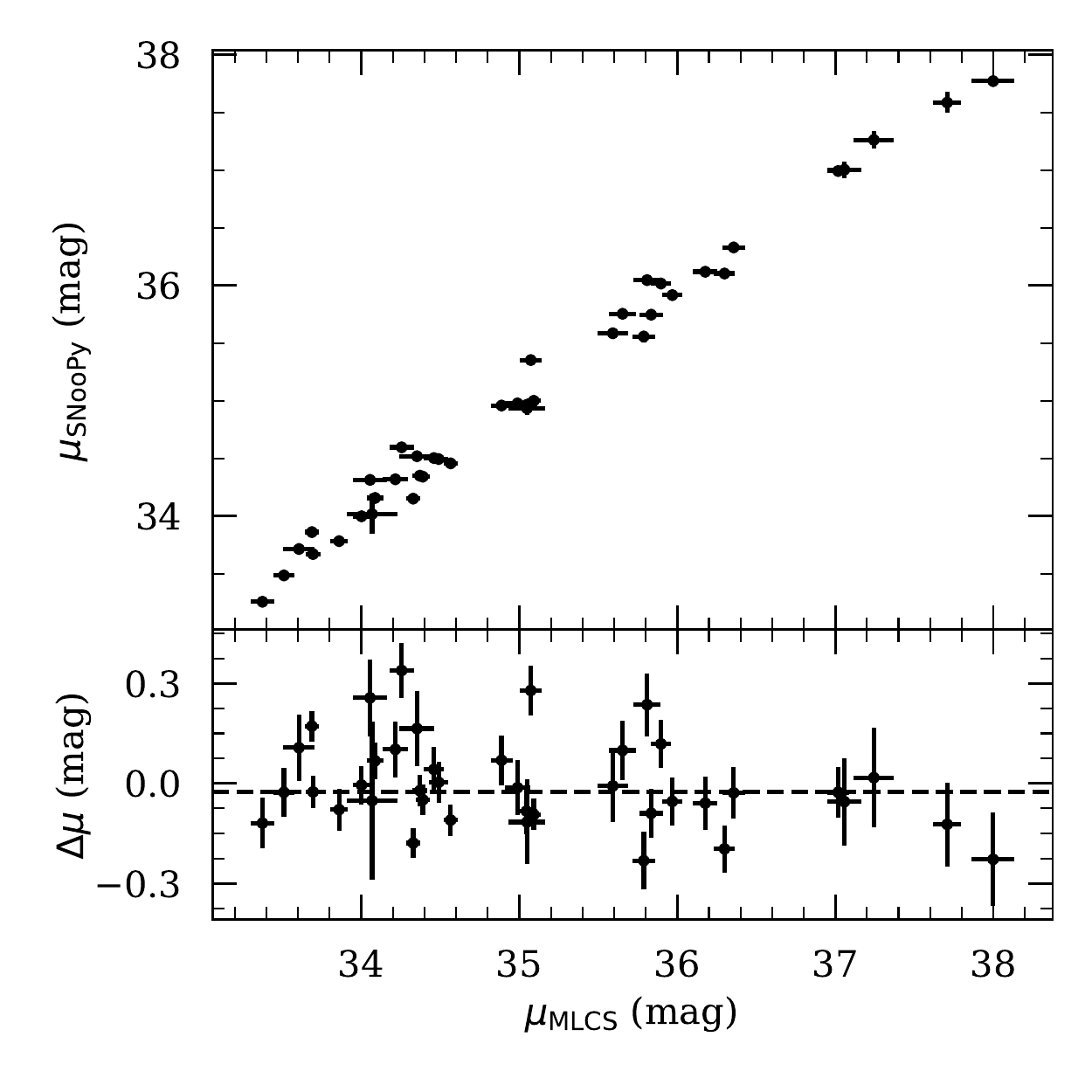}%
\includegraphics[width=0.5\textwidth]{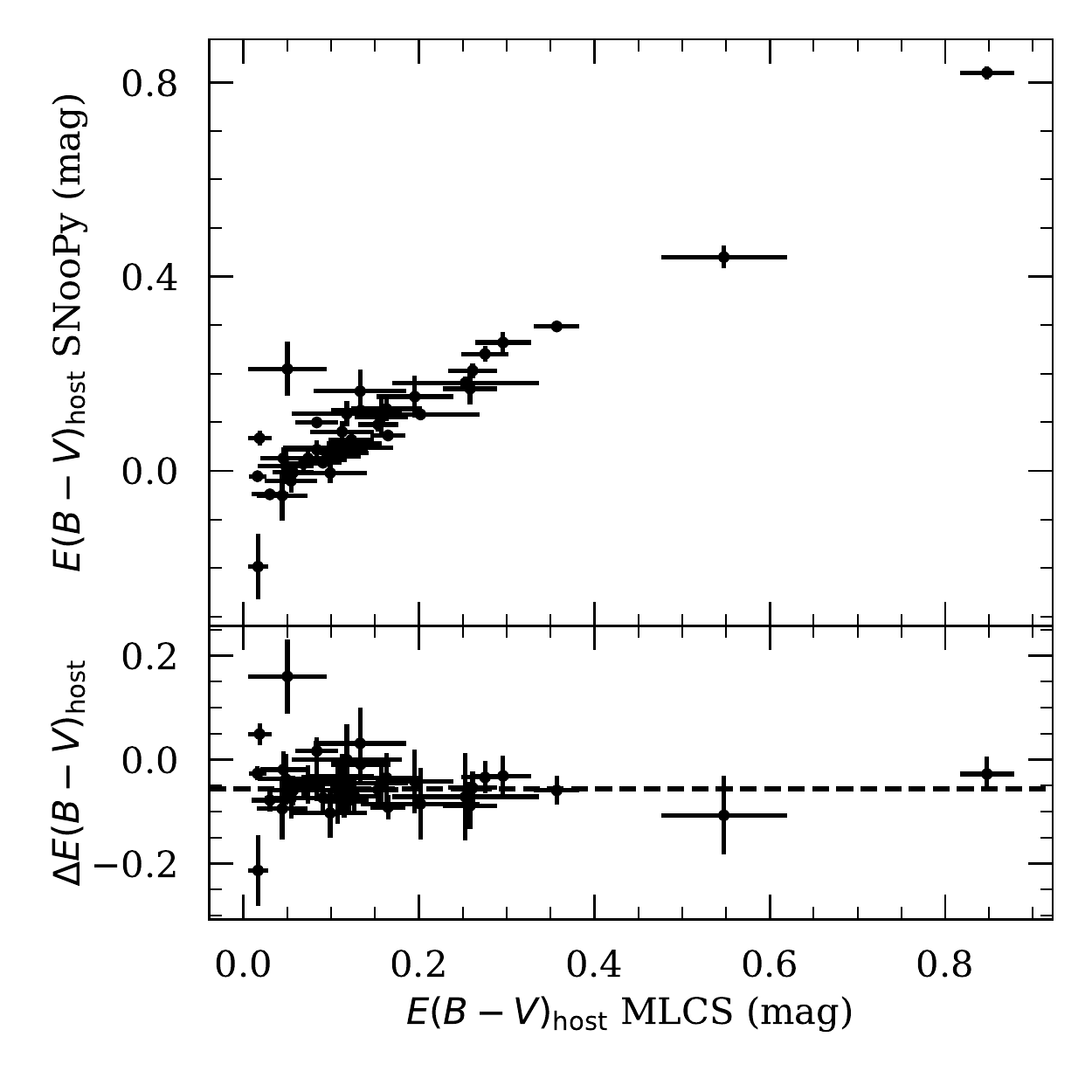}
\caption{Comparison of the (scaled) distance modulus and host-galaxy reddening results from both light-curve fitters for the selected subset of our dataset.
\label{fig:ebvmlcs-comparison}}
\end{figure*}

To make any cosmological statements based on the results in the previous section is beyond the scope of this paper, as this would require a detailed study and justification of the utilised light-curve fitters and their parameters, amongst many other considerations. It is interesting and possible, however, to compare results from the two light-curve fitters we employ to check for consistency. As the principal quantity of interest when fitting the light curves of SNe~Ia is distance, we will focus our comparison on the derived distance moduli.

The left plot in Figure~\ref{fig:ebvmlcs-comparison} compares the distance moduli from {\tt SNooPy} and {\tt MLCS2k2} after correcting to put the measurements on the same scale (so that \emph{relative} distance moduli are compared, independent from assumptions about the Hubble constant). This correction consists of adding an offset to the distance moduli from each fitter such that the value of $H_0$ measured from each set of results yields 65 km~s$^{-1}$~Mpc$^{-1}$. We perform this comparison only for spectroscopically normal SNe~Ia in our sample for which {\tt SNooPy} finds $\Delta m_{15} < 1.7$~mag and for which $z > 0.01$. Of course, further restrictions should be placed when selecting a sample for cosmological purposes, but our selection is reasonable for performing a general comparison. We find strong agreement between the two sets of corrected distance moduli --- a Kolmogorov-Smirnoff test gives a $p$-value of 1.000. The median residual is $-0.026$~mag with a statistical dispersion of $0.135$~mag.

If we were to ensure consistency in choosing the parameters for each light-curve fitter, the residuals would almost certainly decrease. In particular, when fitting with {\tt MLCS2k2}, we place an exponentially decaying prior on $A_V$, but no such prior was imposed with {\tt SNooPy}. This difference may well manifest in statistically and systematically different results for host-galaxy reddening and distance moduli between the two fitters. We compare host-galaxy reddening results in the right panel of Figure~\ref{fig:ebvmlcs-comparison}, where for {\tt MLCS2k2} we have converted to reddening using $E(B - V)_{\rm host} = A_V / R_V$, with $R_V = 1.7$. The agreement is reasonable, with a median residual of $-0.056$~mag and statistical uncertainty of $0.055$~mag. Furthermore, the facts that the median residual ({\tt SNooPy} minus {\tt MLCS2k2}) is negative and that the disagreement is most severe for small $E(B - V)_{\rm host}$ are consistent with what one might expect given the prior imposed by {\tt MLCS2k2}.

\section{Conclusion}
\label{sec:conclusion}

In this paper we present \emph{BVRI} (along with some unfiltered) light curves of 93 SNe~Ia derived from images collected by the LOSS follow-up program primarily over the interval from 2009--2018, but with several instances as early as 2005. Careful and consistent observational and processing techniques ensure that our data is prepared in a homogeneous fashion. We estimate the systematic uncertainty in our dataset to be 0.03~mag in \emph{BVRI}, and we encourage the community to incorporate our light curves in future studies.

In cases where our results overlap with previous reductions of LOSS data, we provide a set of comparisons as a consistency check. In general, we find good agreement, giving us confidence in the quality of our processing and analysis. When combined with the light curves of G10, the resulting dataset spans 20 years of observations of 258 SNe~Ia from the same two telescopes.

We study the properties of the light curves in our dataset, with particular focus on the parameters used in various width-luminosity relationships. Using direct interpolations, we measured $\Delta m_{15}(B)$ and $\Delta m_{15}(V)$. We also apply the light-curve fitters {\tt SNooPy} and {\tt MLCS2k2.v007} to measure $\Delta m_{15}$ and $\Delta$, respectively. We compare results derived from these methods, and find an acceptable degree of agreement given the differences in starting assumptions.

A consideration of the photometric dataset presented here alongside spectra from the Berkeley Supernova Ia Program (BSNIP) database will enable further utility. Our dataset overlaps with 13 SNe from the first BSNIP data release \citep{BSNIP}, with an average of 4.5 spectra each. Furthermore we expect significant overlap between the SNe in our dataset and our upcoming second BSNIP data release of $\sim700$ spectra from $\sim250$ SNe~Ia observed over a similar temporal range (Stahl et al. 2019, in prep.).

\section*{Acknowledgements}

We are grateful to the staff at Lick Observatory for their assistance with the Nickel telescope and KAIT. We thank Brian Dupraw at the U.C. Santa Cruz Optical Shop for providing an extended wavelength measurement of the Nickel2 \emph{I}-band transmission function, and our anonymous referee whose careful reading and constructive comments improved the manuscript. KAIT and its ongoing operation were made possible by donations from Sun Microsystems, Inc., the Hewlett-Packard Company, Auto Scope Corporation, Lick Observatory, the National Science Foundation (NSF), the University of California, the Sylvia \& Jim Katzman Foundation, and the TABASGO Foundation. A major upgrade of the Kast spectrograph on the Shane 3~m telescope at Lick Observatory was made possible through generous gifts from the Heising-Simons Foundation as well as William and Marina Kast. 

Research at Lick Observatory is partially supported by a generous gift from Google. Support for A.V.F.'s supernova group has also been provided by the NSF, 
Marc J. Staley (B.E.S. is a Marc J. Staley Graduate Fellow), the Richard and Rhoda Goldman Fund, the TABASGO Foundation, 
Gary and Cynthia Bengier (T.deJ. is a Bengier Postdoctoral Fellow), the Christopher R. Redlich Fund, and the Miller
Institute for Basic Research in Science (U.C. Berkeley). 
In addition, we greatly appreciate contributions from 
numerous individuals, including
Charles Baxter and Jinee Tao,
George and Sharon Bensch 
Firmin Berta,     
Marc and Cristina Bensadoun, 
Frank and Roberta Bliss,   
Eliza Brown and Hal Candee,
Kathy Burck and Gilbert Montoya,
Alan and Jane Chew,        
David and Linda Cornfield,
Michael Danylchuk,        
Jim and Hildy DeFrisco,   
William and Phyllis Draper,
Luke Ellis and Laura Sawczuk,
Jim Erbs and Shan Atkins,    
Alan Eustace and Kathy Kwan, 
Peter and Robin Frazier 
David Friedberg,             
Harvey Glasser,              
Charles and Gretchen Gooding,
Alan Gould and Diane Tokugawa,
Thomas and Dana Grogan,
Timothy and Judi Hachman 
Alan and Gladys Hoefer,
Charles and Patricia Hunt, 
Stephen and Catherine Imbler,
Adam and Rita Kablanian, 
Roger and Jody Lawler,
Kenneth and Gloria Levy,     
Peter Maier, 
DuBose and Nancy Montgomery,
Rand Morimoto and Ana Henderson,
Sunil Nagaraj and Mary Katherine Stimmler, 
Peter and Kristan Norvig,   
James and Marie O'Brient,  
Emilie and Doug Ogden,   
Paul and Sandra Otellini,     
Jeanne and Sanford Robertson,
Sissy Sailors and Red Conger, 
Stanley and Miriam Schiffman,
Thomas and Alison Schneider,
Ajay Shah and Lata Krishnan, 
Alex and Irina Shubat,     
the Silicon Valley Community Foundation,
Mary-Lou Smulders and Nicholas Hodson,
Hans Spiller,
Alan and Janet Stanford,
the Hugh Stuart Center Charitable Trust,
Clark and Sharon Winslow,
Weldon and Ruth Wood,
David and Angie Yancey, 
and many others.
X.G.W. is supported by the National Natural Science Foundation of China (NSFC grant 11673006) and the Guangxi Science Foundation (grants 2016GXNSFFA380006 and 2017AD22006).

We thank (mostly U.C. Berkeley undergraduate students)
Carmen Anderson,
James Bradley,
Stanley Browne,
Jieun Choi,
Ian Crossfield,
Griffin Foster,
Don Gavel,
Mark Gleed,
Christopher Griffith,
Jenifer Gross,
Andrew Halle,
Michael Hyland,
Anthony Khodanian,
Laura Langland,
Thomas Lowe,
Paul Lynam,
Emily Ma,
Kyle McAllister,
Alekzandir Morton,
Isha Nayak,
Daniel Perley,
Tyler Pritchard,
Andrew Rikhter,
Jackson Sipple,
Costas Soler,
Stephen Taylor,
Jeremy Wayland,
and Dustin Winslow,
for their effort in taking Lick/Nickel data.

This research has made use of the NASA/IPAC Extragalactic Database (NED), which is operated by the Jet Propulsion Laboratory, California Institute of Technology, under contract with NASA. The Pan-STARRS1 Surveys (PS1) and the PS1 public science archive have been made possible through contributions by the Institute for Astronomy, the University of Hawaii, the Pan-STARRS Project Office, the Max-Planck Society and its participating institutes, the Max Planck Institute for Astronomy, Heidelberg and the Max Planck Institute for Extraterrestrial Physics, Garching, The Johns Hopkins University, Durham University, the University of Edinburgh, the Queen's University Belfast, the Harvard-Smithsonian Center for Astrophysics, the Las Cumbres Observatory Global Telescope Network Incorporated, the National Central University of Taiwan, the Space Telescope Science Institute, NASA under Grant No. NNX08AR22G issued through the Planetary Science Division of the NASA Science Mission Directorate, the National Science Foundation Grant No. AST-1238877, the University of Maryland, Eotvos Lorand University (ELTE), the Los Alamos National Laboratory, and the Gordon and Betty Moore Foundation. Funding for the Sloan Digital Sky Survey (SDSS) has been provided by the Alfred P. Sloan Foundation, the Participating Institutions, NASA, the National Science Foundation, the U.S. Department of Energy, the Japanese Monbukagakusho, and the Max Planck Society. The SDSS Web site is \url{http://www.sdss.org/}. The SDSS is managed by the Astrophysical Research Consortium (ARC) for the Participating Institutions. The Participating Institutions are The University of Chicago, Fermilab, the Institute for Advanced Study, the Japan Participation Group, The Johns Hopkins University, Los Alamos National Laboratory, the Max-Planck-Institute for Astronomy (MPIA), the Max-Planck-Institute for Astrophysics (MPA), New Mexico State University, University of Pittsburgh, Princeton University, the United States Naval Observatory, and the University of Washington.








\appendix

\section{Sample Information}

\onecolumn
\begin{landscape}
{\footnotesize
\begin{longtable}{lrrlcccclcrrc}
\caption{SN~Ia sample.\label{tab:sample-information}}\\
\hline
\hline
\multicolumn{1}{l}{SN} & \multicolumn{1}{r}{R.A.$^{a}$} & \multicolumn{1}{r}{Decl.$^{a}$} & \multicolumn{1}{l}{Discovery$^{a}$} & \multicolumn{1}{c}{Discovery} & \multicolumn{1}{c}{Spectroscopic$^{b}$} & \multicolumn{1}{c}{Type$^{c}$} & \multicolumn{1}{c}{Host$^{a}$} & \multicolumn{1}{c}{$z_{\rm helio}$$^d$} & \multicolumn{1}{c}{$E(B-V)_{\rm MW}$$^e$} & \multicolumn{1}{c}{E$^f$} & \multicolumn{1}{c}{N$^f$} & \multicolumn{1}{c}{Host}\\
\multicolumn{1}{l}{Name} & \multicolumn{1}{r}{$\alpha(2000)$} & \multicolumn{1}{r}{$\delta(2000)$} & \multicolumn{1}{l}{Date (UT)} & \multicolumn{1}{c}{Reference} & \multicolumn{1}{c}{Reference} & \multicolumn{1}{c}{} & \multicolumn{1}{c}{Galaxy} & \multicolumn{1}{c}{} & \multicolumn{1}{c}{(mag)} & \multicolumn{1}{c}{($^{\prime \prime}$)} & \multicolumn{1}{c}{($^{\prime \prime}$)} & \multicolumn{1}{c}{Subtr.$^g$}\\
\hline
\endfirsthead
\hline
\hline
\multicolumn{1}{l}{SN} & \multicolumn{1}{r}{R.A.$^{a}$} & \multicolumn{1}{r}{Decl.$^{a}$} & \multicolumn{1}{l}{Discovery$^{a}$} & \multicolumn{1}{c}{Discovery} & \multicolumn{1}{c}{Spectroscopic$^{b}$} & \multicolumn{1}{c}{Type$^{c}$} & \multicolumn{1}{c}{Host$^{a}$} & \multicolumn{1}{c}{$z_{\rm helio}$$^d$} & \multicolumn{1}{c}{$E(B-V)_{\rm MW}$$^e$} & \multicolumn{1}{c}{E$^f$} & \multicolumn{1}{c}{N$^f$} & \multicolumn{1}{c}{Host}\\
\multicolumn{1}{l}{Name} & \multicolumn{1}{r}{$\alpha(2000)$} & \multicolumn{1}{r}{$\delta(2000)$} & \multicolumn{1}{l}{Date (UT)} & \multicolumn{1}{c}{Reference} & \multicolumn{1}{c}{Reference} & \multicolumn{1}{c}{} & \multicolumn{1}{c}{Galaxy} & \multicolumn{1}{c}{} & \multicolumn{1}{c}{(mag)} & \multicolumn{1}{c}{($^{\prime \prime}$)} & \multicolumn{1}{c}{($^{\prime \prime}$)} & \multicolumn{1}{c}{Subtr.$^g$}\\
\hline
\endhead
\hline
\multicolumn{13}{c}{Table \thetable~continued}
\endfoot
\hline
\multicolumn{13}{p{24cm}}{$^a$Basic information for each SN, including its J2000 right ascension and declination (in decimal degrees), its host galaxy, and its discovery date, were sourced from TNS. However, host galaxies marked with a ``$\dagger$'' symbol were obtained from \citet{2012AA...538A.120L}, while those with a ``$\ddagger$'' are from the given discovery reference.}\\
\multicolumn{13}{p{24cm}}{$^b$Spectroscopic classification reference. Ph07 refers to \citet{2005hk}.}\\
\multicolumn{13}{p{24cm}}{$^c$Spectroscopic type as classified in the spectroscopic reference. Super-Chandrasekhar candidates are labeled with ``SC''.}\\
\multicolumn{13}{p{24cm}}{$^d$Host-galaxy heliocentric redshifts are from NED unless otherwise indicated. Those marked with a ``$\ddagger$'' symbol were obtained from their spectroscopic references, and ``$\pm$'' refers to \citet{2010ApJ...713.1073S} and ``$\mp$'' to \citet{2012AA...538A.120L}.}\\
\multicolumn{13}{p{24cm}}{$^e$Extinction is calculated at the SN position using the dust maps of \citet{Schlegel98} subject to the recalibration of \citet{Schlafly2011}.}\\
\multicolumn{13}{p{24cm}}{$^f$Offsets from host-galaxy nuclei are computed using the host location as given by NED (if available) for all SNe except SN~2010hs, whose host coordinates are from the catalog of \citet{2012AA...538A.120L}.}\\
\multicolumn{13}{p{24cm}}{$^g$Indicates whether the SN had its host galaxy subtracted (Y) or not (N).}
\endlastfoot
  2005hk &    $6.96196$ &   $-1.19792$ &       30 Oct 2005 &          IAUC 8625 &     CBET 269, Ph07 &        Iax &                               UGC 272 &             0.013 &   0.020 &   $16.9$ &     $7.5$ &    Y \\
  2005ki &  $160.11758$ &    $9.20233$ &       18 Nov 2005 &           CBET 294 &           CBET 296 &         Ia &                              NGC 3332 &             0.019 &   0.027 &   $-2.2$ &    $71.2$ &    N \\
  2006ev &  $322.74692$ &   $13.98922$ &       12 Sep 2006 &          IAUC 8747 &           CBET 622 &         Ia &                             UGC 11758 &             0.029 &   0.077 &   $23.9$ &    $11.3$ &    Y \\
  2006mq &  $121.55162$ &  $-27.56261$ &       22 Oct 2006 &           CBET 721 &           CBET 724 &         Ia &                          ESO 494--G26 &             0.003 &   0.362 &   $17.3$ &  $-123.1$ &    N \\
   2007F &  $195.81283$ &   $50.61881$ &       11 Jan 2007 &           CBET 803 &           CBET 805 &         Ia &                              UGC 8162 &             0.024 &   0.015 &   $-9.8$ &    $-7.0$ &    Y \\
  2007bd &  $127.88867$ &   $-1.19944$ &        4 Apr 2007 &           CBET 914 &           CBET 915 &         Ia &                              UGC 4455 &             0.031 &   0.029 &    $6.0$ &    $-6.2$ &    Y \\
  2007bm &  $171.25958$ &   $-9.79828$ &       20 Apr 2007 &           CBET 936 &           CBET 939 &         Ia &                              NGC 3672 &             0.006 &   0.035 &   $-2.5$ &   $-10.4$ &    Y \\
  2007fb &  $359.21821$ &    $5.50883$ &        3 Jul 2007 &           CBET 992 &           CBET 993 &         Ia &                             UGC 12859 &             0.018 &   0.048 &   $12.2$ &     $1.5$ &    Y \\
  2007fs &   $330.4185$ &  $-21.50822$ &       15 Jul 2007 &          CBET 1002 &          CBET 1003 &         Ia &                           ESO 601--G5 &             0.017 &   0.029 &   $34.5$ &    $10.6$ &    Y \\
  2007if &   $17.71404$ &   $15.46108$ &       16 Aug 2007 &          CBET 1059 &          CBET 1059 &         SC &                                 Anon. &       0.074$^\pm$ &   0.071 &      ... &       ... &    N \\
  2007jg &   $52.46175$ &    $0.05683$ &       14 Sep 2007 &          CBET 1076 &          CBET 1076 &         Ia &  SDSS J032950.83 + 000316.0$^\dagger$ &             0.037 &   0.091 &   $-0.1$ &     $8.6$ &    Y \\
  2007kk &   $55.59692$ &   $39.24178$ &       28 Sep 2007 &          CBET 1096 &          CBET 1097 &         Ia &                              UGC 2828 &             0.041 &   0.196 &   $-9.1$ &    $-9.9$ &    Y \\
   2008Y &  $169.87737$ &   $54.46283$ &        6 Feb 2008 &          CBET 1240 &          CBET 1246 &         Ia &                       MCG +09--19--39 &             0.070 &   0.011 &   $-2.3$ &     $7.1$ &    Y \\
  2008dh &    $8.79717$ &   $23.25419$ &        8 Jun 2008 &          CBET 1409 &          CBET 1409 &         Ia &                 PGC 1684149$^\dagger$ &             0.037 &   0.026 &   $12.2$ &    $-3.0$ &    Y \\
  2008ds &    $7.46179$ &   $31.39275$ &       28 Jun 2008 &          CBET 1419 &          CBET 1419 &     Ia-pec &                               UGC 299 &             0.021 &   0.055 &  $-33.0$ &    $-2.2$ &    Y \\
  2008eg &   $27.90112$ &   $19.10469$ &       20 Jul 2008 &          CBET 1444 &          CBET 1444 &         Ia &                              UGC 1324 &             0.034 &   0.057 &    $0.3$ &     $4.3$ &    Y \\
  2008ek &  $241.38821$ &   $17.59256$ &       28 Jul 2008 &          CBET 1452 &          CBET 1454 &         Ia &                               IC 1181 &             0.033 &   0.038 &   $-9.7$ &    $-4.1$ &    Y \\
  2008eo &   $10.46683$ &   $32.99033$ &        3 Aug 2008 &          CBET 1459 &          CBET 1465 &         Ia &                               UGC 442 &             0.016 &   0.070 &    $4.4$ &    $-3.5$ &    Y \\
  2008eq &     $255.03$ &   $23.13239$ &        2 Aug 2008 &          CBET 1460 &          CBET 1465 &         Ia &                  PGC 214560$^\dagger$ &             0.057 &   0.063 &    $4.1$ &     $3.6$ &    Y \\
  2008fk &   $38.52108$ &    $1.39514$ &        2 Sep 2008 &          CBET 1494 &          CBET 1499 &         Ia &     2MASX J02340513+0123408$^\dagger$ &             0.072 &   0.020 &   $-1.2$ &     $1.9$ &    Y \\
  2008fu &   $45.61875$ &  $-24.45597$ &       25 Sep 2008 &          CBET 1517 &          CBET 1519 &         Ia &                         ESO 480--IG21 &             0.052 &   0.019 &   $-2.6$ &    $-0.5$ &    Y \\
  2008gg &     $21.346$ &  $-18.17244$ &        9 Oct 2008 &          CBET 1538 &          CBET 1540 &         Ia &                               NGC 539 &             0.032 &   0.021 &   $18.7$ &   $-30.9$ &    N \\
  2008gl &   $20.22842$ &    $4.80531$ &       20 Oct 2008 &          CBET 1545 &          CBET 1547 &         Ia &                               UGC 881 &             0.034 &   0.024 &   $20.2$ &    $14.3$ &    Y \\
  2008go &  $332.68679$ &  $-20.78811$ &       22 Oct 2008 &          CBET 1553 &          CBET 1554 &         Ia &                       Anon.$^\dagger$ &             0.062 &   0.032 &   $11.9$ &     $8.8$ &    N \\
  2008gp &   $50.75304$ &    $1.36189$ &       27 Oct 2008 &          CBET 1555 &          CBET 1558 &         Ia &                        MCG +00--9--74 &             0.033 &   0.104 &   $10.9$ &   $-14.0$ &    Y \\
  2008ha &  $353.71954$ &    $18.2265$ &        7 Nov 2008 &          CBET 1567 &          CBET 1576 &        Iax &                             UGC 12682 &             0.005 &   0.068 &  $-11.5$ &    $-2.6$ &    Y \\
  2008hs &   $36.37342$ &   $41.84308$ &        1 Dec 2008 &          CBET 1598 &          CBET 1599 &         Ia &                               NGC 910 &             0.017 &   0.049 &   $31.7$ &    $67.7$ &    N \\
   2009D &   $58.59512$ &  $-19.18172$ &        2 Jan 2009 &          CBET 1647 &          CBET 1647 &         Ia &                      MCG --03--10--52 &             0.025 &   0.046 &  $-26.1$ &    $30.9$ &    N \\
  2009al &  $162.84196$ &    $8.57853$ &       26 Feb 2009 &          CBET 1705 &          CBET 1708 &         Ia &                    NGC 3425$^\dagger$ &             0.022 &   0.021 &  $-51.3$ &    $41.0$ &    N \\
  2009an &  $185.69779$ &   $65.85117$ &       27 Feb 2009 &          CBET 1707 &          CBET 1709 &         Ia &                              NGC 4332 &             0.009 &   0.016 &    $4.4$ &    $26.6$ &    Y \\
  2009dc &   $237.8005$ &   $25.70778$ &        9 Apr 2009 &          CBET 1762 &          CBET 1776 &         SC &                             UGC 10064 &             0.021 &   0.060 &  $-15.7$ &    $21.1$ &    N \\
  2009ee &  $170.35542$ &   $34.33981$ &        9 May 2009 &          CBET 1795 &          CBET 1802 &         Ia &                               IC 2738 &             0.035 &   0.021 &   $27.7$ &   $-60.7$ &    N \\
  2009eq &  $280.03458$ &   $40.12681$ &       11 May 2009 &          CBET 1805 &          CBET 1817 &     Ia-pec &                              NGC 6686 &             0.024 &   0.053 &   $14.7$ &   $-39.0$ &    N \\
  2009eu &  $247.17137$ &   $39.55347$ &       21 May 2009 &          CBET 1813 &          CBET 1817 &         Ia &                              NGC 6166 &             0.030 &   0.010 &   $30.6$ &     $6.9$ &    Y \\
  2009fv &  $247.43425$ &   $40.81161$ &        2 Jun 2009 &          CBET 1834 &          CBET 1846 &         Ia &                              NGC 6173 &             0.029 &   0.005 &   $-7.7$ &     $0.0$ &    Y \\
  2009hn &   $38.00129$ &    $1.24819$ &       24 Jul 2009 &          CBET 1886 &          CBET 1889 &         Ia &                              UGC 2005 &             0.022 &   0.021 &   $38.1$ &     $6.0$ &    Y \\
  2009hp &   $44.59983$ &    $6.59308$ &       26 Jul 2009 &          CBET 1888 &          CBET 1889 &         Ia &                       MCG +01--08--30 &             0.021 &   0.198 &   $-9.2$ &     $4.6$ &    Y \\
  2009hs &  $268.96221$ &   $62.59975$ &       28 Jul 2009 &          CBET 1892 &          CBET 1909 &  91bg-like &                              NGC 6521 &             0.027 &   0.035 &   $17.2$ &   $-45.0$ &    N \\
  2009ig &   $39.54837$ &   $-1.31253$ &       20 Aug 2009 &          CBET 1918 &          CBET 1918 &         Ia &                              NGC 1015 &             0.009 &   0.028 &    $0.7$ &    $22.2$ &    Y \\
  2009kq &  $129.06288$ &   $28.06714$ &        5 Nov 2009 &          CBET 2005 &          ATEL 2291 &         Ia &                        MCG +05--21--1 &             0.012 &   0.035 &   $-4.2$ &    $24.5$ &    Y \\
  2010ao &  $205.92079$ &    $3.90003$ &       18 Mar 2010 &          CBET 2211 &          CBET 2223 &         Ia &                              UGC 8686 &             0.023 &   0.023 &   $11.8$ &    $14.5$ &    Y \\
  2010hs &   $36.41308$ &   $24.76489$ &       12 Sep 2010 &          CBET 2454 &          CBET 2461 &         Ia &                 PGC 1715790$^\dagger$ &       0.076$^\mp$ &   0.100 &  $-93.4$ &   $-46.4$ &    N \\
  2010ii &  $339.55492$ &   $35.49167$ &       30 Sep 2010 &          CBET 2474 &          CBET 2474 &         Ia &                              NGC 7342 &             0.027 &   0.075 &    $0.4$ &   $-25.9$ &    Y \\
  2010ju &   $85.48329$ &    $18.4975$ &       14 Nov 2010 &          CBET 2549 &          CBET 2550 &         Ia &                              UGC 3341 &             0.015 &   0.361 &    $6.3$ &    $18.5$ &    Y \\
   2011M &   $75.17312$ &   $62.24406$ &       19 Jan 2011 &          CBET 2640 &          CBET 2640 &         Ia &                              UGC 3218 &             0.017 &   0.352 &  $-15.1$ &     $0.1$ &    Y \\
  2011bd &  $266.77633$ &   $57.30131$ &       24 Mar 2011 &          CBET 2685 &          CBET 2685 &         Ia &                              NGC 6473 &  0.028$^\ddagger$ &   0.041 &    $3.3$ &   $-31.0$ &    Y \\
  2011by &  $178.93983$ &   $55.32606$ &       26 Apr 2011 &          CBET 2708 &          CBET 2708 &         Ia &                              NGC 3972 &             0.003 &   0.012 &    $4.0$ &    $19.1$ &    Y \\
  2011df &  $291.89017$ &   $54.38647$ &       21 May 2011 &          CBET 2729 &          CBET 2729 &         Ia &                              NGC 6801 &             0.014 &   0.112 &  $-19.0$ &    $48.9$ &    Y \\
  2011dl &  $244.52071$ &   $21.55111$ &       17 Jun 2011 &          CBET 2744 &          CBET 2744 &         Ia &                             UGC 10321 &  0.026$^\ddagger$ &   0.067 &  $-18.6$ &   $-35.0$ &    N \\
  2011dz &  $243.18675$ &   $28.28422$ &       26 Jun 2011 &          CBET 2761 &          CBET 2761 &         Ia &                             UGC 10273 &             0.025 &   0.044 &   $-2.4$ &   $-61.8$ &    Y \\
  2011ek &   $36.45371$ &   $18.53333$ &        4 Aug 2011 &          CBET 2783 &          CBET 2783 &         Ia &                               NGC 918 &             0.005 &   0.307 &  $-27.7$ &   $133.5$ &    Y \\
  2011fe &  $210.77421$ &   $54.27372$ &       24 Aug 2011 &          CBET 2792 &          CBET 2792 &         Ia &                       M101$^\ddagger$ &             0.001 &   0.008 &  $-59.3$ &  $-270.1$ &    Y \\
  2011fs &  $334.33133$ &   $35.58056$ &       15 Sep 2011 &          CBET 2825 &          CBET 2825 &         Ia &                             UGC 11975 &             0.021 &   0.101 &   $-2.7$ &    $33.8$ &    N \\
   2012E &   $38.34496$ &    $9.58489$ &       14 Jan 2012 &          CBET 2981 &          CBET 2981 &         Ia &                               NGC 975 &             0.020 &   0.063 &    $0.6$ &   $-60.5$ &    N \\
   2012Z &   $50.52229$ &  $-15.38767$ &       29 Jan 2012 &          CBET 3014 &          CBET 3014 &        Iax &                              NGC 1309 &             0.007 &   0.034 &  $-17.5$ &    $44.6$ &    N \\
  2012bh &  $183.40546$ &   $46.48347$ &       11 Mar 2012 &          CBET 3066 &          CBET 3066 &         Ia &                              UGC 7228 &             0.025 &   0.016 &    $5.2$ &   $-37.8$ &    N \\
  2012cg &  $186.80346$ &    $9.42033$ &       17 May 2012 &          CBET 3111 &          CBET 3111 &         Ia &                              NGC 4424 &             0.001 &   0.018 &   $18.1$ &    $-1.2$ &    Y \\
  2012dn &  $305.90108$ &  $-28.27872$ &        8 Jul 2012 &          CBET 3174 &          CBET 3174 &         SC &                  PGC 64605$^\ddagger$ &             0.010 &   0.052 &      ... &       ... &    Y \\
  2012ea &  $266.29333$ &   $18.14078$ &        8 Aug 2012 &          CBET 3199 &          CBET 3199 &  91bg-like &                              NGC 6430 &             0.010 &   0.055 &  $-55.2$ &     $6.6$ &    N \\
  2012gl &  $153.20967$ &   $12.68242$ &       29 Oct 2012 &          CBET 3302 &          CBET 3302 &         Ia &                              NGC 3153 &             0.009 &   0.036 &   $-2.6$ &    $56.7$ &    N \\
  2013bs &  $259.34179$ &   $41.06672$ &       18 Apr 2013 &          CBET 3494 &          CBET 3494 &         Ia &                              NGC 6343 &             0.028 &   0.025 &   $65.1$ &    $50.4$ &    N \\
  2013dh &  $232.50454$ &   $12.98692$ &       12 Jun 2013 &          CBET 3561 &          CBET 3561 &   91T-like &                              NGC 5936 &             0.013 &   0.033 &    $3.8$ &    $-8.7$ &    Y \\
  2013dr &  $259.87608$ &   $47.70128$ &        1 Jul 2013 &          CBET 3576 &          CBET 3576 &         Ia &                  PGC 60077$^\ddagger$ &             0.017 &   0.021 &   $-8.7$ &    $-4.3$ &    Y \\
  2013dy &  $334.57333$ &   $40.56933$ &       10 Jul 2013 &          CBET 3588 &          CBET 3588 &         Ia &                              NGC 7250 &             0.004 &   0.132 &   $-2.3$ &    $25.0$ &    Y \\
  2013ex &   $83.19425$ &  $-14.04594$ &       19 Aug 2013 &          CBET 3635 &          CBET 3635 &         Ia &                              NGC 1954 &             0.010 &   0.123 &  $-24.9$ &    $60.6$ &    N \\
  2013fa &  $310.97321$ &   $12.51436$ &       25 Aug 2013 &          CBET 3641 &          CBET 3641 &         Ia &                              NGC 6956 &             0.016 &   0.086 &   $-2.1$ &     $8.8$ &    Y \\
  2013fw &  $318.43654$ &   $13.57592$ &       21 Oct 2013 &          CBET 3681 &          CBET 3681 &         Ia &                              NGC 7042 &             0.017 &   0.067 &  $-15.9$ &     $3.6$ &    Y \\
  2013gh &    $330.591$ &  $-18.91678$ &        8 Aug 2013 &          CBET 3706 &          CBET 3706 &         Ia &                              NGC 7183 &             0.009 &   0.025 &    $3.1$ &    $-1.0$ &    Y \\
  2013gq &  $124.47275$ &   $23.46958$ &       25 Mar 2013 &          CBET 3730 &          CBET 3730 &         Ia &                              NGC 2554 &             0.014 &   0.049 &   $-0.4$ &    $-9.2$ &    Y \\
  2013gy &   $55.57033$ &   $-4.72181$ &        6 Dec 2013 &          CBET 3743 &          CBET 3743 &         Ia &                              NGC 1418 &             0.014 &   0.050 &   $10.8$ &    $32.2$ &    N \\
   2014J &  $148.92558$ &   $69.67389$ &       21 Jan 2014 &          CBET 3792 &          CBET 3792 &         Ia &                              NGC 3034 &             0.001 &   0.136 &  $-55.2$ &   $-19.8$ &    Y \\
  2014ai &  $139.93404$ &   $33.76378$ &       21 Mar 2014 &          CBET 3838 &          CBET 3838 &         Ia &                              NGC 2832 &             0.023 &   0.015 &  $-33.5$ &    $50.5$ &    N \\
  2014ao &  $128.63883$ &   $-2.54336$ &       17 Apr 2014 &          CBET 3855 &          CBET 3855 &         Ia &                              NGC 2615 &             0.014 &   0.031 &   $-0.4$ &    $12.4$ &    Y \\
  2014bj &  $290.66312$ &   $43.89081$ &       22 May 2014 &          CBET 3893 &          CBET 3893 &         Ia &                                 Anon. &  0.005$^\ddagger$ &   0.091 &      ... &       ... &    N \\
  2014dt &  $185.48987$ &    $4.47181$ &       29 Oct 2014 &          CBET 4011 &          CBET 4011 &        Iax &                              NGC 4303 &             0.005 &   0.019 &   $39.9$ &    $-6.6$ &    Y \\
   2015N &  $325.82037$ &   $43.57989$ &        6 Jul 2015 &          CBET 4124 &          CBET 4124 &         Ia &                             UGC 11797 &             0.019 &   0.456 &  $-36.1$ &    $12.9$ &    Y \\
 2016aew &  $212.86037$ &    $1.28596$ &       12 Feb 2016 &   TNSTR--2016--106 &   TNSCR--2016--114 &         Ia &                               IC 0986 &             0.025 &   0.033 &    $3.9$ &    $-2.0$ &    Y \\
 2016coj &  $182.02833$ &   $65.17729$ &       28 May 2016 &   TNSTR--2016--384 &   TNSCR--2016--386 &         Ia &                              NGC 4125 &             0.005 &   0.016 &    $4.9$ &    $11.3$ &    Y \\
 2016fbk &   $26.02737$ &   $34.38283$ &       16 Aug 2016 &   TNSTR--2016--568 &   TNSCR--2016--572 &         Ia &                             UGC 01212 &             0.036 &   0.042 &  $-19.6$ &   $-16.1$ &    Y \\
 2016ffh &  $227.95617$ &   $46.25089$ &       17 Aug 2016 &   TNSTR--2016--583 &   TNSCR--2016--589 &         Ia &                         CGCG 249--011 &             0.018 &   0.024 &   $11.4$ &   $-10.7$ &    Y \\
 2016gcl &  $354.48592$ &   $27.27715$ &        8 Sep 2016 &   TNSTR--2016--644 &   TNSCR--2016--655 &   91T-like &                            AGC 331536 &             0.028 &   0.063 &   $-2.7$ &    $-1.5$ &    Y \\
 2016gdt &  $328.09396$ &    $3.42181$ &        8 Sep 2016 &   TNSTR--2016--652 &   TNSCR--2016--666 &  91bg-like &                               IC 1407 &             0.029 &   0.072 &  $-13.3$ &   $-19.3$ &    N \\
 2016hvl &    $101.009$ &   $12.39662$ &        4 Nov 2016 &   TNSTR--2016--884 &   TNSCR--2016--892 &         Ia &                              UGC 3524 &             0.013 &   0.377 &   $22.9$ &   $-19.2$ &    N \\
 2017cfd &  $130.20479$ &   $73.48754$ &       16 Mar 2017 &   TNSTR--2017--315 &   TNSCR--2017--325 &         Ia &                                IC 511 &             0.012 &   0.019 &   $-5.5$ &     $3.1$ &    Y \\
 2017drh &  $263.10854$ &     $7.0632$ &        3 May 2017 &   TNSTR--2017--513 &   TNSCR--2017--516 &         Ia &                              NGC 6384 &             0.006 &   0.106 &   $26.1$ &    $10.5$ &    Y \\
 2017dws &  $235.05904$ &   $11.34486$ &        3 May 2017 &   TNSTR--2017--528 &   TNSCR--2017--534 &         Ia &                                 Anon. &  0.082$^\ddagger$ &   0.035 &      ... &       ... &    Y \\
 2017erp &  $227.31171$ &  $-11.33422$ &       13 Jun 2017 &   TNSTR--2017--647 &   TNSCR--2017--655 &         Ia &                              NGC 5861 &             0.006 &   0.093 &  $-18.8$ &   $-45.2$ &    N \\
 2017fgc &   $20.06017$ &    $3.40277$ &       11 Jul 2017 &   TNSTR--2017--753 &   TNSCR--2017--757 &         Ia &                              NGC 0474 &             0.008 &   0.029 &  $116.0$ &   $-45.4$ &    N \\
 2017glx &  $295.91787$ &   $56.11008$ &        3 Sep 2017 &   TNSTR--2017--963 &   TNSCR--2017--970 &   91T-like &                              NGC 6824 &             0.011 &   0.107 &   $-3.4$ &     $2.2$ &    Y \\
 2017hbi &   $38.13154$ &    $35.4836$ &        2 Oct 2017 &  TNSTR--2017--1066 &  TNSCR--2017--1074 &         Ia &                                 Anon. &  0.040$^\ddagger$ &   0.061 &      ... &       ... &    N \\
 2018aoz &  $177.75762$ &  $-28.74406$ &        2 Apr 2018 &   TNSTR--2018--428 &   TNSCR--2018--433 &         Ia &                              NGC 3923 &             0.006 &   0.072 &    $1.8$ &   $223.1$ &    N \\
 2018dem &  $317.99387$ &    $-0.2181$ &        8 Jul 2018 &   TNSTR--2018--947 &  TNSCR--2018--1219 &         Ia &             SDSS J211158.77--001309.9 &             0.060 &   0.072 &   $-3.6$ &     $4.8$ &    Y \\
  2018gv &  $121.39421$ &  $-11.43786$ &       15 Jan 2018 &    TNSTR--2018--57 &    TNSCR--2018--75 &         Ia &                              NGC 2525 &             0.005 &   0.050 &  $-50.4$ &   $-39.0$ &    Y \\
\end{longtable}}
\end{landscape}
\twocolumn

\section{Light Curves}

\begin{figure*}
\includegraphics[width=\textwidth]{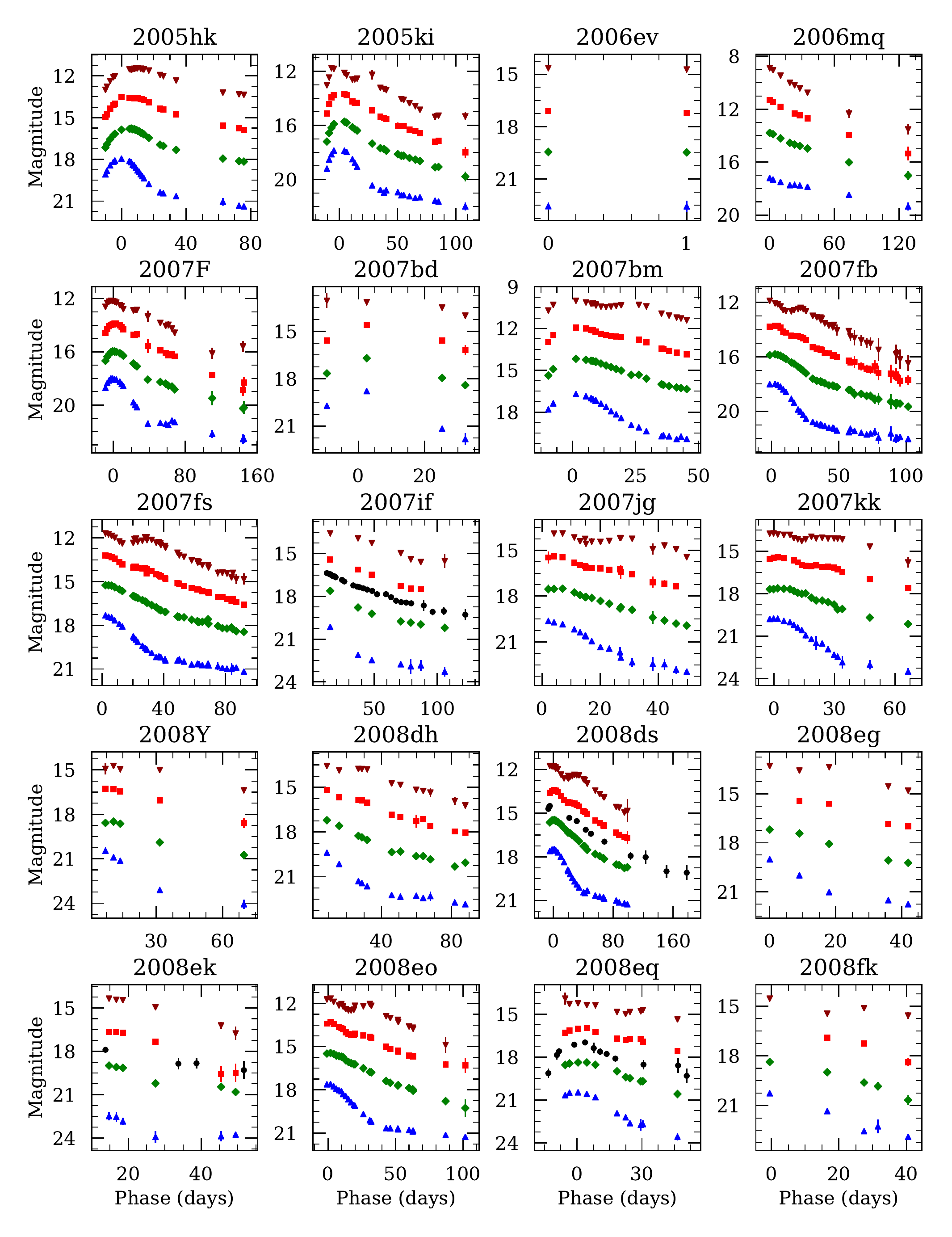}
\caption{Observed \emph{BVRI} and unfiltered light curves of our SN~Ia sample. Blue up-triangles are $B+2$, green diamonds are $V$, red squares are $R - 2$, dark red down-triangles are $I - 4$, and black circles are {\it Clear}$ - 1$. In most cases the error bars are smaller than the points themselves. All dates have been shifted relative to the time of maximum $B$-band brightness, if determined, and relative to the time of the first epoch otherwise.\label{fig:lc-sample}}
\end{figure*}

\begin{figure*}
\addtocounter{figure}{-1}
\includegraphics[width=\textwidth]{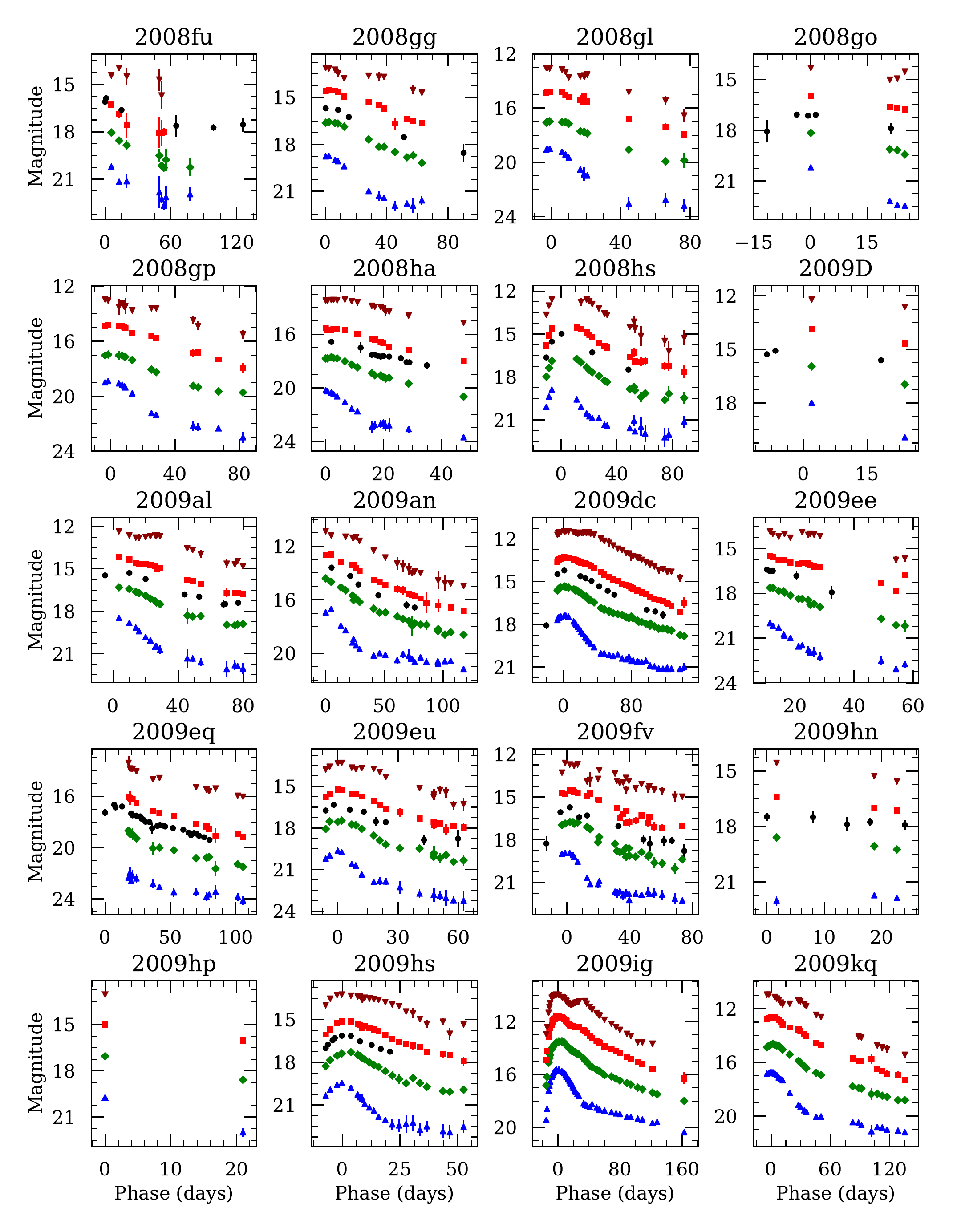}
\caption{Observed \emph{BVRI} and unfiltered light curves of our SN~Ia sample. Blue up-triangles are $B+2$, green diamonds are $V$, red squares are $R - 2$, dark red down-triangles are $I - 4$, and black circles are {\it Clear}$ - 1$. In most cases the error bars are smaller than the points themselves. All dates have been shifted relative to the time of maximum $B$-band brightness, if determined, and relative to the time of the first epoch otherwise.\label{fig:lc-sample}}
\end{figure*}

\begin{figure*}
\addtocounter{figure}{-1}
\includegraphics[width=\textwidth]{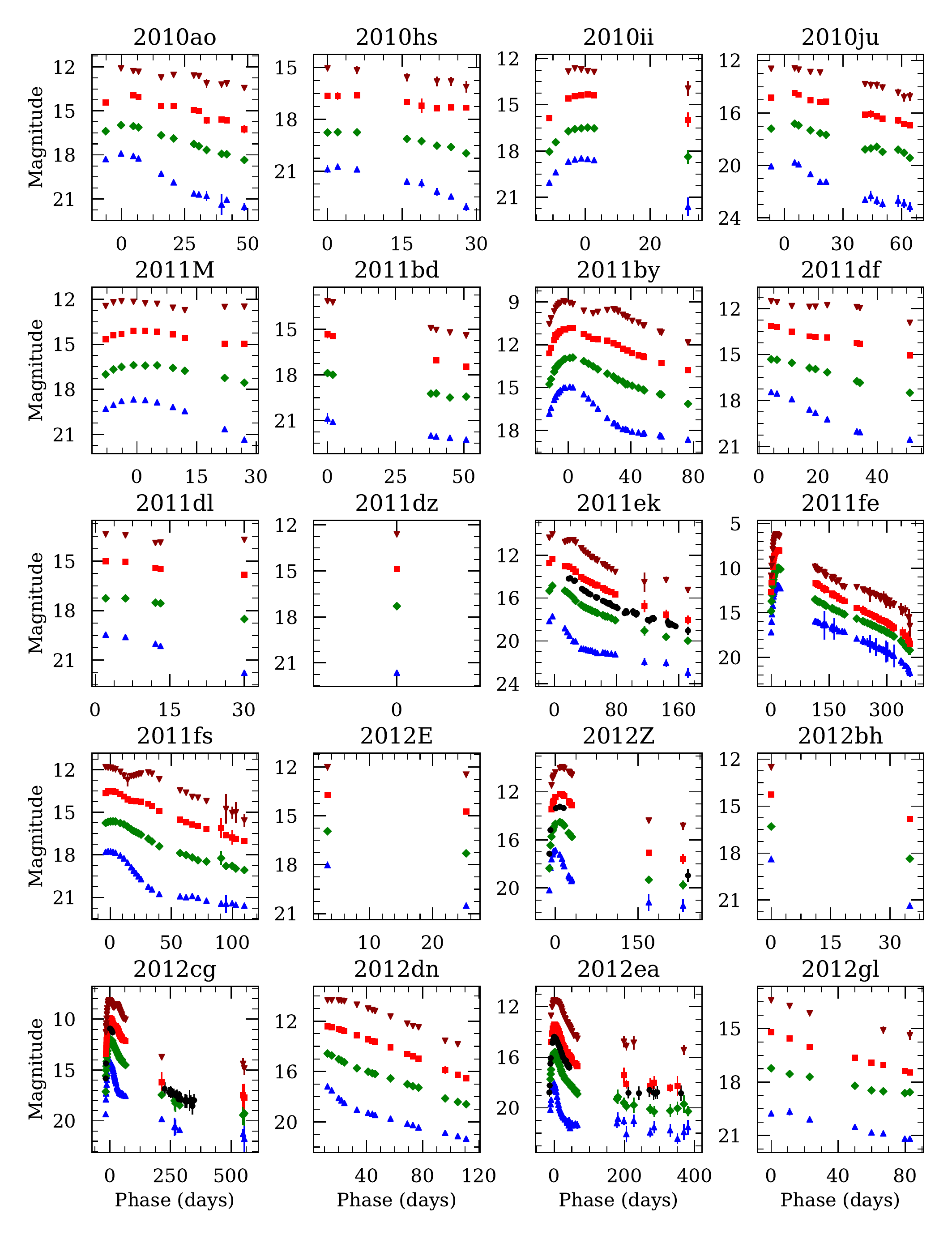}
\caption{Observed \emph{BVRI} and unfiltered light curves of our SN~Ia sample. Blue up-triangles are $B+2$, green diamonds are $V$, red squares are $R - 2$, dark red down-triangles are $I - 4$, and black circles are {\it Clear}$ - 1$. In most cases the error bars are smaller than the points themselves. All dates have been shifted relative to the time of maximum $B$-band brightness, if determined, and relative to the time of the first epoch otherwise.\label{fig:lc-sample}}
\end{figure*}

\begin{figure*}
\addtocounter{figure}{-1}
\includegraphics[width=\textwidth]{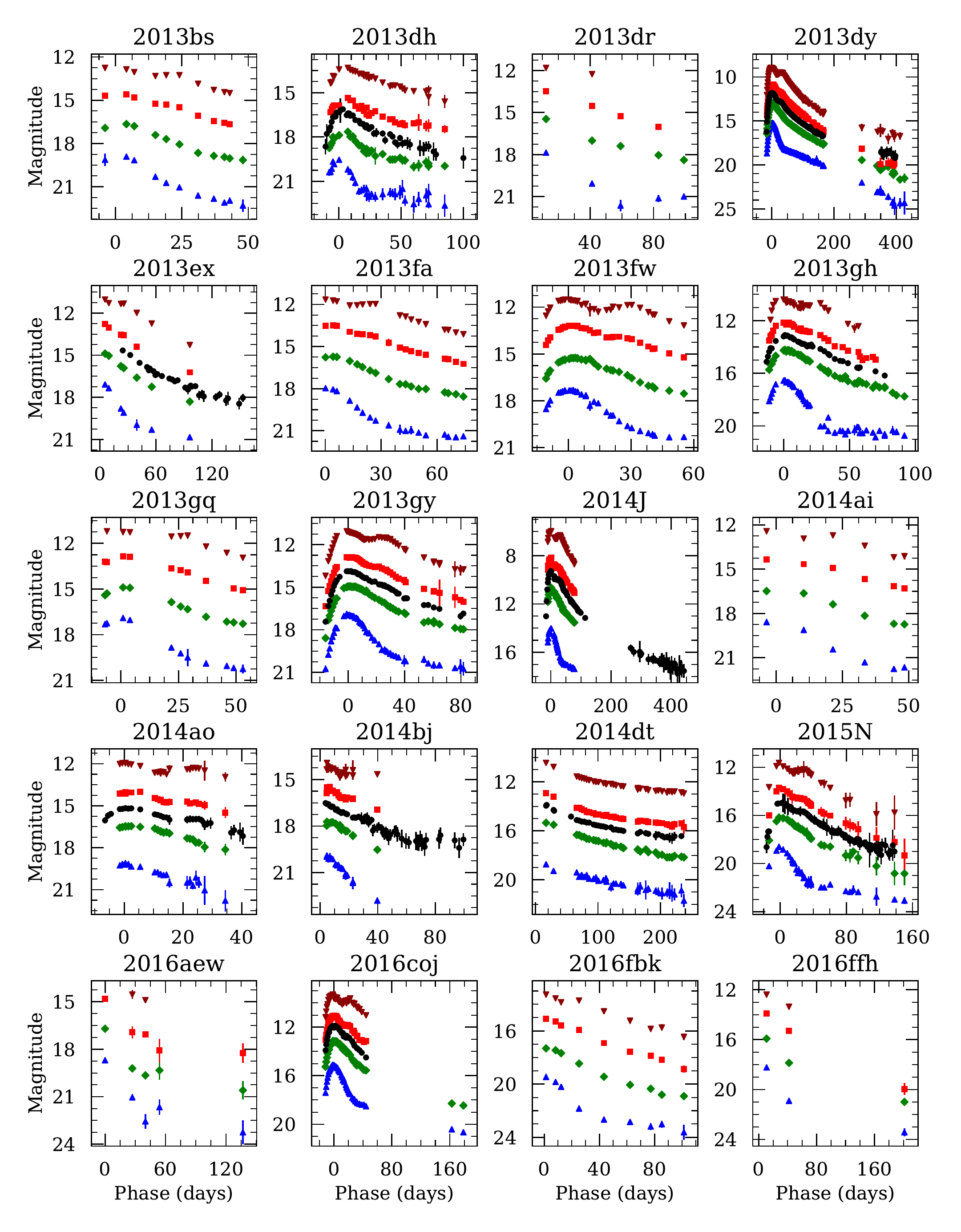}
\caption{Observed \emph{BVRI} and unfiltered light curves of our SN~Ia sample. Blue up-triangles are $B+2$, green diamonds are $V$, red squares are $R - 2$, dark red down-triangles are $I - 4$, and black circles are {\it Clear}$ - 1$. In most cases the error bars are smaller than the points themselves. All dates have been shifted relative to the time of maximum $B$-band brightness, if determined, and relative to the time of the first epoch otherwise.\label{fig:lc-sample}}
\end{figure*}

\begin{figure*}
\addtocounter{figure}{-1}
\includegraphics[width=\textwidth]{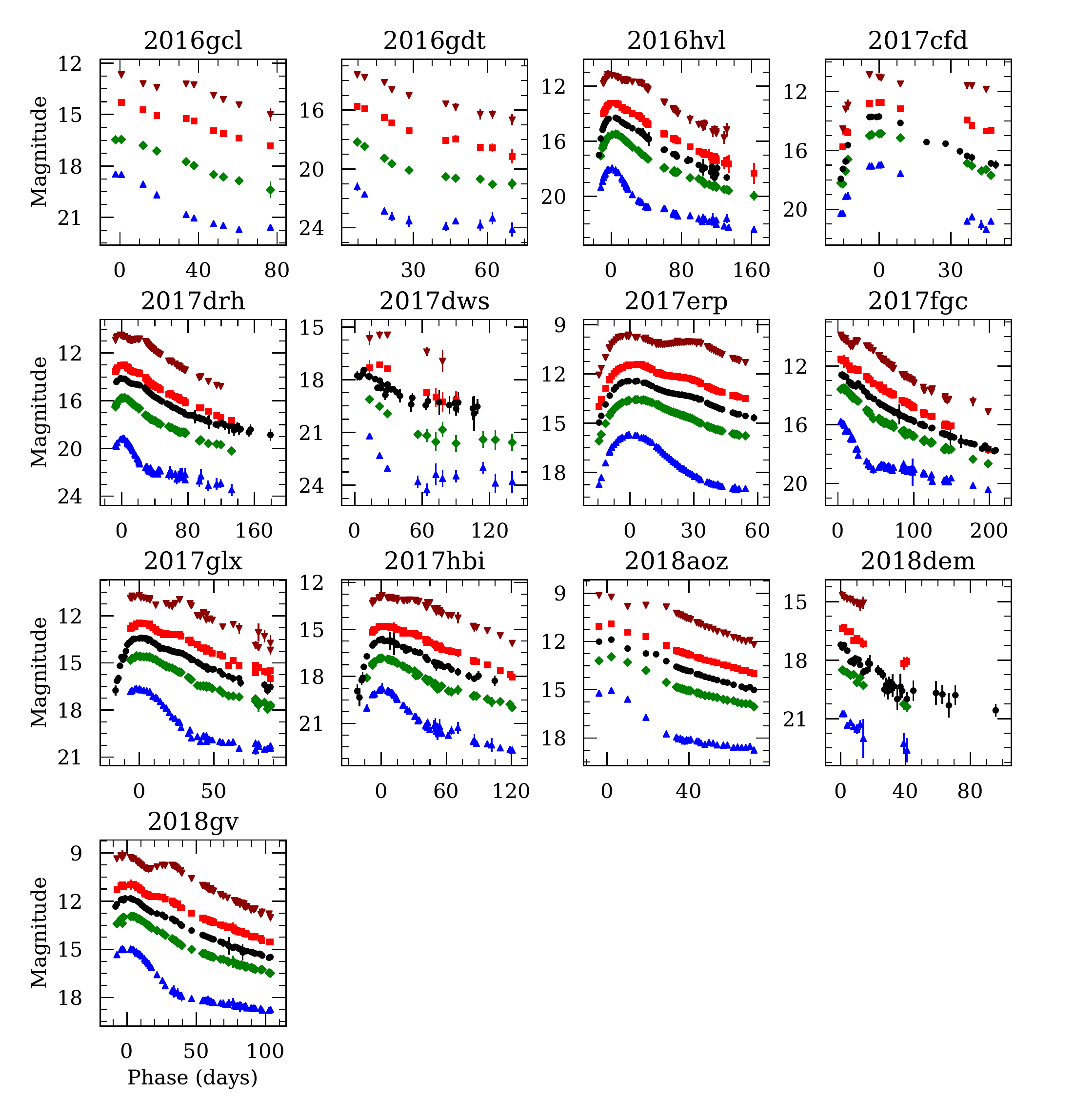}
\caption{Observed \emph{BVRI} and unfiltered light curves of our SN~Ia sample. Blue up-triangles are $B+2$, green diamonds are $V$, red squares are $R - 2$, dark red down-triangles are $I - 4$, and black circles are {\it Clear}$ - 1$. In most cases the error bars are smaller than the points themselves. All dates have been shifted relative to the time of maximum $B$-band brightness, if determined, and relative to the time of the first epoch otherwise.\label{fig:lc-sample}}
\end{figure*}

\subsection{Light-Curve Properties}

\begin{table*}
\caption{Light-curve properties derived from Gaussian Process interpolation.\label{tab:gp-fits}}
\begin{tabular}{lrrrrrrr}
\hline
\hline
\multicolumn{1}{c}{SN} & \multicolumn{1}{c}{$t_{B_{\rm max}}$ (MJD)} & \multicolumn{1}{c}{$B_{\rm max}$ (mag)} & \multicolumn{1}{c}{$\Delta m_{15}(B)$ (mag)} & \multicolumn{1}{c}{$t_{V_{\rm max}}$ (MJD)} & \multicolumn{1}{c}{$V_{\rm max}$ (mag)} & \multicolumn{1}{c}{$\Delta m_{15}(V)$ (mag)} & \multicolumn{1}{c}{$(B - V)_{B_{\rm max}}$ (mag)}\\
\hline
  2005hk &  $53684.32 \pm 0.29$ &  $15.850 \pm 0.022$ &  $1.580 \pm 0.053$ &  $53688.11 \pm 0.56$ &  $15.703 \pm 0.018$ &  $0.799 \pm 0.039$ &   $0.069 \pm 0.029$ \\
  2005ki &  $53704.67 \pm 0.40$ &  $15.572 \pm 0.042$ &  $1.275 \pm 0.080$ &  $53705.97 \pm 0.56$ &  $15.534 \pm 0.043$ &  $0.826 \pm 0.067$ &   $0.021 \pm 0.060$ \\
   2007F &  $54122.32 \pm 0.45$ &  $15.975 \pm 0.016$ &  $0.864 \pm 0.085$ &  $54124.03 \pm 0.54$ &  $15.928 \pm 0.011$ &  $0.550 \pm 0.069$ &   $0.029 \pm 0.020$ \\
  2007bd &  $54210.22 \pm 1.87$ &  $16.680 \pm 0.051$ &  $1.451 \pm 0.248$ &  $54212.58 \pm 1.74$ &  $16.552 \pm 0.074$ &  $0.891 \pm 0.177$ &   $0.095 \pm 0.090$ \\
  2007bm &  $54224.46 \pm 0.50$ &  $14.548 \pm 0.022$ &  $1.232 \pm 0.057$ &  $54225.66 \pm 0.39$ &  $14.057 \pm 0.011$ &  $0.690 \pm 0.025$ &   $0.481 \pm 0.025$ \\
  2007fb &  $54287.92 \pm 0.62$ &  $15.792 \pm 0.021$ &  $1.332 \pm 0.093$ &  $54288.90 \pm 0.65$ &  $15.668 \pm 0.024$ &  $0.726 \pm 0.049$ &   $0.119 \pm 0.032$ \\
  2007kk &  $54382.76 \pm 1.41$ &  $16.953 \pm 0.024$ &  $0.954 \pm 0.169$ &  $54385.52 \pm 0.98$ &  $16.993 \pm 0.017$ &  $0.559 \pm 0.063$ &  $-0.064 \pm 0.030$ \\
  2008ds &  $54651.90 \pm 0.24$ &  $15.263 \pm 0.009$ &  $0.957 \pm 0.030$ &  $54652.49 \pm 0.25$ &  $15.303 \pm 0.004$ &  $0.617 \pm 0.019$ &  $-0.042 \pm 0.010$ \\
  2008eo &  $54688.14 \pm 0.81$ &  $15.311 \pm 0.020$ &  $1.026 \pm 0.070$ &  $54689.74 \pm 0.30$ &  $15.220 \pm 0.006$ &  $0.675 \pm 0.016$ &   $0.074 \pm 0.021$ \\
  2008eq &  $54689.54 \pm 0.93$ &  $18.222 \pm 0.027$ &  $1.029 \pm 0.148$ &  $54691.77 \pm 1.22$ &  $18.141 \pm 0.029$ &  $0.576 \pm 0.092$ &   $0.064 \pm 0.040$ \\
  2008gg &  $54749.80 \pm 1.50$ &  $16.677 \pm 0.033$ &  $0.983 \pm 0.181$ &  $54752.39 \pm 1.20$ &  $16.523 \pm 0.029$ &  $0.570 \pm 0.130$ &   $0.130 \pm 0.044$ \\
  2008gl &  $54767.98 \pm 0.83$ &  $16.882 \pm 0.043$ &  $1.394 \pm 0.158$ &  $54769.51 \pm 1.32$ &  $16.870 \pm 0.039$ &  $0.704 \pm 0.101$ &   $0.005 \pm 0.058$ \\
  2008gp &  $54779.28 \pm 0.85$ &  $16.484 \pm 0.037$ &  $1.136 \pm 0.135$ &  $54780.97 \pm 1.16$ &  $16.610 \pm 0.038$ &  $0.631 \pm 0.108$ &  $-0.136 \pm 0.053$ \\
  2008hs &  $54812.80 \pm 0.52$ &  $15.932 \pm 0.106$ &  $1.991 \pm 0.160$ &  $54814.38 \pm 0.54$ &  $15.769 \pm 0.123$ &  $1.228 \pm 0.161$ &   $0.129 \pm 0.162$ \\
  2009dc &  $54946.34 \pm 0.80$ &  $15.148 \pm 0.014$ &  $0.713 \pm 0.060$ &  $54946.85 \pm 0.85$ &  $15.166 \pm 0.015$ &  $0.294 \pm 0.035$ &  $-0.020 \pm 0.021$ \\
  2009eu &  $54984.59 \pm 0.50$ &  $17.690 \pm 0.054$ &  $1.816 \pm 0.132$ &  $54986.86 \pm 0.68$ &  $17.464 \pm 0.041$ &  $1.006 \pm 0.091$ &   $0.179 \pm 0.068$ \\
  2009fv &  $54994.47 \pm 0.40$ &  $16.887 \pm 0.024$ &  $1.670 \pm 0.090$ &  $54998.15 \pm 1.33$ &  $16.775 \pm 0.022$ &  $0.767 \pm 0.123$ &   $0.069 \pm 0.032$ \\
  2009hs &  $55048.55 \pm 0.34$ &  $17.376 \pm 0.041$ &  $2.090 \pm 0.109$ &  $55051.00 \pm 0.32$ &  $17.170 \pm 0.030$ &  $1.186 \pm 0.058$ &   $0.136 \pm 0.051$ \\
  2009ig &  $55079.70 \pm 1.11$ &  $13.560 \pm 0.032$ &  $0.850 \pm 0.124$ &  $55082.78 \pm 0.44$ &  $13.427 \pm 0.013$ &  $0.682 \pm 0.023$ &   $0.095 \pm 0.034$ \\
  2009kq &  $55155.05 \pm 0.39$ &  $14.591 \pm 0.014$ &  $1.091 \pm 0.067$ &  $55156.49 \pm 0.24$ &  $14.540 \pm 0.010$ &  $0.658 \pm 0.023$ &   $0.037 \pm 0.017$ \\
  2010ao &  $55289.32 \pm 0.57$ &  $15.857 \pm 0.037$ &  $1.329 \pm 0.094$ &  $55290.55 \pm 0.59$ &  $15.921 \pm 0.024$ &  $0.693 \pm 0.053$ &  $-0.073 \pm 0.045$ \\
  2010ii &  $55480.46 \pm 0.21$ &  $16.207 \pm 0.011$ &  $1.034 \pm 0.317$ &  $55481.61 \pm 0.47$ &  $16.248 \pm 0.012$ &  $0.769 \pm 0.241$ &  $-0.052 \pm 0.016$ \\
  2010ju &  $55525.65 \pm 1.04$ &  $16.136 \pm 0.073$ &  $1.315 \pm 0.106$ &  $55526.39 \pm 1.01$ &  $15.628 \pm 0.056$ &  $0.715 \pm 0.053$ &   $0.505 \pm 0.092$ \\
   2011M &  $55593.45 \pm 0.26$ &  $15.225 \pm 0.014$ &  $1.136 \pm 0.050$ &  $55595.27 \pm 0.32$ &  $15.228 \pm 0.013$ &  $0.649 \pm 0.050$ &  $-0.023 \pm 0.019$ \\
  2011by &  $55690.56 \pm 0.68$ &  $12.906 \pm 0.018$ &  $1.085 \pm 0.095$ &  $55692.59 \pm 0.62$ &  $12.874 \pm 0.015$ &  $0.695 \pm 0.052$ &   $0.014 \pm 0.024$ \\
  2011ek &  $55789.58 \pm 0.85$ &  $14.504 \pm 0.123$ &  $1.272 \pm 0.190$ &  $55790.80 \pm 0.67$ &  $13.715 \pm 0.061$ &  $0.795 \pm 0.092$ &   $0.775 \pm 0.137$ \\
  2011fs &  $55832.32 \pm 0.69$ &  $15.357 \pm 0.009$ &  $0.808 \pm 0.071$ &  $55835.04 \pm 0.57$ &  $15.313 \pm 0.008$ &  $0.565 \pm 0.035$ &   $0.018 \pm 0.012$ \\
   2012Z &  $55965.90 \pm 0.38$ &  $14.662 \pm 0.026$ &  $1.199 \pm 0.074$ &  $55973.93 \pm 0.86$ &  $14.377 \pm 0.016$ &  $0.790 \pm 0.066$ &   $0.105 \pm 0.030$ \\
  2012cg &  $56081.36 \pm 0.26$ &  $12.115 \pm 0.012$ &  $0.906 \pm 0.032$ &  $56083.25 \pm 0.24$ &  $11.952 \pm 0.005$ &  $0.631 \pm 0.013$ &   $0.144 \pm 0.013$ \\
  2012ea &  $56157.89 \pm 0.11$ &  $15.848 \pm 0.009$ &  $1.945 \pm 0.028$ &  $56160.18 \pm 0.14$ &  $15.403 \pm 0.007$ &  $1.224 \pm 0.018$ &   $0.387 \pm 0.012$ \\
  2013bs &  $56406.88 \pm 1.68$ &  $16.697 \pm 0.090$ &  $1.533 \pm 0.144$ &  $56409.11 \pm 0.71$ &  $16.589 \pm 0.038$ &  $0.903 \pm 0.049$ &   $0.073 \pm 0.098$ \\
  2013dh &  $56463.02 \pm 0.62$ &  $17.507 \pm 0.069$ &  $1.554 \pm 0.155$ &  $56467.07 \pm 0.54$ &  $17.524 \pm 0.048$ &  $1.014 \pm 0.071$ &  $-0.151 \pm 0.084$ \\
  2013dy &  $56500.40 \pm 0.19$ &  $12.697 \pm 0.008$ &  $0.870 \pm 0.023$ &  $56501.84 \pm 0.34$ &  $12.578 \pm 0.005$ &  $0.609 \pm 0.021$ &   $0.109 \pm 0.010$ \\
  2013fw &  $56601.14 \pm 0.26$ &  $15.078 \pm 0.006$ &  $1.038 \pm 0.037$ &  $56603.53 \pm 0.29$ &  $15.059 \pm 0.006$ &  $0.630 \pm 0.021$ &  $-0.010 \pm 0.008$ \\
  2013gh &  $56527.13 \pm 0.41$ &  $14.434 \pm 0.028$ &  $1.223 \pm 0.050$ &  $56529.24 \pm 0.49$ &  $14.180 \pm 0.011$ &  $0.606 \pm 0.029$ &   $0.225 \pm 0.030$ \\
  2013gq &  $56384.64 \pm 0.66$ &  $14.738 \pm 0.029$ &  $1.229 \pm 0.154$ &  $56386.45 \pm 0.77$ &  $14.753 \pm 0.019$ &  $0.645 \pm 0.072$ &  $-0.035 \pm 0.035$ \\
  2013gy &  $56647.80 \pm 0.65$ &  $14.751 \pm 0.025$ &  $1.247 \pm 0.072$ &  $56650.05 \pm 0.55$ &  $14.803 \pm 0.006$ &  $0.644 \pm 0.034$ &  $-0.071 \pm 0.025$ \\
   2014J &  $56688.93 \pm 0.65$ &  $11.452 \pm 0.020$ &  $0.890 \pm 0.074$ &  $56689.71 \pm 0.50$ &  $10.237 \pm 0.017$ &  $0.553 \pm 0.033$ &   $1.211 \pm 0.026$ \\
   2015N &  $57222.81 \pm 0.27$ &  $14.853 \pm 0.025$ &  $1.109 \pm 0.078$ &  $57225.28 \pm 0.79$ &  $14.768 \pm 0.032$ &  $0.628 \pm 0.054$ &   $0.040 \pm 0.041$ \\
 2016coj &  $57547.15 \pm 0.19$ &  $13.082 \pm 0.007$ &  $1.329 \pm 0.030$ &  $57547.89 \pm 0.18$ &  $13.088 \pm 0.007$ &  $0.681 \pm 0.018$ &  $-0.010 \pm 0.010$ \\
 2016gcl &  $57647.90 \pm 1.63$ &  $16.227 \pm 0.023$ &  $0.741 \pm 0.126$ &  $57650.42 \pm 1.18$ &  $16.251 \pm 0.016$ &  $0.543 \pm 0.069$ &  $-0.044 \pm 0.028$ \\
 2016hvl &  $57709.70 \pm 0.47$ &  $14.392 \pm 0.022$ &  $1.037 \pm 0.055$ &  $57713.43 \pm 0.67$ &  $14.282 \pm 0.011$ &  $0.619 \pm 0.028$ &   $0.058 \pm 0.025$ \\
 2017drh &  $57891.14 \pm 0.44$ &  $16.691 \pm 0.022$ &  $1.370 \pm 0.065$ &  $57891.98 \pm 0.48$ &  $15.396 \pm 0.010$ &  $0.720 \pm 0.032$ &   $1.291 \pm 0.024$ \\
 2017erp &  $57934.53 \pm 0.22$ &  $13.336 \pm 0.008$ &  $1.086 \pm 0.031$ &  $57937.21 \pm 0.35$ &  $13.275 \pm 0.007$ &  $0.667 \pm 0.020$ &   $0.036 \pm 0.010$ \\
 2017glx &  $58007.78 \pm 0.25$ &  $14.228 \pm 0.009$ &  $0.780 \pm 0.026$ &  $58009.73 \pm 0.87$ &  $14.250 \pm 0.007$ &  $0.493 \pm 0.045$ &  $-0.037 \pm 0.011$ \\
 2017hbi &  $58045.80 \pm 0.61$ &  $16.580 \pm 0.019$ &  $0.710 \pm 0.074$ &  $58045.64 \pm 0.76$ &  $16.671 \pm 0.014$ &  $0.310 \pm 0.045$ &  $-0.091 \pm 0.024$ \\
 2018aoz &  $58222.46 \pm 0.58$ &  $12.761 \pm 0.030$ &  $1.305 \pm 0.124$ &  $58223.38 \pm 0.46$ &  $12.730 \pm 0.018$ &  $0.779 \pm 0.077$ &   $0.025 \pm 0.035$ \\
  2018gv &  $58149.38 \pm 0.31$ &  $12.751 \pm 0.015$ &  $0.853 \pm 0.037$ &  $58153.39 \pm 0.32$ &  $12.788 \pm 0.007$ &  $0.740 \pm 0.017$ &  $-0.125 \pm 0.017$ \\
\hline
\multicolumn{8}{l}{\textbf{Note:} Only those SNe from our sample where the fitting process described in Section~\ref{ssec:light-curve-properties} succeeded appear here.}
\end{tabular}
\end{table*}

\onecolumn
\begin{landscape}
{\footnotesize
\begin{longtable}{l|rrrr|rrrr}
\caption{Results of {\tt SNooPy} and {\tt MLCS2k2} fitting.\label{tab:model-fits}}\\
\hline
\hline
\multicolumn{1}{c}{} & \multicolumn{4}{c}{{\tt SNooPy} $E(B-V)$ Fitted Parameters} & \multicolumn{4}{c}{{\tt MLCS2k2} Fitted Parameters}\\
\multicolumn{1}{c}{SN} & \multicolumn{1}{c}{$t_{\rm max}$ (MJD)} & \multicolumn{1}{c}{$\Delta m_{15}$ (mag)} & \multicolumn{1}{c}{$E(B - V)_{\rm host}$ (mag)} & \multicolumn{1}{c}{$\mu$ (mag)} & \multicolumn{1}{c}{$t_0$ (MJD)} & \multicolumn{1}{c}{$\Delta$} & \multicolumn{1}{c}{$A_V$ (mag)} & \multicolumn{1}{c}{$\mu$ (mag)}\\
\hline
\endfirsthead
\hline
\hline
\multicolumn{1}{c}{} & \multicolumn{4}{c}{{\tt SNooPy} $E(B-V)$ Fitted Parameters} & \multicolumn{4}{c}{{\tt MLCS2k2} Fitted Parameters}\\
\multicolumn{1}{c}{SN} & \multicolumn{1}{c}{$t_{\rm max}$ (MJD)} & \multicolumn{1}{c}{$\Delta m_{15}$ (mag)} & \multicolumn{1}{c}{$E(B - V)_{\rm host}$ (mag)} & \multicolumn{1}{c}{$\mu$ (mag)} & \multicolumn{1}{c}{$t_0$ (MJD)} & \multicolumn{1}{c}{$\Delta$} & \multicolumn{1}{c}{$A_V$ (mag)} & \multicolumn{1}{c}{$\mu$ (mag)}\\
\hline
\endhead
\hline
\multicolumn{9}{c}{Table \thetable~continued}
\endfoot
\hline
\multicolumn{9}{l}{\textbf{Note:} Only those SNe from our sample where the fitting process described in Sections~\ref{sssec:ebv} or \ref{sssec:mlcs} succeeded appear here.}
\endlastfoot
  2005ki &  $53705.23 \pm 0.06$ &  $1.419 \pm 0.013$ &  $-0.011 \pm 0.009$ &  $34.666 \pm 0.013$ &  $53705.21 \pm 0.11$ &   $0.373 \pm 0.052$ &  $0.027 \pm 0.017$ &  $34.719 \pm 0.065$ \\
   2007F &  $54123.83 \pm 0.09$ &  $1.096 \pm 0.012$ &   $0.041 \pm 0.010$ &  $35.163 \pm 0.011$ &  $54123.13 \pm 0.10$ &  $-0.179 \pm 0.033$ &  $0.204 \pm 0.036$ &  $35.351 \pm 0.046$ \\
  2007bd &  $54207.12 \pm 0.50$ &  $1.351 \pm 0.067$ &   $0.010 \pm 0.037$ &  $35.748 \pm 0.050$ &  $54206.65 \pm 0.23$ &   $0.209 \pm 0.103$ &  $0.082 \pm 0.054$ &  $35.851 \pm 0.097$ \\
  2007bm &  $54225.02 \pm 0.15$ &  $1.224 \pm 0.014$ &   $0.588 \pm 0.011$ &  $32.635 \pm 0.019$ &  $54223.94 \pm 0.08$ &   $0.057 \pm 0.038$ &  $1.109 \pm 0.036$ &  $32.389 \pm 0.048$ \\
  2007fb &  $54287.48 \pm 0.16$ &  $1.353 \pm 0.016$ &   $0.100 \pm 0.009$ &  $34.657 \pm 0.017$ &  $54286.73 \pm 0.31$ &   $0.285 \pm 0.055$ &  $0.142 \pm 0.042$ &  $34.749 \pm 0.059$ \\
  2007fs &  $54293.70 \pm 0.42$ &  $0.879 \pm 0.015$ &   $0.015 \pm 0.013$ &  $34.505 \pm 0.014$ &  $54295.17 \pm 0.35$ &  $-0.161 \pm 0.028$ &  $0.116 \pm 0.032$ &  $34.649 \pm 0.044$ \\
  2007if &  $54338.39 \pm 0.86$ &  $0.768 \pm 0.029$ &   $0.034 \pm 0.026$ &  $36.133 \pm 0.033$ &  $54343.02 \pm 1.17$ &  $-0.350 \pm 0.062$ &  $0.384 \pm 0.068$ &  $36.245 \pm 0.123$ \\
  2007jg &  $54365.35 \pm 0.47$ &  $1.199 \pm 0.022$ &  $-0.021 \pm 0.024$ &  $36.493 \pm 0.032$ &  $54364.35 \pm 0.55$ &  $-0.025 \pm 0.060$ &  $0.092 \pm 0.051$ &  $36.616 \pm 0.071$ \\
  2007kk &  $54383.83 \pm 0.26$ &  $1.088 \pm 0.035$ &  $-0.004 \pm 0.022$ &  $36.267 \pm 0.025$ &  $54382.59 \pm 0.44$ &  $-0.340 \pm 0.040$ &  $0.168 \pm 0.071$ &  $36.558 \pm 0.066$ \\
   2008Y &  $54499.62 \pm 1.52$ &  $0.939 \pm 0.126$ &   $0.164 \pm 0.045$ &  $37.425 \pm 0.078$ &  $54498.33 \pm 1.66$ &  $-0.110 \pm 0.112$ &  $0.226 \pm 0.090$ &  $37.503 \pm 0.127$ \\
  2008dh &  $54625.56 \pm 0.67$ &  $0.924 \pm 0.035$ &   $0.026 \pm 0.024$ &  $36.282 \pm 0.020$ &  $54626.31 \pm 0.66$ &  $-0.124 \pm 0.052$ &  $0.077 \pm 0.044$ &  $36.436 \pm 0.077$ \\
  2008ds &  $54651.45 \pm 0.15$ &  $0.865 \pm 0.010$ &  $-0.013 \pm 0.007$ &  $34.746 \pm 0.011$ &  $54652.06 \pm 0.18$ &  $-0.270 \pm 0.023$ &  $0.045 \pm 0.027$ &  $34.975 \pm 0.039$ \\
  2008ek &  $54668.63 \pm 2.52$ &  $1.813 \pm 0.033$ &   $0.669 \pm 0.145$ &  $36.434 \pm 0.096$ &  $54662.46 \pm 1.63$ &   $1.213 \pm 0.141$ &  $0.220 \pm 0.135$ &  $35.997 \pm 0.120$ \\
  2008eo &  $54686.91 \pm 0.38$ &  $0.884 \pm 0.018$ &   $0.095 \pm 0.015$ &  $34.513 \pm 0.015$ &  $54688.23 \pm 0.30$ &  $-0.197 \pm 0.028$ &  $0.261 \pm 0.039$ &  $34.630 \pm 0.045$ \\
  2008eq &  $54689.59 \pm 0.28$ &  $0.971 \pm 0.032$ &   $0.207 \pm 0.015$ &  $37.155 \pm 0.036$ &  $54689.46 \pm 0.34$ &  $-0.227 \pm 0.053$ &  $0.444 \pm 0.048$ &  $37.277 \pm 0.068$ \\
  2008fk &  $54722.03 \pm 1.02$ &  $1.263 \pm 0.074$ &  $-0.197 \pm 0.067$ &  $37.749 \pm 0.091$ &  $54719.62 \pm 0.99$ &  $-0.229 \pm 0.084$ &  $0.028 \pm 0.020$ &  $37.967 \pm 0.087$ \\
  2008gg &  $54750.61 \pm 0.58$ &  $1.087 \pm 0.060$ &   $0.111 \pm 0.036$ &  $35.720 \pm 0.050$ &  $54749.06 \pm 0.72$ &  $-0.350 \pm 0.046$ &  $0.267 \pm 0.051$ &  $36.047 \pm 0.071$ \\
  2008gl &  $54766.97 \pm 0.27$ &  $1.178 \pm 0.027$ &   $0.124 \pm 0.012$ &  $35.917 \pm 0.024$ &  $54767.32 \pm 0.37$ &   $0.189 \pm 0.109$ &  $0.227 \pm 0.058$ &  $35.913 \pm 0.086$ \\
  2008go &  $54765.09 \pm 1.09$ &  $1.158 \pm 0.101$ &   $0.081 \pm 0.022$ &  $37.167 \pm 0.073$ &  $54764.78 \pm 0.65$ &   $0.002 \pm 0.118$ &  $0.191 \pm 0.062$ &  $37.317 \pm 0.109$ \\
  2008gp &  $54779.01 \pm 0.08$ &  $1.087 \pm 0.011$ &  $-0.048 \pm 0.008$ &  $35.909 \pm 0.009$ &  $54778.92 \pm 0.35$ &  $-0.106 \pm 0.064$ &  $0.051 \pm 0.035$ &  $36.094 \pm 0.073$ \\
  2008hs &  $54813.07 \pm 0.11$ &  $1.720 \pm 0.012$ &   $0.103 \pm 0.017$ &  $34.836 \pm 0.033$ &  $54812.83 \pm 0.08$ &   $1.181 \pm 0.042$ &  $0.011 \pm 0.010$ &  $34.297 \pm 0.058$ \\
   2009D &  $54841.02 \pm 0.54$ &  $0.932 \pm 0.041$ &   $0.026 \pm 0.017$ &  $35.140 \pm 0.018$ &  $54841.93 \pm 1.47$ &  $-0.138 \pm 0.108$ &  $0.125 \pm 0.056$ &  $35.248 \pm 0.080$ \\
  2009al &  $54896.75 \pm 0.35$ &  $1.106 \pm 0.029$ &   $0.264 \pm 0.022$ &  $35.127 \pm 0.023$ &  $54894.38 \pm 0.79$ &  $-0.264 \pm 0.043$ &  $0.503 \pm 0.054$ &  $35.305 \pm 0.066$ \\
  2009dc &                  ... &                ... &                 ... &                 ... &  $54945.34 \pm 0.16$ &  $-0.693 \pm 0.017$ &  $0.348 \pm 0.031$ &  $34.687 \pm 0.037$ \\
  2009ee &  $54951.64 \pm 0.80$ &  $1.273 \pm 0.021$ &   $0.210 \pm 0.056$ &  $36.209 \pm 0.042$ &  $54949.75 \pm 0.71$ &   $0.466 \pm 0.086$ &  $0.085 \pm 0.076$ &  $36.068 \pm 0.085$ \\
  2009eu &  $54984.30 \pm 0.12$ &  $1.787 \pm 0.013$ &   $0.279 \pm 0.021$ &  $35.924 \pm 0.025$ &  $54984.38 \pm 0.20$ &   $1.199 \pm 0.058$ &  $0.056 \pm 0.046$ &  $35.606 \pm 0.063$ \\
  2009hs &  $55048.76 \pm 0.11$ &  $1.798 \pm 0.013$ &   $0.269 \pm 0.025$ &  $35.728 \pm 0.024$ &  $55048.51 \pm 0.13$ &   $1.259 \pm 0.034$ &  $0.018 \pm 0.012$ &  $35.404 \pm 0.048$ \\
  2009ig &                  ... &                ... &                 ... &                 ... &  $55079.47 \pm 0.09$ &  $-0.354 \pm 0.023$ &  $0.123 \pm 0.029$ &  $33.167 \pm 0.039$ \\
  2009kq &  $55154.71 \pm 0.15$ &  $1.103 \pm 0.018$ &   $0.017 \pm 0.011$ &  $33.834 \pm 0.014$ &  $55154.69 \pm 0.20$ &  $-0.062 \pm 0.035$ &  $0.154 \pm 0.036$ &  $33.954 \pm 0.047$ \\
  2010ao &  $55288.84 \pm 0.30$ &  $1.129 \pm 0.031$ &   $0.037 \pm 0.019$ &  $35.122 \pm 0.028$ &  $55288.75 \pm 0.26$ &   $0.009 \pm 0.056$ &  $0.195 \pm 0.048$ &  $35.147 \pm 0.069$ \\
  2010ii &                  ... &                ... &                 ... &                 ... &  $55481.48 \pm 0.19$ &   $0.315 \pm 0.116$ &  $0.031 \pm 0.022$ &  $35.457 \pm 0.096$ \\
  2010ju &  $55524.52 \pm 0.29$ &  $1.175 \pm 0.032$ &   $0.440 \pm 0.023$ &  $34.477 \pm 0.044$ &  $55524.07 \pm 0.23$ &  $-0.044 \pm 0.070$ &  $0.931 \pm 0.122$ &  $34.315 \pm 0.107$ \\
   2011M &  $55593.49 \pm 0.12$ &  $1.119 \pm 0.025$ &   $0.048 \pm 0.012$ &  $34.482 \pm 0.019$ &  $55593.14 \pm 0.15$ &  $-0.008 \pm 0.060$ &  $0.183 \pm 0.107$ &  $34.475 \pm 0.082$ \\
  2011by &  $55690.78 \pm 0.09$ &  $1.091 \pm 0.010$ &   $0.094 \pm 0.011$ &  $32.077 \pm 0.011$ &  $55690.33 \pm 0.09$ &  $-0.037 \pm 0.029$ &  $0.300 \pm 0.028$ &  $32.071 \pm 0.042$ \\
  2011df &  $55715.10 \pm 0.30$ &  $0.943 \pm 0.019$ &   $0.056 \pm 0.010$ &  $34.161 \pm 0.013$ &  $55716.02 \pm 0.41$ &  $-0.162 \pm 0.038$ &  $0.215 \pm 0.053$ &  $34.261 \pm 0.056$ \\
  2011dl &  $55738.35 \pm 0.50$ &  $1.089 \pm 0.046$ &   $0.169 \pm 0.033$ &  $36.079 \pm 0.031$ &  $55736.95 \pm 0.77$ &  $-0.278 \pm 0.060$ &  $0.439 \pm 0.053$ &  $36.228 \pm 0.064$ \\
  2011ek &  $55789.74 \pm 0.10$ &  $1.522 \pm 0.021$ &   $0.503 \pm 0.012$ &  $32.250 \pm 0.026$ &  $55789.14 \pm 0.15$ &   $0.562 \pm 0.073$ &  $0.979 \pm 0.101$ &  $31.821 \pm 0.090$ \\
  2011fe &  $55815.22 \pm 0.06$ &  $1.096 \pm 0.005$ &  $-0.006 \pm 0.005$ &  $29.228 \pm 0.006$ &                  ... &                 ... &                ... &                 ... \\
  2011fs &  $55833.25 \pm 0.19$ &  $0.911 \pm 0.016$ &   $0.064 \pm 0.012$ &  $34.620 \pm 0.013$ &  $55832.95 \pm 0.26$ &  $-0.310 \pm 0.026$ &  $0.209 \pm 0.044$ &  $34.825 \pm 0.045$ \\
   2012E &  $55949.73 \pm 0.79$ &  $1.343 \pm 0.051$ &   $0.117 \pm 0.026$ &  $34.682 \pm 0.018$ &  $55948.57 \pm 1.67$ &   $0.343 \pm 0.162$ &  $0.200 \pm 0.107$ &  $34.612 \pm 0.111$ \\
  2012cg &  $56082.40 \pm 0.06$ &  $1.060 \pm 0.006$ &   $0.173 \pm 0.007$ &  $31.054 \pm 0.006$ &  $56081.62 \pm 0.06$ &  $-0.254 \pm 0.021$ &  $0.543 \pm 0.026$ &  $31.120 \pm 0.035$ \\
  2012dn &  $56132.44 \pm 0.00$ &  $0.940 \pm 0.028$ &   $0.458 \pm 0.025$ &  $32.725 \pm 0.023$ &  $56134.14 \pm 0.57$ &  $-0.181 \pm 0.050$ &  $0.841 \pm 0.044$ &  $32.744 \pm 0.076$ \\
  2012ea &  $56158.17 \pm 0.06$ &  $1.821 \pm 0.000$ &   $0.389 \pm 0.009$ &  $33.883 \pm 0.009$ &  $56158.11 \pm 0.07$ &   $1.396 \pm 0.022$ &  $0.048 \pm 0.029$ &  $33.496 \pm 0.034$ \\
  2013bs &  $56406.52 \pm 0.22$ &  $1.507 \pm 0.027$ &   $0.067 \pm 0.016$ &  $35.516 \pm 0.026$ &  $56406.38 \pm 0.23$ &   $0.686 \pm 0.062$ &  $0.031 \pm 0.023$ &  $35.332 \pm 0.071$ \\
  2013dr &  $56486.38 \pm 1.02$ &  $0.987 \pm 0.089$ &   $0.153 \pm 0.044$ &  $34.181 \pm 0.174$ &  $56486.04 \pm 1.44$ &  $-0.217 \pm 0.080$ &  $0.332 \pm 0.075$ &  $34.328 \pm 0.160$ \\
  2013dy &  $56501.48 \pm 0.07$ &  $0.995 \pm 0.006$ &   $0.123 \pm 0.005$ &  $31.773 \pm 0.008$ &  $56500.13 \pm 0.06$ &  $-0.325 \pm 0.016$ &  $0.420 \pm 0.046$ &  $31.923 \pm 0.040$ \\
  2013ex &  $56529.48 \pm 0.40$ &  $1.013 \pm 0.030$ &   $0.044 \pm 0.019$ &  $33.648 \pm 0.033$ &  $56530.06 \pm 0.60$ &  $-0.066 \pm 0.051$ &  $0.142 \pm 0.061$ &  $33.770 \pm 0.067$ \\
  2013fa &  $56536.19 \pm 0.23$ &  $1.140 \pm 0.023$ &   $0.297 \pm 0.011$ &  $34.320 \pm 0.018$ &  $56535.17 \pm 0.46$ &  $-0.114 \pm 0.037$ &  $0.607 \pm 0.044$ &  $34.347 \pm 0.053$ \\
  2013fw &  $56601.68 \pm 0.10$ &  $1.085 \pm 0.014$ &   $0.031 \pm 0.009$ &  $34.314 \pm 0.014$ &  $56600.81 \pm 0.09$ &  $-0.277 \pm 0.027$ &  $0.189 \pm 0.038$ &  $34.588 \pm 0.043$ \\
  2013gh &  $56529.10 \pm 0.32$ &  $1.142 \pm 0.032$ &   $0.366 \pm 0.021$ &  $33.146 \pm 0.028$ &  $56528.32 \pm 0.08$ &   $0.112 \pm 0.036$ &  $0.798 \pm 0.034$ &  $32.808 \pm 0.046$ \\
  2013gq &  $56385.29 \pm 0.18$ &  $1.233 \pm 0.015$ &  $-0.003 \pm 0.016$ &  $33.946 \pm 0.029$ &  $56384.22 \pm 0.17$ &   $0.008 \pm 0.043$ &  $0.096 \pm 0.040$ &  $34.119 \pm 0.056$ \\
  2013gy &  $56649.21 \pm 0.12$ &  $1.125 \pm 0.011$ &   $0.073 \pm 0.012$ &  $34.024 \pm 0.012$ &  $56648.36 \pm 0.07$ &   $0.026 \pm 0.032$ &  $0.280 \pm 0.034$ &  $33.947 \pm 0.044$ \\
   2014J &  $56690.04 \pm 0.13$ &  $0.952 \pm 0.020$ &   $1.179 \pm 0.014$ &  $28.415 \pm 0.025$ &  $56689.20 \pm 0.09$ &  $-0.219 \pm 0.025$ &  $2.194 \pm 0.048$ &  $27.865 \pm 0.047$ \\
  2014ai &  $56745.96 \pm 0.23$ &  $1.490 \pm 0.058$ &   $0.128 \pm 0.025$ &  $35.097 \pm 0.054$ &  $56744.75 \pm 0.51$ &   $0.191 \pm 0.124$ &  $0.277 \pm 0.069$ &  $35.308 \pm 0.115$ \\
  2014ao &  $56766.17 \pm 0.34$ &  $0.977 \pm 0.032$ &   $0.820 \pm 0.014$ &  $34.759 \pm 0.033$ &  $56765.77 \pm 0.61$ &  $-0.204 \pm 0.088$ &  $1.441 \pm 0.053$ &  $34.515 \pm 0.076$ \\
  2014bj &  $56796.73 \pm 0.55$ &  $1.108 \pm 0.038$ &   $0.044 \pm 0.021$ &  $36.632 \pm 0.024$ &  $56795.74 \pm 0.65$ &  $-0.168 \pm 0.071$ &  $0.169 \pm 0.062$ &  $36.824 \pm 0.078$ \\
   2015N &  $57223.19 \pm 0.15$ &  $1.087 \pm 0.015$ &   $0.181 \pm 0.012$ &  $33.877 \pm 0.017$ &  $57222.89 \pm 0.21$ &  $-0.134 \pm 0.045$ &  $0.430 \pm 0.142$ &  $33.865 \pm 0.098$ \\
 2016coj &  $57547.89 \pm 0.23$ &  $1.131 \pm 0.034$ &   $0.121 \pm 0.018$ &  $32.306 \pm 0.026$ &  $57547.83 \pm 0.06$ &   $0.613 \pm 0.033$ &  $0.024 \pm 0.017$ &  $31.969 \pm 0.042$ \\
 2016fbk &  $57624.94 \pm 0.41$ &  $0.993 \pm 0.034$ &   $0.241 \pm 0.016$ &  $36.180 \pm 0.036$ &  $57625.09 \pm 0.51$ &  $-0.046 \pm 0.049$ &  $0.468 \pm 0.046$ &  $36.156 \pm 0.062$ \\
 2016gcl &  $57649.84 \pm 0.53$ &  $0.849 \pm 0.024$ &   $0.025 \pm 0.032$ &  $35.608 \pm 0.056$ &  $57649.62 \pm 0.38$ &  $-0.366 \pm 0.029$ &  $0.126 \pm 0.041$ &  $35.852 \pm 0.050$ \\
 2016gdt &  $57641.53 \pm 1.08$ &  $1.822 \pm 0.001$ &   $0.677 \pm 0.064$ &  $35.832 \pm 0.051$ &  $57640.11 \pm 0.91$ &   $1.499 \pm 0.077$ &  $0.147 \pm 0.103$ &  $35.492 \pm 0.076$ \\
 2016hvl &  $57711.00 \pm 0.12$ &  $1.123 \pm 0.014$ &   $0.116 \pm 0.012$ &  $33.420 \pm 0.014$ &  $57709.48 \pm 0.11$ &  $-0.281 \pm 0.026$ &  $0.343 \pm 0.115$ &  $33.634 \pm 0.075$ \\
 2017cfd &                  ... &                ... &                 ... &                 ... &  $57844.39 \pm 0.13$ &   $0.093 \pm 0.048$ &  $0.504 \pm 0.041$ &  $33.693 \pm 0.058$ \\
 2017drh &  $57890.60 \pm 0.09$ &  $1.340 \pm 0.011$ &   $1.601 \pm 0.014$ &  $32.687 \pm 0.013$ &  $57889.72 \pm 0.10$ &   $0.112 \pm 0.036$ &  $2.558 \pm 0.045$ &  $32.169 \pm 0.053$ \\
 2017dws &  $57867.60 \pm 1.20$ &  $0.882 \pm 0.030$ &  $-0.051 \pm 0.051$ &  $37.935 \pm 0.038$ &  $57869.18 \pm 1.29$ &  $-0.339 \pm 0.091$ &  $0.075 \pm 0.049$ &  $38.258 \pm 0.135$ \\
 2017erp &  $57935.15 \pm 0.06$ &  $1.118 \pm 0.006$ &   $0.099 \pm 0.006$ &  $32.405 \pm 0.006$ &  $57933.88 \pm 0.06$ &  $-0.234 \pm 0.021$ &  $0.444 \pm 0.039$ &  $32.503 \pm 0.039$ \\
 2017fgc &  $57955.52 \pm 0.38$ &  $0.840 \pm 0.008$ &   $0.081 \pm 0.016$ &  $32.775 \pm 0.035$ &  $57955.78 \pm 0.41$ &  $-0.324 \pm 0.026$ &  $0.305 \pm 0.033$ &  $32.866 \pm 0.049$ \\
 2017glx &                  ... &                ... &                 ... &                 ... &  $58009.16 \pm 0.16$ &  $-0.196 \pm 0.025$ &  $0.174 \pm 0.044$ &  $33.684 \pm 0.044$ \\
 2017hbi &                  ... &                ... &                 ... &                 ... &  $58044.44 \pm 0.14$ &  $-0.692 \pm 0.017$ &  $0.186 \pm 0.035$ &  $36.347 \pm 0.041$ \\
 2018aoz &  $58221.43 \pm 0.14$ &  $1.283 \pm 0.008$ &  $-0.079 \pm 0.011$ &  $32.001 \pm 0.014$ &  $58221.27 \pm 0.19$ &   $0.187 \pm 0.040$ &  $0.018 \pm 0.012$ &  $32.107 \pm 0.053$ \\
  2018gv &  $58150.11 \pm 0.08$ &  $1.006 \pm 0.011$ &  $-0.046 \pm 0.006$ &  $32.164 \pm 0.013$ &  $58149.59 \pm 0.11$ &  $-0.169 \pm 0.024$ &  $0.035 \pm 0.020$ &  $32.363 \pm 0.038$ \\
\end{longtable}}
\end{landscape}
\twocolumn

\onecolumn
\subsection{Natural-System Light Curves}
\label{app:nat-sys-light-curves}

SN light curves have long been released on the Landolt system (e.g., CfA1, CfA2, G10), thus allowing for easy comparison between datasets from different telescopes. Indeed, we analysed our light curves only after transforming to the Landolt system --- a decision motivated largely by the fact that our dataset is derived from observations collected with four distinct telescope/CCD/filter combinations. However, there are instances where natural-system light curves are more attractive. Since the stellar SEDs that are used to derive colour terms do not accurately reflect those of SNe~Ia, SN photometry transformed using such colour terms will not necessarily be on the Landolt system. Conventionally, second-order ``$S$-corrections'' are performed to properly account for the SN SED by using a selected spectral series~\citep{scorr}, but many groups are now releasing their low-$z$ SN~Ia photometry datasets in the natural systems of their telescopes along with the transmission curves of their photometry systems (e.g., CfA3, CfA4, CSP1-3). Thus, given a spectral series \citep[e.g.,][]{Hsiao} and transmission functions, one can transform photometry from one system to another without the need for colour corrections. In turn, this should provide less scatter in SN flux measurements. 

The aforementioned benefits motivate us to release our photometric dataset (see Section~\ref{sec:results}) in the relevant natural systems in addition to the Landolt system. A table of natural-system magnitudes analogous to Table~\ref{tab:phot-sample} is available for our entire dataset, with a sample given in Table~\ref{tab:phot-sample-natural}. We reiterate that owing to changes in the observing equipment, there are four transmission curves (KAIT3, KAIT4, Nickel1, Nickel2) for each bandpass. Any analysis of the dataset as a whole should therefore be done either on the Landolt system or after transforming all of the data to a common system \citep[see Appendix A of][]{G12}. Transmission curves for all filter and system combinations covered by our dataset are archived with the journal and available online in our SNDB.

\begin{table*}
\caption{Natural-System Photometry of SN 2008ds.\label{tab:phot-sample-natural}}
\begin{tabular}{lccccccr}
\hline
\hline
SN & MJD & $B$ (mag) & $V$ (mag) & $R$ (mag) & $I$ (mag) & {\it Clear} (mag) & System\\
\hline
 2008ds &  54645.47 &                 ... &                 ... &                 ... &                 ... &  $15.700 \pm 0.033$ &    kait4 \\
 2008ds &  54646.47 &                 ... &                 ... &                 ... &                 ... &  $15.574 \pm 0.024$ &    kait4 \\
 2008ds &  54647.46 &  $15.615 \pm 0.012$ &  $15.629 \pm 0.010$ &  $15.597 \pm 0.011$ &  $15.742 \pm 0.018$ &  $15.501 \pm 0.010$ &    kait4 \\
 2008ds &  54650.47 &  $15.501 \pm 0.014$ &  $15.488 \pm 0.010$ &  $15.476 \pm 0.012$ &  $15.762 \pm 0.015$ &                 ... &    kait4 \\
 2008ds &  54653.13 &  $15.482 \pm 0.009$ &  $15.474 \pm 0.005$ &  $15.418 \pm 0.005$ &  $15.768 \pm 0.008$ &                 ... &  nickel1 \\
 2008ds &  54653.44 &  $15.489 \pm 0.018$ &  $15.471 \pm 0.010$ &  $15.439 \pm 0.010$ &  $15.823 \pm 0.016$ &                 ... &    kait4 \\
 2008ds &  54655.13 &  $15.565 \pm 0.008$ &  $15.515 \pm 0.006$ &  $15.456 \pm 0.006$ &  $15.840 \pm 0.009$ &                 ... &  nickel1 \\
 2008ds &  54655.48 &  $15.559 \pm 0.016$ &  $15.510 \pm 0.012$ &  $15.471 \pm 0.013$ &  $15.919 \pm 0.022$ &                 ... &    kait4 \\
 2008ds &  54658.13 &  $15.695 \pm 0.008$ &  $15.611 \pm 0.006$ &  $15.548 \pm 0.005$ &  $15.978 \pm 0.008$ &                 ... &  nickel1 \\
 2008ds &  54662.16 &  $15.975 \pm 0.011$ &  $15.785 \pm 0.005$ &                 ... &                 ... &                 ... &  nickel1 \\
 \hline
 \multicolumn{8}{p{15cm}}{\textbf{Note:} First 10 epochs of natural-system \emph{BVRI} + unfiltered photometry of SN 2008ds. This table shows the form and content organisation of a much larger table that covers each epoch of photometry for each SN in our dataset. The full table is available in the online version of this article.}
\end{tabular}
\end{table*}


\bsp	
\label{lastpage}
\end{document}